\documentclass[11pt,a4paper]{article}
\usepackage{jheppub,amsmath,  amssymb,slashed,url,bm,textgreek,upgreek}
\usepackage{graphicx}
\usepackage{epstopdf}
\def\t{{ \sf t}} 

\def\star{\diamond}
\def\eO{{\eusm O}}

\def\J{\mathcal J}
\def\neg{\negthinspace}
\def\a{a}

\def\be{\begin{equation}}
\def\ee{\end{equation}}

\def\XX{{\mathcal Y}}
\def\NS{{\mathrm {NS}}}
\def\Ra{{\mathrm R}}

\def\BBer{{\textit{Ber}}}
\def\X{{\eusm X}}

\def\hat{\widehat}

\def\fN{{\mathfrak N}}
\def\frak{\mathfrak}
\def\h{\widehat}

\def\D{{\mathcal D}}
\def\S{{\mathcal S}}
\def\SIgma{\Sigma}
\def\YY{{\eusm Y}}
\def\V{{\mathcal V}}
\def\O{{\mathcal O}}

\def\red{{\mathrm{red}}}

\def\d{{\mathrm d}}

\def\R{{\mathbb R}}
\def\C{{\mathbb C}}

\def\D{{\mathcal D}}
\def\[{\bigl [}
\def\]{\bigr ]}

\def\N{{\mathcal N}}
\def\T{{\mathcal T}}
\def\F{{\mathcal F}}
\def\tr{{\mathrm {tr}}}
\def\Z{{\mathbb Z}}

\def\sQ{{\sf Q}_{\sf B}}

\def\L{{\mathcal  L}}
\def\t{\widetilde }
\def\h{\widehat}

\def\V{{\mathcal V}}
\def\I{{\mathcal I}}

\def\M{{\mathcal M}}

\def\MM{{\mathfrak M}}

\def\W{{\mathcal W}}

\def\Ra{{\mathrm{R}}}

\def\tphi{\text{\textphi}}
\def\trho{\text{\textmu}}

\def\st{{\mathrm{st}}}

\def\bar{\overline}

\def\neg{\negthinspace}

\def\Y{{\eusm Y}}
\font\teneurm=eurm10 \font\seveneurm=eurm7  \font\fiveeurm=eurm5
\newfam\eurmfam
\textfont\eurmfam=\teneurm \scriptfont\eurmfam=\seveneurm
\scriptscriptfont\eurmfam=\fiveeurm

\font\teneusm=eusm10 \font\seveneusm=eusm7 \font\fiveeusm=eusm5
\newfam\eusmfam
\textfont\eusmfam=\teneusm \scriptfont\eusmfam=\seveneusm
\scriptscriptfont\eusmfam=\fiveeusm
\def\eusm#1{{\fam\eusmfam\relax#1}}
\font\tencmmib=cmmib10 \skewchar\tencmmib='177
\font\sevencmmib=cmmib7 \skewchar\sevencmmib='177
\font\fivecmmib=cmmib5 \skewchar\fivecmmib='177
\newfam\cmmibfam
\textfont\cmmibfam=\tencmmib \scriptfont\cmmibfam=\sevencmmib
\scriptscriptfont\cmmibfam=\fivecmmib

\title{Superstring Perturbation Theory Via Super Riemann Surfaces: An Overview\footnote{The arXiv version of this article was originally entitled ``More on
Superstring Perturbation Theory.''}}

 \author{Edward Witten}
\affiliation{School of Natural Sciences, Institute for Advanced Study,\\ 1 Einstein Drive, Princeton, NJ 08540 USA}
\abstract{  This article is devoted to an overview of superstring perturbation theory from the point of view of super Riemann surfaces. 
We aim to elucidate some of the subtleties of superstring perturbation that caused difficulty in the early literature, 
 focusing on a concrete example -- the $SO(32)$ heterotic string compactified on a  Calabi-Yau
manifold, with the spin connection embedded in the gauge group.  This model is known to be a significant
test case for superstring perturbation theory. Supersymmetry is spontaneously broken at 1-loop order, and to treat correctly the 
supersymmetry-breaking effects that arise at 1- and 2-loop order requires a precise formulation of the procedure for integration
over supermoduli space.  In this paper, we aim as much as possible for an informal explanation, though at some points we provide
more detailed explanations that can be omitted on first reading.}

\begin{document}\maketitle

\section{Introduction}\label{intro}

String perturbation theory is based on a generalization from point particles and Feynman graphs to strings and Riemann surfaces.  
It has the remarkable property of preserving the general properties of relativistic quantum field theory, while eliminating the ultraviolet
region and forcing the inclusion of gravity.  For historical references, see \cite{Birth}.

The generalization from bosonic string theory to superstrings eliminates  infrared instabilities and leads to a theory with
a well-behaved perturbation theory, describing quantum gravity unified with other fields and forces.   

The basic foundations of superstring
perturbation theory -- including superconformal symmetry, modular invariance, worldsheet anomaly cancellation, and fermion vertex
operators -- were all well established by the mid-1980's.  
Roughly speaking, to complete that story in a natural way only requires a couple of steps:

{\it (A)} One should formulate all essential arguments, and especially those that involve integration by parts, 
on the moduli space of super Riemann surfaces, and not on the moduli space  of ordinary Riemann surfaces.

{\it (B)} One needs a careful treatment of integrals that are only conditionally convergent in the infrared region. 
The supersymmetric version of the Deligne-Mumford compactification of moduli space provides a natural infrared regulator.

To explain these points in the abstract can be rather dry, and, if one chooses to fill in details, also long \cite{Revisited}.
The purpose of this paper\footnote{The article is based on a lecture presented at the conference {\it The Search For 
Fundamental Physics:
Higgs Bosons And Supersymmetry}, in honor of Michael Dine and Howard Haber 
(University of California at Santa Cruz,
January 4-6, 2013).}  is to give a more informal explanation in the context of a model -- or more precisely a class of models --
 that is  known to give a significant test case
for superstring perturbation theory.  In the most basic case, we consider the $SO(32)$ heterotic string 
compactified on a Calabi-Yau
threefold, with the spin connection embedded in the gauge group in the standard fashion; this was first studied in \cite{DIS,ADS},
with subsequent work in \cite{GrSe,AS,DD}, following an earlier analysis of the associated effective field theory \cite{DSW}.  
This example
is a prototype for a large class of heterotic string compactifications to four dimensions that are supersymmetric at tree level but
have an anomalous $U(1)$ gauge field.  
 The loop corrections
that cancel the anomaly also trigger the spontaneous breaking of supersymmetry, giving 
 the only known method of supersymmetry breakdown by loop effects in superstring perturbation theory.
Such models turn out to  provide an important test case for arguments that claim
to show why supersymmetry is valid in loops.  Oversimplified arguments can easily give the wrong answer
when applied to these models. 

In analyzing this class of models, we will treat three topics:
\begin{enumerate}
\item the mass splitting between bosons and fermions that arises at one-loop order;
\item the vacuum energy that arises at two-loop order;
\item
 the mechanism by which a Goldstone fermion appears in supersymmetric Ward identities, signaling the spontaneous
breakdown of supersymmetry.\end{enumerate}

The three points are treated respectively in sections \ref{mass}, \ref{dilaton}, and \ref{supersym}.
In section \ref{mass}, we begin by summarizing the insights of the original papers \cite{DIS,ADS,GrSe,AS,DSW}.
Then we go on and explain how this example illustrates a general procedure
to regularize conditionally convergent integrals in superstring perturbation theory.  
In section \ref{dilaton}, we show that the two-loop vacuum energy in the same model can be understood by the same
methods.  In section \ref{supersym}, we first explain the general formulation of a supersymmetric Ward identity in superstring
perturbation theory, along the lines of section 8 of \cite{Revisited},
 and then implement this in detail at the one-loop level in our illustrative class of models, showing the appearance of a Goldstone
 fermion contribution. 

We aim in this paper for an informal explanation, though sections \ref{dilaton} and \ref{supersym} both have concluding
sections with technical details.  
The reader who works through the present paper -- or even most of it, without the more technical  parts --
should emerge with a fairly clear picture of some of the points that caused difficulty in the literature of the 1980's on superstring
perturbation theory.  
  Among other things, it should become clear  that the  phenomena are best described in 
terms of the full set of
bosonic and fermionic variables. 

The short version of this paper consists of section 2, which suffices for an overview of many of the ideas.
The mid-length version consists of omitting the technical sections \ref{details} and \ref{goldstonefermion}.

The models considered here, since they do have a dilaton tadpole at two-loop order,
are not models in which superstring perturbation theory works to all orders, at least not in its usual form.   But they illustrate some  essential subtleties
of superstring perturbation theory in a particularly simple way.  Once one understands these subtleties,
one is well-placed to demonstrate that superstring
perturbation theory works to all orders in those models in which tadpoles and supersymmetry-breaking effects do not arise.

A procedure to generalize superstring perturbation theory to describe vacuum shifts that are necessary when tadpoles appear has been developed recently
\cite{ASen}.  (This procedure can certainly be restated in terms of super Riemann surfaces, though this has not yet been done.)  We do not consider such issues here.
Our goal is only to explain how ``integration over moduli space'' when implemented in the superworld resolves various issues that caused confusion in the literature of the 1980's.

\section{The Mass Splitting}\label{mass}

\subsection{Review Of Effective Field Theory}\label{effreview}

We begin by reviewing the models of interest in the context of effective field theory \cite{DSW}.
We consider compactification of the heterotic string to four dimensions on a 
Calabi-Yau three-fold
$\XX$, with the $SU(3)$ holonomy group embedded in the gauge group in the usual fashion.    In the case of the 
$E_8\times E_8$ heterotic
string, the embedding identifies the $SU(3)$ holonomy group with the first factor of the subgroup
$SU(3)\times E_6\times E_8\subset E_8\times E_8$.  The unbroken subgroup in four dimensions
is $E_6\times E_8$.  With minor modifications, this construction leads to semirealistic models of
particle physics.

We will consider instead the same construction in the $SO(32)$ (or more precisely $\mathrm{Spin}(32)/\Z_2$)
heterotic string.  In this case, the relevant subgroup is $SU(3)\times U(1)\times SO(26)\subset SO(32)$, and the unbroken
subgroup is $U(1)\times SO(26)$.   Generically, this $U(1)$ is anomalous.  For
example, there is a $U(1)\cdot SO(26)^2$ anomaly with a coefficient that is a multiple of the Euler characteristic of $\XX$.  
The anomaly leads to the issues examined in this paper.  
Such anomalous $U(1)$'s frequently arise in supersymmetric compactifications of the heterotic string, including 
semirealistic ones.  The models that we have described illustrate the relevant issues in a simple context.

In string theory, the anomaly is canceled by the Green-Schwarz mechanism.  This depends upon the fact that
at one-loop order, Green-Schwarz interactions such as $I=\frac{1}{2\cdot 4!(2\pi)^5}\int_{\R^4\times \XX} B\wedge \tr\,F^4$ are generated.
Here $B$ is the usual two-form field of the Neveu-Schwarz sector, $F$ is the $SO(32)$ gauge field strength,
$\R^4\times \XX$ is the ten-dimensional spacetime, and the trace is taken in the fundamental representation of 
$SO(32)$.  Assuming that \begin{equation}\label{volf}p=\frac{1}{48\pi^3}\int_\XX\tr_{SU(3)}\,F^3\end{equation}
 is nonzero (this integral is one-half  the Euler characteristic of
$\XX$), the interaction $I$ reduces in four dimensions to $I_4=(p/(2\pi)^2)\int_{\R^4}B\wedge F$, where
henceforth $F$  is the field strength of the anomalous
$U(1)$.  

The effect of the interaction $I_4$ is to cause the $U(1)$ photon, which we will call
$A$, to become massive.  To understand this mass generation in a possibly more familiar way,
we can dualize the purely four-dimensional part of $B$ to a periodic 
spin-zero field $a$.  The $B\wedge F$ interaction
dualizes to $\partial_\mu a\cdot A^\mu$.  This means that including one- and two-loop effects the kinetic
energy of $a$ is not $\partial_\mu a\partial^\mu a$, but
\begin{equation}\label{zelbo} D_\mu a\cdot D^\mu a =(\partial_\mu a+pA_\mu)(\partial^\mu a+pA^\mu). \end{equation}
Accordingly, the field $a$ is not gauge-invariant; a gauge transformation $A_\mu\to A_\mu-\partial_\mu s$ must be accompanied
by $a\to a+ps$, and the field $e^{ia}$ has charge $p$.

From the standpoint of spacetime supersymmetry, the field $a$ is the imaginary part of a chiral multiplet
\begin{equation}\label{elbo} S(x^\mu|\theta^\alpha)=e^{-2\phi}-ia+\theta^\alpha \kappa_\alpha+\dots,\end{equation}
where the four-dimensional string coupling constant is $g_\st=e^\phi$;  we write 
$x^\mu, \,\mu=0,\dots,3$ and $\theta^\alpha$,
 $\alpha=1,2$ for bose and fermi coordinates
of chiral superspace; and $\kappa_\alpha$ is a fermi field of spin $1/2$.  $Z=e^{-S}$ is a charged chiral multiplet of $U(1)$
charge $p$.   We call $\phi$ and $\kappa$ the dilaton and dilatino.
There is no way to make $S$ or $Z$ vanish in the context of superstring perturbation theory, so inevitably
the $U(1)$ gauge symmetry is spontaneously broken.  
 
This symmetry-breaking mechanism has the important and unusual property
that in perturbation theory, though $U(1)$
is spontaneously broken as a gauge symmetry, it survives as a global symmetry.  The reason  is that
an amplitude that violates the global $U(1)$ conservation law would arise from a term in the effective
action that is proportional to a nonzero power of the charged field $e^{ia}$.  Such terms are not generated
in perturbation theory, because in perturbation theory, $a$ decouples at zero momentum.   At the nonperturbative
level, the global $U(1)$ symmetry is broken (at least down to a subgroup of finite order) by spacetime
instanton effects.  For a recent analysis, see \cite{DiM}.

This mechanism of $U(1)$ gauge symmetry breaking also leads to spontaneous breaking of supersymmetry.
Indeed, the vector multiplet that contains the $U(1)$ gauge field $A$ also
contains an auxiliary field $D$.  The expectation value of $D$ receives a contribution from the
expectation value of $S$ (or $Z$) as well as from the massless charged chiral multiplets 
$\varrho_a=\rho_\a+\theta^\alpha\psi_{a\,\alpha}+\dots$ that
arise in the four-dimensional expansion of the ten-dimensional $SO(32)$ vector multiplet.    The potential
energy of the theory has a contribution  
\begin{equation}\label{imoxip} \frac{D^2}{2g_\st^2},\end{equation}
where
\begin{equation}\label{mimoxip} D= \frac{p}{\mathrm {Re}\, S}+\sum_\a e_\a |\rho_\a|^2=pg_\st^2
+\sum_\a e_\a |\rho_\a|^2.\end{equation}
(Here $e_\a$ is the $U(1)$ charge of the chiral multiplet $\varrho_\a$, which is normalized so that its kinetic energy is canonical.)  As explained in \cite{DSW}, the dependence of $D$ on $S$ follows entirely from the dilaton Kahler
potential $K=-\log (\mathrm{Re}\,S)$ and the fact that a $U(1)$ gauge transformation $A_\mu\to A_\mu+\partial_\mu s$
transforms $S$ to $S-ips$.  Thus, the expectation value of $D$ is of order $g_\st^2$ relative to the classical
contribution $\sum_\a e_\a |\rho_\a|^2$, and this effect must arise at one-loop order. This should come as no 
surprise, since the contribution of the
multiplet $S$ to $D$ is related by supersymmetry to the Green-Schwarz interaction that triggers $U(1)$ breaking.
  Since the expectation value of $D$ is a one-loop effect, the resulting contribution to the vacuum energy (or dilaton tadpole)
 $D^2/2g_\st^2$ will have to arise in two-loop order.    How this happens was investigated in the 1980's
 \cite{DIS,ADS,AS,GrSe} and will be further explored in the rest of this paper.

Models such as we have described  are the only known superstring models in which 
supersymmetry is spontaneously broken in perturbation theory despite being unbroken at tree level. 
That makes them an interesting test
case for superstring perturbation theory.  Oversimplified treatments of superstring perturbation theory
tend to predict that the behavior seen in models of this class is impossible.

\subsubsection{Two Classes Of Vacua}\label{shifting}

The statement that supersymmetry breaks down in perturbation theory in these models refers specifically to
  the vacua with $\rho_\a=0$.  In these vacua, the global $U(1)$ symmetry is 
  conserved in perturbation theory, but supersymmetry is expected
to break down.    Alternatively, we could give expectation values of order $g_\st$ to the 
$\rho_\a$, so as to make $D$ vanish
even though $S\not=0$.   Then the global $U(1)$ symmetry is violated perturbatively,
but supersymmetry is maintained.  One expects that such a vacuum will lead to a stable perturbation theory,
with the 
property -- unusual among supersymmetric models --  that perturbative stability of the vacuum depends on a cancellation between effects of different orders in perturbation theory,
as discussed qualitatively in \cite{FS}.  

Tools to analyze superstring perturbation theory in such a situation have
been developed recently \cite{ASen} (this work was expressed in the language of picture-changing operators, though we anticipate that it can be straightforwardly expressed in terms of super Riemann
surfaces).  The analysis in \cite{Revisited} was limited to more straightforward models
(supersymmetric compactifications above four dimensions and four-dimensional ones
without anomalous $U(1)$'s) in which 
effective field theory predicts that supersymmetry is maintained in perturbation
theory without shifting the values of massless fields.   Even then, superstring perturbation theory involves
subtleties that caused some difficulty in the 1980's.  We will gain experience with those subtleties in the present
paper by studying a class of models in which supersymmetry-breaking effects (requiring a shift in the vacuum to maintain supersymmetry)  {\it do} arise in perturbation theory.  
In these models, the subtleties of superstring perturbation theory arise in low orders in a particularly visible way.

\subsection{A First Look At The Mass Splitting}\label{took}

Following \cite{DIS,ADS}, we will now take a first look at the one-loop mass splittings of charged chiral multiplets.   We write 
$x^\mu$, $\mu=1,\dots,4$ for coordinates on $\R^4$,  and $y^i$, $y^{\bar i}$, $i,\bar i=1,\dots,3$ for local holomorphic and antiholomorphic
coordinates on the Calabi-Yau manifold $\XX$.  Similarly, we denote the right-moving RNS worldsheet fermions as $\psi^\mu$,
$\psi^i$, and $\psi^{\bar i}$.  The $SO(32)$ current algebra of the heterotic string is carried by 32 left-moving fermions in the
vector representation of $SO(32)$.  Upon making the embedding $U(1)\times SU(3)\times SO(26)\subset SO(32)$, 
the left-moving fermions  transform as 
$(\mathbf{3},\mathbf{1})^{\mathbf 1}\oplus (\bar{\mathbf{3}},\mathbf{1})^{\mathbf{-1}}\oplus (\mathbf 1,\mathbf{26})^{\mathbf 0}$, where the exponent
is the $U(1)$ charge.
We denote these components respectively as $\lambda^i$, $\lambda^{\bar i}$, $i=1,\dots,3$, and $\lambda^T$, $T=1,\dots,26$.
Massless  charged chiral multiplets arise in four dimensions from the Kaluza-Klein expansion of the ten-dimensional 
gauge field $A$.  The relevant ansatz, suppressing $SO(32)$ indices, is
\begin{equation}\label{zongo}A_{\bar i}(x;y)=\sum_{a} \rho_a(x) w_{a\, \bar i}(y)+\dots,\end{equation}
where $a$ runs over the set of chiral multiplets,
$w_{a\,\bar i}(y)$ is for each $a$ a harmonic $(0,1)$-form on $\XX$ (valued in the $SO(32)$ bundle),  and $\rho_a(x)$ is a massless scalar field in 
spacetime, part of a chiral supermultiplet $\varrho_a=\rho_a+\theta^\alpha \chi_{a,\alpha}+\dots.$  We
write $\bar\varrho_a=\bar\rho_a+\bar \theta^{\dot\alpha}\bar\chi_{a\,\dot\alpha}+\dots$ for the conjugate antichiral multiplet.

We consider the case that the supermultiplet $\varrho_a$ has a nonzero $U(1)$ charge $e_a$.  In this case, a one-loop
$D$-term will generate a $\bar\rho_a\rho_a$ coupling.  If such a term is generated, it will represent a mass splitting for
bosons and fermions in the $\varrho_a$ multiplet, since no corresponding mass term is possible for the fermions.
(A $\bar\chi_a\chi_a$ term is not Lorentz invariant because $\chi_a$ and $\bar\chi_a$ have opposite chirality,
while $\chi_a\chi_a$ or $\bar\chi_a\bar\chi_a$ terms do not conserve the $U(1)$ charge.)  
Our goal is to understand that the one-loop superstring amplitude does generate the $\bar\rho_a\rho_a$ coupling.  

For brevity, we will do this for chiral superfields transforming as ${\bf {26}}^{\mathbf 1}$ under the unbroken $U(1)\times SO(26)$.
Such modes arise from the part of the adjoint representation of $SO(32)$ that transforms under $U(1)\times SU(3)\times SO(26)$
as $(\bf 3,\bf {26})^1$.  For these modes, with $SO(32)$ indices included, the ansatz (\ref{zongo}) becomes
\begin{equation}\label{longo}A_{\bar i\, iT}(x;y)=\sum_a \rho_{a\,T}(x) w_{a,\,\bar i i}(y),~~ T=1,\dots,26.\end{equation}
Here for each $a$, $w_{a\,\bar i i}$ is a harmonic $(1,1)$-form on $\XX$, and now the four-dimensional wavefunction
$\rho_{a,T}$ carries the $SO(26)$ index $T$.  

In the RNS description of the heterotic string, we describe the string worldsheet by even and odd holomorphic local coordinates
$z|\theta$ and an even antiholomorphic local coordinate\footnote{What we call $\t z$ is commonly called $\bar z$, but we prefer
to avoid this notation because identifying $\t z$ as the complex conjugate of $z$ is not invariant under superconformal
transformations of the pair $z|\theta$. It is best to simply think of $\t z$ as a bosonic coordinate that is close to the
complex conjugate of $z$, and to be more precise only when necessary.  In the derivation
below, this will be near $z=\t z=0$.}
 $\t z$.   The map of the string to spacetime is described by superfields $\X^\mu$, $\Y^i$, $\Y^{\bar i}$ which
 have expansions such as $\X^\mu(\t z;\neg z|\theta)=X^\mu(\t z;\neg z)+\theta\psi^\mu(\t z;\neg z)$.  To construct supersymmetric
 expressions, one uses the superspace derivative $D_\theta=\partial_\theta+\theta\partial_z$, and, for example,
 one has
 \begin{equation}\label{ido}D_\theta \X^\mu=\psi^\mu+\theta\partial X^\mu. \end{equation}
 
 A vertex operator for a Neveu-Schwarz state is a superfield $W(\t z;\neg z|\theta)$.
In the case of a mode  $\rho_T$ of momentum $k$  (for brevity we pick a particular multiplet
and suppress the label $a$), the appropriate superfield is
\begin{equation}\label{urkey}W_{T,k}(\t z;\neg z|\theta)=\exp(ik\cdot \X)\Lambda_T\Lambda^i
w_{i \bar i }(\Y)D_\theta \Y^{\bar i}.\end{equation}
The vertex operator for the conjugate mode $\bar\rho_T$ is similar:
\begin{equation}\label{murkey}\t W_{T,k}(\t z;\neg z|\theta)=\exp(ik\cdot \X)\Lambda_T\Lambda^{\bar i}
w_{\bar i i}(\Y)D_\theta \Y^i.\end{equation}
In these formulas,  $\Lambda_T=\lambda_T+\theta G_T$ is a superfield that reduces to $\lambda_T$ by the equations of motion;
$G_T$ is an auxiliary field that vanishes on-shell.
Similarly 
$\Lambda^i=\lambda^i+\theta G^i$ and $\Lambda^{\bar i}=\lambda^{\bar i}+\theta G^{\bar i}$ are superfields 
that reduce on-shell to $\lambda^i$ and $\lambda^{\bar i}$.

A genus 1 mass shift will be derived from the two-point function 
\begin{equation}\label{kolo} \bigl\langle W_{T,k}(\t z;\neg z|\theta)\t 
W_{T,-k}(\t z';\neg z'|\theta')\bigr\rangle\end{equation}
on a super Riemann surface $\Sigma$ of genus 1.  The effect we are looking for is parity-conserving, so the
relevant case is that $\Sigma$ has an even spin structure.  This means that holomorphically $\Sigma$ can be described, up to isomorphism, by
even and odd coordinates $z|\theta$ with equivalences
\begin{align}\label{mune} z\cong &z+1 \cr
                    \theta  \cong & -\theta\end{align}
and
\begin{align}\label{zune} z\cong & z+\tau \cr
                    \theta\cong & \theta.\end{align}
while antiholomorphically there is a single even coordinate $\t z$ with equivalences                     
\begin{equation}\label{une} \t z\cong \t z+1 \cong \t z+\bar \tau.\end{equation}
Here $\tau$ is a complex modulus. As shown in \cite{DIS,ADS}, this modulus plays no important role in the analysis,
except that one has to integrate over it at the end.  So we can just think of $\tau$ as a complex constant.

A genus 1 super Riemann surface $\Sigma$ with an even spin structure has no odd moduli (until we include punctures).  
So apart from $\tau$, the only parameters in the problem are the positions $\t z;\neg z|\theta$ and $\t z';\neg z'|\theta'$ at which the two
vertex operators are inserted.  Moreover, we can set $z'=\t z'=0$ using the translation symmetry of $\Sigma$.  With an even spin structure,
there is no such translation symmetry for the $\theta$'s.
So finally, the genus 1 mass shift will be derived from the integral
\begin{equation}\label{derfrom}I_{TT'}=\int \d^2z\d\theta\d\theta'\,\langle W_{T,k}(\t z;\neg z|\theta) \,\t W_{T',-k}(0;\neg
0|\theta')\rangle. 
\end{equation}
(Here $\d^2z $ is short for $-i \d\t z\wedge \d z$.)
We can think of this as the integral over the moduli space of super Riemann surfaces of genus 1 with 2 NS punctures
(except that we also need
to integrate over $\tau$ at the end).
It remains just to learn how to perform this integral. 

The traditional approach in superstring perturbation theory is to first integrate over the odd moduli, which in the present
context are $\theta$ and $\theta'$, and then try to perform the bosonic integral. The integral over the odd variables
can be evaluated using 
\begin{equation}\label{izzo}\int \d \theta \,W_{T,k}(\t z;\neg z|\theta)=V_{T,k}(\t z;\neg z),\,\,\int \d\theta\,\t W_{T',-k}(\t z;\neg z|\theta)=
\t V_{T',-k}(\t z;\neg z),\end{equation}
with
\begin{equation}\label{turkey} V_{T,k}=\exp(ik\cdot X) \lambda_T\lambda^i \left(w_{i\bar i }(Y)
\left(\partial_z Y^{\bar i}+i k_\mu\psi^\mu \psi
^{\bar i}\right)+D_j w_{i\bar i}\psi^j\psi^{\bar i}\right)\end{equation} 
and a similar formula for $\t V_{T',-k}$.  (The auxiliary fields have been set to zero by their equations of motion.)
The term proportional to $D_j w_{i\bar i}$ is a sort of $\alpha'$ correction, since the zero-mode wavefunction $w_{i\bar i}$
is nearly constant when $\XX$ is much larger than the string scale.  Our interest here is really in string-loop corrections, not 
$\alpha'$ corrections.  A convenient way to avoid  issues that are not really relevant for our purposes is to consider
the special case that $\XX$ is a Calabi-Yau orbifold\footnote{By such an orbifold, we mean the quotient of a torus $T=\R^6/\Lambda$
(here $\Lambda$ is a lattice in $\R^6$ of maximal rank) by a finite subgroup of $SU(3)\subset SO(6)$, in other words by a finite
group of symmetries of $\Lambda$ that preserves $\N=1$ supersymmetry in four dimensions.} and the chiral multiplet of
interest comes from the untwisted sector, so that $w_{i\bar i}$ is a constant matrix and $D_j w_{i\bar i}=0$.
 This case was analyzed in the present context in \cite{ADS} and suffices to illustrate
the ideas we wish to explore.  We explain in section \ref{alphadetail} why the general case behaves similarly.

Dropping the $D_jw_{i\bar i}$ term in $V_{T,k}$, the 
integration over the odd moduli  via (\ref{izzo}) leads to the bosonic integral
\begin{align}\label{mizzo}
\int_{\Sigma} \d^2z \left\langle \left. e^{ik\cdot X} \lambda_T\lambda^i w_{i\bar i }(Y)
(\partial_z Y^{\bar i}+i k\cdot\psi \psi
^{\bar i})\right|_{\t z;z} \left. e^{-ik\cdot X} \lambda_{T'}\lambda^{\bar j} w_{\bar jj}(Y)
\left(\partial_zY^{j}-i k\cdot\psi \psi
^j\right)\right|_{0; 0} \right\rangle  . \end{align}
This expression must then be summed over the three even spin structures on 
$\Sigma$ and integrated over $\tau$.

If we drop the terms in the vertex operator that depend explicitly on $\psi$,
the expression (\ref{mizzo}) vanishes after summing over spin structures
(even before integration over $z$).  This is explained in \cite{DIS,ADS}.  If $\psi^i$ and $\psi^{\bar i}$ are treated
as free fields, the claim is true because of the usual GSO cancellation \cite{GOS} 
between spin structures that leads to vanishing of the 1-loop
cosmological constant for superstrings in $\R^{10}$.  If $\XX$ is a Calabi-Yau orbifold, then
its path integral is a sum of contributions 
of different sectors in each of which $\psi^i$ and $\psi^{\bar i}$
are free fields with $SU(3)$-valued twists.  $SU(3)$-valued twists do not disturb the 
GSO cancellation, so the contribution
of each sector to (\ref{mizzo}) vanishes after summing over spin structures if one drops 
the terms that depend explicitly on $\psi$.
In section \ref{alphadetail}, we  explain that the same is true if $\XX$ is a generic Calabi-Yau threefold, rather than an orbifold.

The contribution of the terms in (\ref{mizzo}) that do explicitly depend on the $\psi$'s is proportional to $\langle
k\cdot\psi(\t z;\neg z)k\cdot \psi(0;\neg 0)\rangle$, which in turn is proportional to $k^2$.  Since $k^2$ vanishes on-shell
for the massless scalar fields whose mass renormalization we are exploring, it seems at first sight that these terms are
not relevant.  However, it is shown in \cite{DIS,ADS} that what multiplies $k^2$ is an integral that diverges as $1/k^2$ for $k^2\to 0$,
because of singular behavior near $z=\t z=0$. 
As a result, the $k^2$ factor in the numerator is illusory.

To understand this, we analyze the small $z$ behavior of the integrand in (\ref{mizzo}) using the operator product expansion.  
The worldsheet fields $\psi^\mu$ really are free fields, with leading singularity $\psi^\mu(z)\psi^\nu(0)\sim \eta^{\mu\nu}/z$.  
Other contributions to this OPE are not singular enough to be relevant to what we are about to say.\footnote{Only Lorentz-invariant
operators appearing in the $\psi^\mu(z)\psi^\nu(0)$ OPE can contribute to the integral (\ref{mizzo}).  After the identity operator, the lowest dimension Lorentz-invariant operator
in this channel is $\psi^\lambda\partial_z\psi_\lambda$, of dimension 2.  Its contribution
$\psi^\mu(z)\psi^\nu(0)\sim \dots -(z/4)\eta^{\mu\nu}\psi^\lambda\partial_z\psi_\lambda$ is much too soft near $z=0$ to be relevant
in what follows.  Similar remarks apply for other
OPE's that we consider momentarily.}
Similarly the $SO(26)$ fermions
$\lambda_T$ are free fields with leading singularity $\lambda_T(\t z)\lambda_{T'}(0)\sim \delta_{TT'}/\t z$, and $X^\mu(\t z;\neg z)$
are free fields with leading singularity $\exp(ik\cdot X(\t z;\neg z))\exp(-ik\cdot X(0;\neg 0))\sim |\t z z|^{-k^2}$.  In each of these
cases the less singular terms are not relevant.  The remaining operators whose OPE's we need to understand are
 $\O=w_{i\bar i}\lambda^i\psi^{\bar i}$ and
$\O^*=w_{\bar i i}\lambda^{\bar i}\psi^{ i}$.  These are   primaries of dimension $(1/2,1/2)$ for the left-moving conformal and right-moving superconformal algebras of the sigma-model with target
$\XX$.  To be more precise, $\O$ and $\O^*$ are  respectively chiral and
antichiral  primaries for the $\N=2$ superconformal algebra of this sigma-model.  The $\O\cdot \O^*$ operator 
product expansion is not simple,
since a whole tower of Kaluza-Klein modes on $\XX$ can appear.  However (assuming that $\XX$ is such that compactification on 
$\XX$ preserves
 $\N=1$ supersymmetry and no more), there is just one operator $V_D$ of dimension $(1,1)$  that contributes to this
expansion.
Its contribution is nonsingular:
\begin{equation}\label{inzo}\O(\t z;\neg z)\O^*(0;\neg 0)\sim \dots + V_D(0;\neg 0).\end{equation}
This operator is
\begin{equation}\label{winzo}V_D=J_\ell J_r,\end{equation}
where
\begin{equation}\label{irox} J_\ell =g_{\bar i i}\lambda^{\bar i}\lambda^{ i},~~J_r=g_{\bar i i}\psi^{\bar i}\psi^i,\end{equation}
with $g_{i\bar i}$  the Kahler metric of $\XX$.  $J_\ell$ is the antiholomorphic  current associated to the anomalous
$U(1)$ gauge symmetry whose $D$-term we are investigating, and $J_r$ is the holomorphic current that generates the $U(1)$
subalgebra of the $\N=2$ superconformal algebra.  Because $J_\ell$ and $J_r$ are antiholomorphic and holomorphic conserved
currents, the operator $V_D$ has dimension precisely $(1,1)$, and
the coefficient with which $V_D$ appears in the product (\ref{inzo}) depends only on the $U(1)$ charges of the operator
$\O$ (namely 1 and $-1$).   

The contribution of $V_D$ to our operator product is thus
\begin{equation}\label{zenno}V_{T,k}(\t z;\neg z)\,V_{T',-k}(0;\neg 0)\sim k^2\frac{V_D}{|\t z z|^{1+k^2}}\end{equation}
Since
\begin{equation}\label{enno}\int \d^2z\frac{1}{|\t z z|^{1+k^2}}\sim \frac{2\pi}{k^2}, ~~k^2\to 0,\end{equation}
the explicit factor of $k^2$ in (\ref{zenno}) disappears, and the integrated two-point function comes out to be
\begin{equation}\label{wrenno}2\pi \langle V_D\rangle. \end{equation}  
This gives the expected supersymmetry-violating one-loop mass shift.

Many other operators apart from $V_D$ appear in the $V_{T,k}(\t z;\neg z)\cdot V_{T',-k}(0;\neg 0)$ operator product, but $V_D$ is the
only one whose contribution to the integral has a pole at $k^2=0$.  So it is the only operator that contributes to the mass shift.

It is further shown in \cite{DIS,ADS} that the expectation value $\langle V_D\rangle$ on a torus can be computed 
just in terms of the spectrum of massless  charged chiral multiplets in space-time.  This is analogous to what happens in supersymmetric
field theory, where likewise the one-loop shift in the expectation value of the auxiliary field $D$
comes entirely from the contribution of massless
chiral multiplets.
 In fact, the final integration over $\tau$ that must be performed to complete the computation in string theory
coincides with an analogous Schwinger parameter integration in field theory, with just the one  usual difference.  In string theory,
 modular invariance removes the ultraviolet region of small $\mathrm{Im}\,\tau$, making the effect finite, while the analogous computation in field theory is ultraviolet-divergent.

\begin{figure}
 \begin{center}
   \includegraphics[width=3.5in]{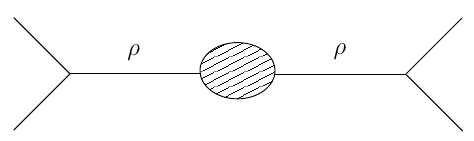}
 \end{center}
\caption{\small The mass shift of a massless particle $\rho$ can be computed slightly off-shell by treating $\rho$ as a resonance
in a scattering amplitude with four external particles. This process is affected by the one-loop mass shift of $\rho$, but now the $\rho$
particle whose mass is shifted is generically off-shell, giving a sound framework for the $k^2/k^2$ computation. }
 \label{resonance}
\end{figure}

One may worry at first whether it is valid to continue away from the mass-shell $k^2=0$
and cancel powers of $k^2$, as assumed in the above calculation. But actually
\cite{DIS,ADS}, one can put all this on a firm foundation by considering a scattering amplitude in which the 
particle $\rho$ appears as  a resonance or
 intermediate state.  Such an amplitude  (fig. \ref{resonance}) is affected by the mass shift of the $\rho$ particle, but now the particle
whose mass is shifted can be slightly off-shell, giving a clear basis for the calculation sketched above.   The
mass shift $\delta m^2$ appears in the perturbative computation of the scattering amplitude as the coefficient of a double pole, because of the usual
expansion
\begin{equation}\label{lummoz}\frac{1}{k^2+\delta m^2}=\frac{1}{k^2}-\frac{1}{k^2}\delta m^2\frac{1}{k^2}+\dots. \end{equation}

In \cite{DIS}, arguments were given for interpreting 
the operator $V_D$ as the vertex operator for the auxiliary field $D$ in the vector multiplet associated
to the anomalous $U(1)$ gauge field.  There is not a systematic theory of correlation functions with insertions of vertex operators
for auxiliary fields (as opposed to vertex operators associated to physical states).  However, $V_D$ does appear in a number of
interesting calculations, including the two-loop vacuum energy, which we explore in section \ref{dilaton}.

.

\subsection{More On The Mass Splitting}\label{moresplit}

Now we will explain an alternative perspective \cite{GrSe} on the same calculation. In this approach, we set the momentum
$k$ to zero from the beginning.  The vertex operators thus reduce to
\begin{equation}\label{lurkey}W_{T}(\t z;\neg z|\theta)=\Lambda_T\Lambda^i
w_{i \bar i }(\Y)D_\theta \Y^{\bar i}\end{equation}
and
\begin{equation}\label{wurkey}\t W_{T}(\t z;\neg z|\theta)=\Lambda_T\Lambda^{\bar i}
w_{\bar i i}(\Y)D_\theta \Y^i\end{equation}
and the mass shift is supposed to be computed from
\begin{equation}\label{durkey}\int_\SIgma\d^2 z \d\theta\d\theta'\,\langle \W_T(\t z;\neg z|\theta) \t 
W_{T'}(0;\neg 0|\theta')\rangle.\end{equation}

Since we have set $k$ to 0, there will be no $k^2/k^2$ terms.  So
how can we possibly get a nonzero result?

In view of our previous experience, the
 answer must somehow come from the appearance in the $W_T\cdot \t W_{T'}$ operator product of the operator $V_D$.
The relevant operator product coefficient is 
\begin{equation}\label{iodi}W_T(\t z;\neg z|\theta)\cdot \t W_{T'}(0;\neg 0|\theta')\sim 
\frac{\delta_{TT'}V_D(0;\neg 0)}{\t z}+\dots.\end{equation}
Thus we must contemplate the integral
\begin{equation}\label{bodi}\int \d^2z\, \d\theta\,\d\theta'\,\frac{1}{\t z} .\end{equation}
Since we obtained it from the operator product expansion, this is the right form of the integral only near $z=\t z=0$.  When
$\t z$ is not small, the integrand has to be modified to be consistent with the doubly-periodic nature of the torus. See Appendix D
of \cite{Revisited} for more on this, but the details  are not important in what follows.  We need some
sort of cutoff at large $z$ for the following analysis, since otherwise the integral (\ref{bodi}) has a problem at large
$z$ analogous to the problem we will describe at small $z$, but it does not matter if this comes from 
the compactification of the $z$-plane to a torus or from a sharp cutoff $\t z z\leq 1$.  The latter is simpler, so the integral that we will consider 
is\footnote{\label{melbo} Integration
on a supermanifold with boundary is subtle -- see section 3.5 of \cite{Supermanifold} for an introduction.  
To ensure that none of these
subtleties are relevant in what follows, we take the relation between $\t z$ and $z$ to be precisely  
$\t z=\bar z$ near $|z|=1$, and we only
make changes of variables that are trivial near $|z|=1$.  With this procedure, the sharp cutoff at $|z|=1$ produces
equivalent results to compactification of the $z$-plane to a torus. }
\begin{equation}\label{nodi}\J=\int_{\t z z\leq 1}\d^2 z\,\d\theta\,\d\theta'\, \frac{1}{\t z}.\end{equation}

The idea in \cite{GrSe} is that instead of integrating over $\theta$ and $\theta'$ at fixed $\t z $ and $z$, after which we integrate
over $\t z$ and $z$, we should integrate over
$\theta$ and $\theta'$ at fixed values of $\t z$ and $\h z=z-\theta\theta'$, after which we integrate over $\t z$ and $\h z$. 
A heuristic explanation of why this may be the right thing to do is that $\h z$, rather than $z$, is invariant under global supersymmetry
transformations.  In other words, let $z|\theta$ and $z'|\theta'$ be two points on the complex superplane $\C^{1|1}$.  A global
supersymmetry transformation $\delta\theta=\epsilon$, $\delta z=-\epsilon\theta$ (and likewise $\delta\theta'=\epsilon$,
$\delta z'=-\epsilon\theta'$) with constant $\epsilon$ leaves fixed not $z-z'$ but $z-z'-\theta\theta'$.  For $z'=0$,
this reduces to $\h z=z-\theta\theta'$, and this is some indication that $\h z$ may be the right variable to use.
   
The first question that comes to mind concerning this proposal may be 
why it is necessary to specify what is held fixed when we integrate over $\theta$ and $\theta'$.  The reason 
is that the integral in (\ref{nodi})  is a supermanifold analog of what in the bosonic world is a conditionally convergent integral.  
An example of a bosonic integral that is only conditionally convergent is
\begin{equation}\label{tonico} \int_{\bar z z\leq 1} \d^2 z \frac{1}{\bar z^2} .\end{equation}
This integral is not absolutely convergent, since replacing the integrand by its absolute value gives a divergent integral.
However, if we write $z=re^{i\varphi}$ and integrate first over $\varphi$, the integral converges (and in fact vanishes).   
With a different procedure, it may diverge or may converge to a different value.

A corresponding bosonic integral with only a simple pole
\begin{equation}\label{wonico}\int_{\bar z z\leq 1}\d^2z \frac{1}{\bar z}\end{equation}
is absolutely convergent.  This is ensured by the fact that under a scaling $z\to \lambda z$, $\bar
 z\to \bar\lambda \bar z$,
the measure $\d^2z/\bar z$ scales with a positive power of $\lambda$, making the singularity ``soft'' near $z=0$.
An integral -- like that in eqn. (\ref{tonico}) -- that scales as a zero or negative power of $\lambda$ is at most only conditionally convergent.
In a supersymmetric context, the natural scaling of the odd variables is $\theta\to \lambda^{1/2}\theta$, $\theta' \to \lambda^{1/2}\theta'$.
The measures $\d\theta$ and $\d\theta'$ thus scale as $\lambda^{-1/2}$.  
With this scaling, the singularity $\d^2 z\,\d\theta\,\d\theta'/\t z$
is scale-invariant, corresponding to a supersymmetric version of a conditionally convergent integral.  

The following is a  procedure to calculate the integral by keeping fixed $\h z=z-\theta\theta'$, 
rather than $z$, when integrating over the odd variables near $z=0$.
 We define a new coordinate 
\begin{equation}\label{urcott} z^\star=z-h(\t z;\neg z)\theta\theta',\end{equation}
where $h(\t z;\neg z)$ is any smooth function that is 1 in a neighborhood of $z=0$ and 0 in a 
neighborhood of $z=1$.  (The first condition ensures that $z^\star=\h z$ near $z=0$.  The second condition  avoids any subtlety near $|z|=1$, as discussed in footnote \ref{melbo}. 
Alternatively, as explained in Appendix
D of \cite{Revisited} and in section \ref{gluehol} below, if we compactify the $z$-plane to a torus, we need to introduce a function playing the role of $h$ to respect
the double-periodicity of the torus. The fact that the torus has no boundary then ensures that there is no boundary term at large $z$.)  
Our method of defining the conditionally convergent integral is to integrate first over $\theta$ and $\theta'$ keeping fixed
$\t z$ and $z^\star$ rather than $\t z$ and $z$.  

To complete the explanation of how to do the integral, we also need to specify the relationship between the antiholomorphic
variable $\t z$ and the holomorphic variable $z$.  Here instead of the naive relationship $\bar{\t z}=z$, we take
$\bar{\t z}= z^\star$.  With this definition, we will perform our integral by expressing everything in terms of 
$\t z,$ $\bar {\t z}$, $\theta$ and $\theta'$,
after which, following the recipe of \cite{GrSe}, we integrate first over the odd variables $\theta$ and $\theta'$ and only then over
$\t z$ and $\bar{\t z}$.

To explicitly perform the integral, we have to write $\d^2z \,\d\theta\,\d\theta'=-i\d\t z\,\d z\,\d\theta\,\d\theta'$ in terms of
$\t z$, $\bar {\t z}$, $\theta$, and $\theta'$ only.  Eqn. (\ref{urcott}) is equivalent to 
\begin{equation}\label{meogo} z=z^\star+h(\t z;\neg z^\star)\theta\theta'=\bar {\t z}+h(\t z;\neg \bar{\t z})\theta\theta',\end{equation} 
where in the second step we set $z^\star=\bar{\t z}$.  So 
\begin{equation}\label{mulgo} \d z=\d\bar{\t z}\left(1+\theta\theta'\frac{\partial h(\t z;\neg\bar{\t z})}{\partial\bar{\t z}}\right) 
+\dots.\end{equation}
where on the right hand side, we compute the coefficient of $\d\bar{\t z}$  only. This enables us to express the integral
$\J$ in terms of $\t z, \bar{\t z},\theta,\theta'$; in the measure $\d\t z\,\d z\,\d\theta\,\d\theta'$, we simply substitute for $\d z$
using (\ref{mulgo}).   The integral then becomes
\begin{equation}\label{uritz}\J=-i\int_{|\t z|\leq 1}\d\t z\,\d\bar{\t z}\,\d\theta\,\d\theta'\left(1+\theta\theta'\frac{\partial h(\t z;\neg\bar{\t z})}{\partial\bar{\t z}}\right)\frac{1}{\t z} .\end{equation}
Now we integrate over $\theta$ and $\theta'$ with $\int \d\theta\,\d\theta'\,\theta\theta'=1$, 
to get
\begin{equation}\label{wuritz}\J=-i\int_{|\t z|\leq 1}\d\t z\,\d\bar{\t z}\frac{\partial h(\t z;\neg \bar{\t z})}{\partial\bar{\t z}}\frac{1}{\t z}.
\end{equation}
Integration by parts gives
\begin{equation}\label{ritz}\J=i\int_{|\t z|\leq 1}\d\t z\,\d\bar{\t z}\,h(\t z;\neg\bar{\t z})\frac{\partial}{\partial\bar{\t z}}\frac{1}{\t z}.
\end{equation}
There is no surface term at $|\t z|=1$ since $h=0$ there.  With
\begin{equation}\label{mufato}\frac{\partial}{\partial\bar{\t  z}}\frac{1}{\t z}=2\pi \delta^2(\t z) \end{equation}
and also  $\int \d^2\t z\delta^2(\t z)=1$ and $\d\t z\,\d\bar{\t z}=-i\d^2\t z$, we get
\begin{equation}\label{ifato} \J=2\pi h(0;\neg 0)= 2\pi, \end{equation}
where we use the condition
\begin{equation}\label{nifato}h(0;\neg 0)=1. \end{equation}

The moral of the story is that if one first integrates over $\theta$ and $\theta'$ and then tries to decide what to do next,
it is already too late.   A simple statement has to be made in terms of the full set of variables.

We will explain in section \ref{genless} how what we have just done is a special case of a general procedure.
But here we make
some further remarks on this calculation.
After integration by parts, the result for $\J$ seems to come from a delta function at $z=0$.  But the existence of a natural
integration by parts was special to this particular problem.  To draw a general lesson, it is better
to look at the formulas (\ref{uritz}) or (\ref{wuritz}) before integration by parts, and here we see that  
there actually is no contribution at all near $z=0$ since $h$ is constant near $z=0$. (There should not be a contribution
from $z=0$, since the $D$-auxiliary field vertex operator
$V_D$ that played the starring role is not the vertex operator of a physical field.)
The question of what values of $z$ contribute to the integral
is not well-defined, since it depends on the choice of the arbitrary function $h$, though the value of the integral was independent
of this choice.  We will give a more systematic explanation
of the meaning of the choice of the function $h$ in section \ref{gluehol}.

\subsubsection{A Variant}\label{variant}

A variant of the above procedure to evaluate the integral for $\J$ is to say that, after setting $\bar{\t z}=z^\star,$ 
we pick a lower cutoff $|\t z|\geq \eta$ for some small
positive $\eta$, integrate over all bosonic and fermionic variables, and then take the limit $\eta\to 0$.  
For $\eta>0$, we are evaluating a smooth measure on a compact supermanifold (with boundary) and
there is no need to say anything about the order of integration.
Performing the integral for  $\eta>0$ and then taking $\eta\to 0$ gives
the same result as integrating first over $\theta$ and $\theta'$ and then over $\t z$ and $\bar{\t z}$, since in that latter
procedure, there was no contribution near $\t z=0$.  The formulation
with the lower cutoff $\eta$ is  useful  for the generalization that we discuss in section \ref{genless}.

\subsubsection{Justification}\label{justification}

One justification \cite{GrSe} of the procedure just described  is that
it agrees with the reasoning of \cite{DIS,ADS} that we summarized is section \ref{took}, and in 
particular with the unitarity-based reasoning of fig. \ref{resonance}.

Another justification (also described in \cite{GrSe}) is as follows.   The 1-loop mass renormalization of the field $\rho_T$ arises from  the term of order 
$\langle\rho_T\rangle\langle\bar\rho_T\rangle$ in the expansion of the 1-loop partition function of the sigma-model with target $\XX$.   Let us analyze
how to compute that term.

At zero momentum, the vertex operator of $\rho_T$ is $W_T=\Lambda_T \Lambda^i w_{i\bar i}(\Y)D_\theta \Y^{\bar i}$. 
We have written this expression in terms of superfields since we want to make worldsheet supersymmetry
manifest.
If we give $\rho_T$ and its complex conjugate $\bar\rho_T$ 
an expectation value, the effect of this, to first order, is to add the appropriate terms to the worldsheet
action:
\begin{equation}\label{mezzo}\langle\rho_T\rangle\int \d^2z\,\d\theta\, W_T(\t z;\neg z|\theta)
+\langle\bar\rho_T\rangle\int\d^2 z\,\d\theta \,\t W_T(\t z;\neg z|\theta). \end{equation}
The vertex operators $W_T $ and $\t W_T$ have terms linear in auxiliary fields 
(see the comment following eqn. (\ref{murkey})).  If we integrate out the auxiliary fields, we get
a four-fermion term in the action that we will schematically denote as\footnote{This term arises by expanding the usual four-fermion coupling $\lambda^2F\psi^2$ of the sigma-model (where $F$ is
the Yang-Mills field strength, which becomes the Riemann tensor if the spin connection is embedded in the gauge group) in powers of $\rho_T$ and $\bar\rho_T$.  We omit an explicit formula as it is not
illuminating.} $\lambda^2(\dots)\psi^2$.  This is the familiar four-fermion
term of the supersymmetric nonlinear sigma-model -- or more exactly, it is the part of that term proportional
to $|\langle\rho_T\rangle|^2$.  So after integrating out the odd coordinate $\theta$ and also the auxiliary fields,
the part of the action that depends on $\langle\rho_T\rangle$ becomes
\begin{equation}\label{yezzo}\langle\rho_T\rangle\int\d^2 z \,V_T +\langle\bar\rho_T\rangle
\int \d^2 z\,\t V_T +|\langle\rho_T\rangle|^2\int\d^2 z \,\lambda^2(\dots)\psi^2. \end{equation}

Now  to study the genus 1
heterotic string path integral in the presence of a background field, we have to expand the integrand
of the worldsheet path integral in powers of this field.  This integrand is the exponential of minus the action or
\begin{equation}\label{doome}\exp\left( -\langle\rho_T\rangle\int\d^2 z \,V_T -\langle\bar\rho_T\rangle
\int \d^2 z\,\t V_T -|\langle\rho_T\rangle|^2\int\d^2 z\, \lambda^2(\dots)\psi^2\right),\end{equation}
where we show  only the terms that depend on  $\langle\rho_T\rangle,$ $\langle\bar\rho_T\rangle$.

In that order, 
there   is a bilinear expression involving the two-point function of
the vertex operators $V_T$ and $\t V_T$:
\begin{equation}\label{zumly} \int \d^2 z\, \bigl\langle V_T(\t z;\neg z)\,\t V_T(0;\neg 0)\bigr\rangle.\end{equation}
(As usual, we factor out the translation symmetry of the torus to set the insertion point of $\t V_T$ to
$\t z=z=0$.) This is the ``obvious'' contribution to the mass shift.
 But there is also a ``contact'' term coming from an insertion of the four-fermi
interaction.  As we know by now, the obvious contribution vanishes, so this contact term must give the full answer.

This gives a straightforward explanation of why the obvious expression (\ref{zumly}) needs to be corrected.
The alternative explanation in which we start with the supersymmetric version of (\ref{zumly}), namely
\begin{equation}\label{umly} \int \d^2 z\,\d\theta\,\d\theta' \,\langle W_T(\t z;\neg z|\theta)
\t W_T(0;\neg 0|\theta') \rangle,\end{equation}
and regularize in an appropriate way the resulting 
conditionally convergent integral, has two advantages.  Technically, it is straightforward to get the right
answer this way; it is clear that only the $(1,1)$ part of the relevant OPE matters.  (The $\lambda^2(\dots)\psi^2$
term in (\ref{yezzo})  is a linear combination of a  $(1,1)$ operator and various irrelevant operators that do not contribute.)
Also this approach generalizes to all of the conditionally convergent integrals of superstring perturbation
theory, as we explain next.

\subsection{General Lessons}\label{genless}

\subsubsection{Preliminaries}\label{preliminaries}
\begin{figure}
 \begin{center}
   \includegraphics[width=4.5in]{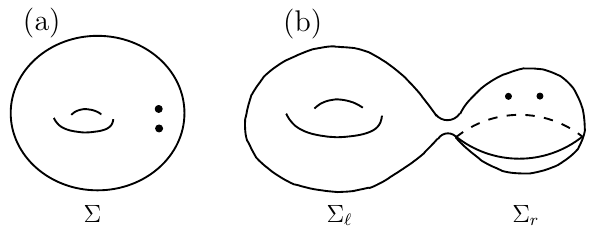}
 \end{center}
\caption{\small   A process (a) in which two punctures on a Riemann surface $\Sigma$ -- here of genus 1 -- approach
each other is equivalent conformally to a process (b) in which $\Sigma$ splits into two components $\Sigma_\ell$ and $\Sigma_r$,
connected by a narrow neck, with one of them a genus 0 surface that contains the two punctures.}
 \label{Short}
\end{figure}
Now we are going to look in yet another way\footnote{Some of the ideas were known in the early literature \cite{DD}.} at the phenomenon studied in sections \ref{took} and \ref{moresplit}.
This phenomenon involved the behavior as two points $z|\theta$ and $z'|\theta'$ on a genus 1 super Riemann surface
$\Sigma$ approach each other.  However (fig. \ref{Short}), up to a conformal transformation, it is equivalent to say
that $\Sigma$ splits into two components, separated by a narrow neck, of which one is a genus 0 surface
that contains the two points in question, while the other has genus 1.

\begin{figure}
 \begin{center}
   \includegraphics[width=3in]{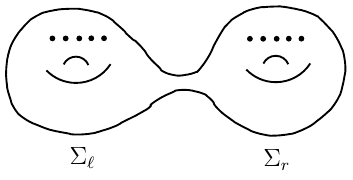}
 \end{center}
\caption{\small    A generalization of fig. \ref{Short} in which a surface $\Sigma$ splits into a pair of components $\Sigma_\ell$
   and $\SIgma_r$ of arbitrary genus, 
   joined via a narrow neck. In the example shown, $\Sigma_\ell$ and $\Sigma_r$ are both genus 1
   surfaces with punctures.}
 \label{Split}
\end{figure}

\begin{figure}
 \begin{center}
   \includegraphics[width=5.5in]{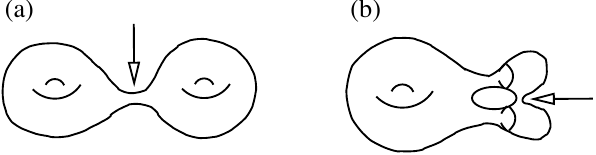}
 \end{center}
\caption{\small A Riemann surface or super Riemann surface can undergo either a separating degeneration as in (a)
or a nonseparating one as in (b).  In each case the degeneration involves the collapse of a narrow neck, labeled
by the arrow.  The singular configurations that arise when the neck collapses are sketched in fig. \ref{Limiting}.  }
 \label{Narrow}
\end{figure}

  This is a special case of a more general type of degeneration
(fig. \ref{Split}) in which a super Riemann surface $\Sigma$ of any genus $g$, containing any number of punctures,
splits into a pair of components $\Sigma_\ell $ and $\Sigma_r$. 
In general, the  punctures are distributed
between the two sides in an almost\footnote{The only restriction is that if $g_\ell$ or 
$g_r$ vanishes, then one requires $\SIgma_\ell$ or $\Sigma_r$
to contain at least two punctures, in addition to the narrow neck.   The moduli space of Riemann
surfaces or super Riemann surfaces
can be compactified without allowing a degeneration in which this is not the case.} 
arbitrary fashion, and the genera $g_\ell$ and $g_r$ are constrained only by
$g_\ell+g_r=g$.   This
 is actually called a separating degeneration (fig. \ref{Narrow}(a)).
There is also an analogous nonseparating degeneration in which $\Sigma$ develops a narrow neck, but the surface
obtained by cutting this neck is still connected (fig. \ref{Narrow}(b)). 

From a conformal point of view, a narrow neck is equivalent to a long tube.  Let us recall how this comes about.
We consider a long tube parametrized
by a complex variable $z$ with an equivalence relation $z\cong z+2\pi i$ and an inequality $0\leq \mathrm{Re}\,z\leq T$,
for some large $T$.  This describes propagation of a closed string of circumference $2\pi$ through an imaginary time $T$.
Now introduce new variables $x=e^{-z}$, $y=e^{-T+z}$, and let $q=e^{-T}$.  The inequalities $0\leq \mathrm{Re}\,
z\leq T$ imply $|x|,|y|\leq 1$.  $x$ and $y$ are related by
\begin{equation}\label{monkey} xy=q.  \end{equation}
One way to describe this gluing is to remove the regions $|x|<|q|^{1/2}$ and $|y|<|q|^{1/2}$ from the unit discs parametrized by $x$ and $y$ and then glue the boundary circles $|x|=|q|^{1/2}$ and
$|y|=|q|^{1/2}$ via $xy=q$.  In this description, for small $q$, the two unit discs have small open balls removed and are glued together along a narrow neck at $|x|,|y|\sim |q|^{1/2}$.

The advantage of the ``long tube'' description is that it makes the physical interpretation clear.  The long tube represents
the propagation of a closed string through a large proper time $T$, so it represents an infrared effect.  
$T$ is analogous to a Schwinger parameter in an ordinary
Feynman diagram.  Integration over $T$
will produce a pole when the closed string propagating down the tube is on-shell. (See section \ref{obso}.) Such a pole is analogous to the pole
in a Feynman propagator $i/(p^2-m^2+i\epsilon)$ in field theory.  

The advantage of the ``narrow neck'' description is that it makes it clear how to compactify the moduli space.  In
terms of the long tube, it is not clear that there is a meaningful limit for $T\to\infty$, but in the narrow neck description
by eqn. (\ref{monkey}), there is no problem in taking the limit $q\to 0$.  The limiting equation
\begin{equation}\label{wonkey} xy=0, \end{equation}
with the restriction $|x|,|y|\leq 1,$
describes two discs, the disc $|x|\leq 1$ and the disc $|y|\leq 1$, glued together at the common point $x=y=0$.

\begin{figure}
 \begin{center}
   \includegraphics[width=5.5in]{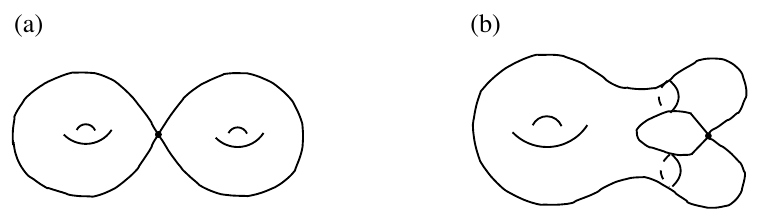}
 \end{center}
\caption{\small Collapse of the narrow necks in fig. \ref{Narrow} leads to these limiting configurations.  The singularities depicted
here are known as ordinary double points.
The local picture is that two branches meet at a common point.  The fact that the only singularities that occur in the Deligne-Mumford
compactification are ordinary double point singularities (which have a long distance or infrared interpretation in spacetime) is the reason
that there are no ultraviolet divergences in superstring perturbation theory.}
 \label{Limiting}
\end{figure}

One of the most fundamental facts about superstring perturbation theory is the existence of the Deligne-Mumford
compactification of the moduli space $\MM$ of Riemann surfaces  or super Riemann
surfaces. (When the context is sufficiently clear,
we use the symbol $\MM$ to denote either of these spaces; we also write $\h\MM$ for the Deligne-Mumford compactification.) Moduli space or supermoduli space can be compactified
by adding limiting configurations  (fig. \ref{Limiting}) that correspond to the collapse of a narrow neck.  
Apart from limiting configurations
of this particular kind, possibly with more than one collapsed neck, the compactified moduli space
parametrizes smooth surfaces only.  

This is actually the fundamental reason that there are no ultraviolet divergences in 
superstring perturbation theory.  The measures that have to be integrated are always smooth measures on the appropriate
moduli spaces.  Integration of a smooth measure on a compact space can never pose a problem.  So a difficulty in superstring
perturbation theory can only arise from noncompactness of the relevant moduli spaces.  The existence of the Deligne-Mumford
compactification is a precise statement that the only pertinent noncompactness is associated to the ``narrow neck'' or
``long tube'' limit.  Physically, singularities arising from integration in this region are on-shell or infrared singularities.
Such on-shell and infrared singularities are crucial in the physical interpretation of the theory, just as they are in ordinary
quantum field theory.

\subsubsection{Conditionally Convergent Integrals}\label{condcon}

Although there are no ultraviolet issues in superstring perturbation theory, one often runs into integrals that are
only conditionally convergent in the infrared region, that is, in the region in which a narrow neck is collapsing.
We have already discussed an example at length, and a large class of additional examples is described in section \ref{tadpole}. 
A general method of treating conditionally convergent integrals is needed.

Such integrals can always be treated by a simple generalization of the procedure explained in section \ref{moresplit}.
What one needs to know is that, roughly speaking (see the last paragraph of this section for a clarification),
there is a distinguished parameter controlling the collapse of a narrow neck in a super Riemann surface.  This parameter 
is the superanalog of the parameter $q$ in eqn. (\ref{monkey}).   For the case of a Neveu-Schwarz (NS) 
degeneration -- that is, the case that the closed string state propagating through the narrow neck is in the NS
sector -- the analog of eqn. (\ref{monkey}) is as follows.  A superdisc parametrized by $x|\theta$ can be glued to
a superdisc parametrized by $y|\psi$ via
\begin{align}\label{melk} xy& = -\varepsilon^2\cr
                                             x\psi&=\varepsilon\theta\cr
                                              y\theta&=-\varepsilon \psi\cr
                                               \theta\psi&=0. \end{align}
 This change of coordinates from $x|\theta$ to $y|\psi$ is superconformal.                                              
(This formula and its Ramond sector analog, which is presented in eqn. (\ref{medzo}) below, are originally due to P. Deligne.  For more, see section 6.2 of  \cite{Surfaces}.)  The closest
analog of the gluing parameter $q$ of bosonic string theory is $q_\NS=-\varepsilon^2$.  (For a given value of $q_\NS$,
there are two choices of $\varepsilon$, which correspond to two possible ways of gluing together the spin structures
on the two superdiscs; the sum over the two choices leads to the GSO projection on the string state propagating
through the neck.)

As an example of this, consider the case that the degeneration arises from two points $z|\theta$ and $z'|\theta'$ approaching
each other on a super Riemann surface $\Sigma$.  In this case, the parameter $q_\NS$ turns out to coincide with the
supersymmetric combination $\h z=z-z'-\theta\theta'$:
\begin{equation}\label{udonk}q_\NS=\h z=z-z'-\theta\theta'. \end{equation}
(This is shown in section 6.3.2 of \cite{Surfaces}.)  The use of this parameter, rather than of the more naive $z-z'$, was
the key insight in \cite{GrSe}, as described in section \ref{moresplit} above.  
With $\h z$ replaced by $q_\NS$, the procedure described
there generalizes almost immediately to treat all of the conditionally convergent integrals of superstring perturbation theory.
We spell this out in section \ref{regu}.

The above explanation has been oversimplified in one important respect.  
The gluing parameters $q$ or $\varepsilon$ are not well-defined as complex numbers, because their definition
depends on local parameters that vanish to first order at the points at which the gluing occurs.
For example, if we glue the point $x=0$ in $\Sigma_\ell$ to the point $y=0$ in $\SIgma_r$ and then deform via
\begin{equation}\label{pelmo} xy = q,\end{equation}
then obviously the definition of $q$ depends on the choice of the functions  $x$ and $y$.  If we rescale $x$ and $y$,
$q$ is also rescaled.  Similarly,
in the example of the last paragraph, a superconformal transformation acting by
$z|\theta\to \lambda z|\lambda^{1/2}\theta$,
$z'|\theta'\to \lambda z'|\lambda^{1/2}\theta'$ (where $\lambda$ is nonzero and independent of $z|\theta$
and $z'|\theta'$, but may depend on other moduli),
will multiply $q_\NS$ by $\lambda$.
But crucially it does not change the relevant coefficient between the $z-z'$ and $\theta\theta'$ terms.
The precise statement here is not that there is a distinguished parameter $q$ or $\varepsilon$ but that compactification of
the moduli space or supermoduli space $\MM$ 
is achieved by adding a divisor $\frak D$, along which $q$ or $\varepsilon$ has
a simple zero.  This condition determines $q$ or $\varepsilon$ only up to multiplication by an invertible function, that is
\begin{equation}\label{mitto} \varepsilon\to e^f \varepsilon\end{equation}
or a similar rescaling of $q$,
and in general there is no way to be more precise.
One may say that (modulo $q^2$ or $\varepsilon^2$)  $q$ or $\varepsilon$ is not a complex number but a linear 
function on the fiber of a complex
line bundle, namely the normal bundle $\frak N$ to the divisor $\frak D\subset\MM$.   See section 6.3 of \cite{Surfaces} for more. 

\subsubsection{Regularization}\label{regu}
Now we will state what we claim is the appropriate  procedure for regularizing conditionally convergent integrals.
For definiteness, let us consider a degeneration of a heterotic string worldsheet $\Sigma$.  The procedure for other
superstring theories is similar.  From a holomorphic point of view, we describe a heterotic string worldsheet near 
the degeneration by local parameters $x|\theta$ and $y|\psi$, glued together as in (\ref{melk}) with a gluing parameter $q_\NS$.  Antiholomorphically, 
we use local parameters $\t x$ and $\t y$, glued by
\begin{equation}\label{zelk} \t x\t y = \t q. \end{equation}
Roughly speaking, $\t x$, $\t y$, and $\t q$ are the complex conjugates of $x,y$, and $q_\NS$.  We will be more precise
about this in section \ref{gluehol}.

We regularize the conditionally convergent integrals of superstring perturbation theory as in section \ref{variant}.
We pick a small positive $\eta$, and restrict the integral to $|\t qq_\NS|^{1/2}\geq \eta$, and then finally take the limit
as $\eta\to 0$.  This generalizes the procedure that we explained in a special case in section \ref{moresplit}.  It is a satisfactory
procedure because it is a conformally- or superconformally-invariant procedure that
makes all conditionally convergent integrals well-defined and is compatible with any further degenerations and thus with unitarity. 

There are two points on which what we have said is incomplete or oversimplified.  First, we  have not explained the
relation between $\t q$ and $q_{\NS}$.  Roughly speaking, one treats $\t q$ as the complex conjugate of $q_{\NS}$.
For a fuller explanation, see section \ref{gluehol}.  Second, the fact that $\t q$ and $q_\NS$ are linear functions on
the appropriate normal bundles  rather than complex numbers means that $\eta$ is really a hermitian metric 
on an appropriate
line bundle (or more accurately, a sesquilinear form)
rather than a positive real number.  This leads to some subtleties, which have nothing
to do with worldsheet or spacetime supersymmetry, and appear already in bosonic string theory.   See \cite{Revisited},
especially sections 7.6 and 7.7.

\subsubsection{Gluing Holomorphic And Antiholomorphic Coordinates}\label{gluehol}

In the language of section 5 of \cite{Supermanifold}, what we will describe next is the integration
cycle $\varGamma$ of superstring perturbation theory.  We aim for an informal explanation.

A point on a heterotic string worldsheet has holomorphic coordinates $z|\theta$ and an antiholomorphic
coordinate $\t z$.  Naively, $\t z$ is the complex conjugate of $z$, but this is oversimplified since a statement
$\bar{\t z}=z$ is not invariant under odd superconformal transformations, which act by
\begin{equation}\label{indu} \delta z=-\alpha(z) \theta,~~\delta\theta=\alpha(z),~~\delta\t z=0. \end{equation}

So what is the relation between $\t z$ and $z$?  For many purposes, one does not need to specify the precise relationship.
The classical action, the vertex operators, and the correlation functions are all real-analytic. So to some
extent one may think of $\t z$ and $z$ as independent complex variables, as long as one does not go too far
away from $\bar{\t z}=z$.  

It is really when one wants to integrate over moduli space that one needs to specify a relationship between holomorphic
and antiholomorphic variables.  For example, in section \ref{moresplit}, to evaluate a 1-loop mass shift, we needed to
compute an integral
\begin{equation}\label{murok}\int\d\t z\,\d z\,\d\theta\,\d\theta' \,F(\t z;\neg z|\theta,\theta'), \end{equation}
where $\t z;\neg z|\theta$ were the coordinates of one puncture and $\theta'$ was the odd coordinate of another puncture.
To evaluate such an integral, one cannot just vaguely treat $\t z $ and $z$ as independent complex variables.
One needs to specify a relationship between them.  What is a natural relationship?  As we have explained,
setting $\bar{\t z}=z$ is not really natural, since it is not consistent with superconformal symmetry.  It is natural, however,
to assert that $\bar{\t z}=z$ modulo nilpotent terms.  In the present example, with only two odd moduli $\theta$ and $\theta'$,
the most general nilpotent term is $h(\t z; \neg z)\theta\theta'$ for some function $h(\t z;\neg z)$, and therefore it is natural
to say that the relationship between ${\t z}$ and $z$ should take the form
\begin{equation}\label{urok}\bar{\t z}=z-h(\t z;\neg z)\theta\theta' \end{equation}
for some $h$.  The procedure of section \ref{moresplit} tells us that we will want $h(0;\neg 0)=1$.  The function $h(\t z;\neg z)$ 
must also obey
\begin{equation}\label{melmo}h(\t z;\neg z) = -h(\t z+1;\neg z+1)=h(\t z+\bar\tau;\neg z+\tau) \end{equation}
to respect the equivalences (\ref{mune}) and (\ref{zune}) (note that $\theta'$ is invariant under those equivalences 
since it
is the odd coordinate of a point whose even coordinates are $\t z'=z'=0$).

We do not try to pick a particular $h$ because no preferred function $h$ obeying $h(0;\neg 0)=1$ and also consistent
with (\ref{melmo}) presents itself.  Instead we proceed by showing that, as long as some general conditions are imposed,
the precise choice of $h$ does not affect the integral.  If the measures that we are integrating extend as
smooth measures over the Deligne-Mumford compactification $\h\MM$ of $\MM$ (so that there is no possible problem in
integration by parts), then the integrals are entirely independent of $h$.  This follows from the supermanifold
version of Stokes's theorem.  In essence, a change in $h$ by $h\to h+\delta h$ can be compensated 
by a change of coordinates on moduli
space
\begin{equation}\label{pumelo} z\to z+\delta h(\t z;\neg z) \theta\theta' \end{equation} 
(with no change in $\theta,\theta',$ or $\t z$).  Integration of a smooth measure
on a compact supermanifold is invariant under any change of
coordinates, so if the integration measure extends smoothly over $\h\MM$, the integral is
completely independent of $h$.   

We are not in the situation assumed in the last paragraph, because the measures we want to integrate do not extend
smoothly over the compactification.  For instance, in section \ref{moresplit}, we wanted to integrate a measure that
behaves near $z=\t z=0$ as
\begin{equation}\label{urwu} \J\sim -i\d\t z\,\d z\,\d\theta\,\d\theta'  \,\frac{1}{\t z}.  \end{equation} 
Such an integral is only conditionally convergent, and its evaluation depends on an infrared regulator.  To preserve the
regularization, which depended on the function $\h z=z-\theta\theta'$ (or more precisely on the divisor in $\h\M$ determined
by vanishing of this function)  we must require $\delta h(0;\neg 0)=0$.
Indeed, we saw in section \ref{moresplit} that the integral depends on $h(\t z;\neg z)$ only via $h(0;\neg 0)$.

Why is $h(0;\neg 0)=1$ the correct condition?  Apart from what was explained in section \ref{moresplit},
one answer is that this is the only condition
that can be stated just in terms of the natural gluing parameters, and which therefore is superconformally-invariant and
capable of generalization.
We recall that holomorphically, the natural gluing parameter in this example is $q_\NS=\h z=z-\theta\theta'$, while
antiholomorphically, the natural gluing parameter is $\t q = \t z$.  So the condition  $h(0;\neg 0)=1$ is equivalent
to the statement that the relation between $\t q$ and $q_\NS$ near $\t q=q_\NS=0$ is
\begin{equation}\label{usfu} \bar{\t q}=q_\NS(1+\dots)\end{equation}
where the ellipses represent nilpotent terms.  (It is not possible to specify the relation between $\t q$
and $q$ more precisely than this, because of facts noted at the end of section \ref{condcon}.)

Now we have all the ingredients to explain the  general relation between holomorphic and antiholomorphic moduli
in the context of heterotic string perturbation theory.  (See section 6.5 of \cite{Revisited} for a fuller explanation
and generalization to the other superstring theories.)  Let us consider a situation in which the moduli space of
bosonic Riemann surfaces has complex dimension $r$ and the moduli space of super Riemann surfaces has
dimension $r|s$.  Thus, antiholomorphically, a heterotic string worldsheet has even moduli $\t m_1,\dots,\t m_r$,
while holomorphically it has even and odd moduli $m_1,\dots,m_r|\eta_1\dots\eta_s$.  What is the relation between
holomorphic and antiholomorphic moduli?  Naively the $\t m_\alpha$ are the complex conjugates of the $m_\alpha$.
But this is too strong a claim because it is not invariant under general reparametrizations of the supermoduli
space.  If one shifts the $m_\alpha$ by functions that depend on the $\eta$'s, for example
\begin{equation}\label{tolfo} m_\alpha\to m_\alpha+\sum_{ij}\eta_i \eta_j f^{ij}_\alpha(m_1,\dots,m_p),\end{equation}
then there is nothing one can do\footnote{We are not allowed to shift the $\t m_\alpha$ by a function antiholomorphic
in the $\eta_i$, since the complex conjugates of the $\eta_i$ are not present in the formalism of the heterotic string.}
 to the $\t m_\alpha$'s to preserve a hypothetical relationship $\bar{\t m_\alpha}=m_\alpha$.
So unless one can identify a distinguished set of even functions that one wants to call the $m_\alpha$ -- in a way compatible
with modular invariance -- one does not want to claim that $\bar{\t m_\alpha}=m_\alpha$.    

In general, we do not have a distinguished set of even coordinates on supermoduli space.  Given one set of local coordinates,
another set differing as in (\ref{tolfo}) is equally natural.   Therefore it is not natural
to pick particular coordinates and impose $\bar{ \t m_\alpha}=m_\alpha$ in that coordinate system.  But it is certainly
natural to insist that this is true modulo nilpotent terms:
\begin{equation}\label{mezmo}\bar{\t m_\alpha}=m_\alpha+\mathrm{nilpotent}\,\,\mathrm{corrections}.\end{equation}
The nilpotent corrections generalize $h(\t z;\neg z)\theta\theta'$ in eqn. (\ref{urok}).  To define the cycle $\varGamma$ over which we will integrate to compute a heterotic string scattering amplitude, we have to make some choice
of the nilpotent terms; as there is no natural choice in general, the goal has to be to show the choice does
not matter.  Now we can repeat everything
that we have said  in our illustrative example.  If the measures we want
to integrate extend smoothly over the compactification $\h\MM$ of the supermoduli space, then by the supermanifold
version of Stokes's theorem, the choice of the nilpotent terms -- or in other words the precise choice of 
the integration cycle $\varGamma$ -- would not matter.  Actually, we
want to integrate measures that are singular along certain divisors $\frak D\subset \h\MM$ along which $\Sigma$ degenerates.
Because of this, we need to impose a condition on how the nilpotent terms in (\ref{mezmo}) behave near $\frak D$.
Along  $\frak D$, one of the ${\t m}_\alpha$ plays a special role, namely the antiholomorphic gluing parameter $\t q$
that has a simple zero along $\frak D$.  Similarly, one of the holomorphic moduli plays a special role near $\frak D$, namely
the holomorphic gluing parameter $q_\NS$.  While placing no condition on the other $\t m_\alpha$'s beyond (\ref{mezmo}),
we need to be more precise about how $\t q$ is related to the holomorphic moduli along $\frak D$.   The condition
we need is that of eqn. (\ref{usfu}):
\begin{equation}\label{busfu} \bar{\t q}=q_\NS(1+\dots), \end{equation}
where again the ellipses represent nilpotent terms.  This is the only general condition that one can formulate in terms
of the available data.    It  suffices (when combined with a condition $|\t q|\geq \eta$, followed by a limit $\eta\to 0$) to regularize all conditionally convergent integrals
of superstring perturbation theory, since the singular behavior of the integration measure near 
$\frak D$ is always controlled by the natural gluing
parameters $\t q$ and $q_\NS$.

We conclude with one further comment that is useful background for section \ref{simplesplit}.  Let us consider
a problem in which the appropriate moduli space $\MM$ has  only one odd modulus.  Then there are no (nonzero) even nilpotent functions on $\MM$, so we cannot
make a change of variables as in (\ref{tolfo}), and there is no way to include nilpotent terms in the relationship (\ref{mezmo}).
Hence none of the characteristic subtleties of superstring perturbation theory
can arise.  Superstring calculations in problems with only one odd modulus (or none at all)
can be subtle, but the subtleties always
involve issues that could arise in bosonic string theory.

\subsection{A Much Simpler 1-Loop Mass Splitting}\label{simplesplit}

Returning to the $SO(32)$ heterotic string on a Calabi-Yau manifold, we are now going to compute a 1-loop mass shift
for a different set of fields.  There actually are three reasons to do this calculation: it is interesting; we will need the result
in section \ref{dilaton}; and it will illustrate our last assertion, namely that superstring perturbation theory is straightforward
when there is only one odd modulus. 

We will consider a genus one Riemann surface $\Sigma$ with two Ramond punctures -- that is, two insertions of Ramond
vertex operators.   For a review of the basics of super Riemann surfaces with punctures, see section 4 of \cite{Surfaces}.
The key point for us is that while 
adding a Neveu-Schwarz puncture increases the odd dimension of supermoduli space by 1 (the extra odd
modulus being the odd coordinate of the puncture), adding a Ramond puncture only increases the dimension of
supermoduli space by 1/2.   So a super Riemann surface of genus 1 with 2 Ramond vertex operators (and none of NS type)
has only 1 odd modulus, ensuring that superstring perturbation theory will be straightforward.  
To calculate, we should use Ramond vertex operators with the canonical picture number\footnote{At tree
level, one can compute with vertex operators of any picture number \cite{FMS}, but in positive genus, the picture-number of the vertex
operators  must be correlated with how the supermoduli space is defined.  See section 4.3 of \cite{Surfaces} and sections
4.1 and 5.1 of \cite{Revisited}.
 There is a systematic procedure to compute with
(unintegrated) NS and Ramond vertex operators with any negative value of the picture number, but the minimal procedure is based on the simplest
definition of the supermoduli space, in conjunction with
NS vertex operators of picture number $-1$ (such as we used in section \ref{took}) and Ramond vertex operators of picture number $-1/2$.
  We will be able in what follows to effectively convert one of the Ramond vertex
operators to picture number $+1/2$ using a picture-changing operator, but the justification for this depends on the fact that in the particular problem we consider,
there is only one odd modulus.} $-1/2$.

Now we will explain the term in the effective action that we aim to compute here.  We let $W_\alpha$
be the chiral superfield on $\R^4$ that contains the field strength $F_{\mu\nu}$ of the anomalous $U(1)$ gauge field.  Its expansion is
\begin{equation}\label{melbot}W_\alpha(x^\mu|\theta^\beta)=\zeta_\alpha+F_{\mu\nu}\sigma^{\mu\nu}_{\alpha\beta}\theta^\beta+
\theta_\alpha D+\dots\end{equation}
where $\zeta_\alpha$ is the fermion field in this multiplet and $D$ 
is the auxiliary field whose expectation value breaks supersymmetry.  The kinetic energy of this multiplet at tree level is
\begin{equation}\label{welbo}\int \d^4 x\,\d^2\theta \,S W_\alpha W^\alpha,\end{equation}
where $S=e^{-2\phi}-ia+\theta^\alpha \kappa_\alpha+\dots$ is the chiral superfield containing the four-dimensional
dilaton; it was introduced
in eqn. (\ref{elbo}).   In view of the $\theta$ expansions of $W_\alpha$ and $S$, the interaction (\ref{welbo}) contains
a term $\kappa_\alpha \zeta^\alpha D$, and at 1-loop order we expect to generate a $\kappa_\alpha\zeta^\alpha$ mass
term that will be proportional to  $\langle V_D\rangle$.  We want to explain here how this comes about. We call $\kappa$
and $\zeta$ the dilatino and gaugino, respectively.

We will do the calculation directly at zero momentum in spacetime. There will be none of the subtleties familiar from
\cite{DIS,ADS} as well as sections \ref{took} and \ref{moresplit} above, because we are now considering a problem with only
one odd modulus.  

Concretely, what we gain from the fact that the worldsheet $\Sigma$ has only two Ramond punctures is the following. 
(Here we are more or less restating in the present context what was already explained in section \ref{gluehol}.)
Topologically, $\Sigma$ is a torus with a 
holomorphic even modulus $\tau$.  With two or more odd moduli, say $\eta_1,\eta_2$, it is subtle to explain what one means
by $\tau$ as opposed to, say,\footnote{An exception, which was important in section \ref{took}, is that if the odd moduli
are positions of NS vertex operators, then this particular difficulty does not arise.  That is because an NS vertex operator
is inserted at a point in a pre-existing super Riemann surface, whose moduli can be defined independently of the position
of the NS puncture.}  $\tau+\eta_1\eta_2$.  There consequently is not a natural operation of integrating over the odd moduli
at fixed $\tau$.  A meaningful answer in general emerges only after integrating over all even and odd variables.  With only one odd modulus,
$\tau$ is uniquely defined and there is a natural notion of integrating first over the odd modulus  and only at the 
end  over $\tau$.  (In this final integration, one takes $\tau$ and the antiholomorphic modulus $\t\tau$ to be 
complex conjugate.)

Similarly, with two or more odd moduli, it is subtle to define what one means by the positions at which vertex operators
are inserted, as $z$ can be confused with $z+\eta_1\eta_2$.  In either of these cases, a BRST transformation that changes
the position at which a picture-changing operator is inserted  can shift $\tau$ to $\tau+\eta_1\eta_2$
or $z$ to $z+\eta_1\eta_2$.  However, with only one odd modulus, such shifts are not possible, and
one can think of the vertex operators as being inserted at
well-defined positions on an underlying bosonic Riemann surface.  Moreover, there is no subtlety in the standard arguments \cite{FMS} stating that the position of a picture-changing insertion
is irrelevant.  

Given this, we can straightforwardly use the familiar formalism of fermion vertex operators and picture-changing
operators.  In this formalism, the basic  fermion emission vertex of picture number $-1/2$ is written as the product of a spin
field of the $\beta\gamma$ ghost system -- which is written as $e^{-\tphi/2}$ in the language of \cite{FMS} --  times
a spin field of the matter system.   In compactification on $\R^4\times \XX$, the matter spin fields we will need are products
\begin{equation}\label{dworf} \Sigma_{\alpha,\pm}=\SIgma_\alpha\cdot \Sigma_\pm,~~~~~\Sigma_{\dot\alpha,\pm}=\Sigma_{\dot\alpha}\cdot\Sigma_\pm,
\end{equation}
where the two factors are as follows.  $\Sigma_\alpha$ and  $\Sigma_{\dot \alpha}$,  $\alpha,\dot\alpha=1,2$, 
are spin fields of positive or negative chirality for the sigma-model with target  $\R^4$.  And  $\Sigma_+ $ and $\Sigma_-$
are spin fields of the sigma-model with target $\XX$ that are associated to a covariantly constant spinor on 
$\XX$ of  positive or negative chirality.
 Of the combined spin fields defined in eqn. (\ref{dworf}), 
 $\Sigma_{\alpha,+}$ and $\Sigma_{\dot\alpha,-}$ are GSO-even and the others are GSO-odd.  

The vertex operator of a zero-momentum gaugino, in the $-1/2$ picture, is
\begin{equation}\label{numbo} V^\zeta_\alpha =J_\ell\cdot e^{-\tphi/2}\Sigma_{\alpha,+}. \end{equation}
The holomorphic factor $e^{-\tphi/2}\Sigma_{\alpha,+}$ was described in the last paragraph, while
the antiholomorphic factor is    the antiholomorphic worldsheet
current
 $J_\ell = g_{\bar i i}\lambda^{\bar i}\lambda^i$,  introduced in eqn. (\ref{irox}), which is associated to the anomalous $U(1)$ gauge symmetry in spacetime.  
Similarly, the vertex operator of the dilatino at zero momentum, again in the $-1/2$ picture, is
\begin{equation}\label{lumbo} V^\kappa_\alpha=\partial_{\t z}X^\mu\gamma_{\mu\alpha\dot\alpha}\epsilon^{\dot\alpha\dot\beta}
  e^{-\tphi/2}\Sigma_{\dot\beta,-}, \end{equation}
  where $\gamma_\mu$ are the four-dimensional Dirac matrices.  
  
  A consequence of setting the spacetime momentum to zero is that the field $X^\mu$ associated to motion of the string
  in $\R^4$ appears in the vertex operators only via the factor $\partial_{\t z}X^\mu$ in $V^\kappa_\alpha$.  Since
  the one-point function of $\partial_{\t z}X^\mu$ certainly vanishes, for instance by Lorentz symmetry, the expectation value of the product
  $V^\zeta_\alpha V^\kappa_\beta$ is trivially zero in the absence of additional insertions.  But one more operator
  must be inserted, namely the picture-changing operator ${\Y}=\{Q,\xi\}=e^\tphi(\psi_\mu\partial X^\mu+\dots),$ where the omitted
  terms are not relevant since they do not depend on $X^\mu$. 
  
 For fixed $\tau$, the integral that we have to evaluate to compute the $\zeta\kappa$ mass term is
 \begin{equation}\label{melfry}\J_{\alpha\beta}=
 -i\int \d\bar z\,\d z\,\bigl\langle V^\kappa_\alpha(\bar z;\neg z) \cdot e^\tphi \psi_\nu \partial_w X^\nu(w)\cdot V^\zeta_\beta(0;\neg 0)   \bigr\rangle ,\end{equation}
where because there is only one odd modulus, there is no need to distinguish $\t z$ from $\bar z$.  We have replaced $\Y$ by its relevant
piece $e^\tphi \psi_\mu\partial X^\mu$, and the point $w$ at which it is inserted is completely arbitrary.    
The $X^\mu$ correlator that we have to evaluate
is $\langle \partial_{\t z}X^\mu(\bar z;\neg z) \partial_w X^\nu(\bar w;\neg w)\rangle$.  This can be evaluated in a simple way because, for the free fields
$X^\mu$, the holomorphic and antiholomorphic operators $\partial_z X^\nu$ and $\partial_{\t z}X^\mu$ decouple except for the
effects of zero-modes.  The result (with $\alpha'=1/2$) is $\langle \partial_{\t z}X^\mu(\bar z;\neg z) \partial_w X^\nu(\bar w;\neg w)\rangle=\eta^{\mu\nu}/2\,\mathrm{Im}\,\tau$, independent of $z$ and $w$.    Moreover, we are free to make a convenient choice of $w$, and we choose
to take the limit that $w$ approaches $z$.  To take this limit, we need the operator product relation
\begin{equation}\label{curmy}e^\tphi \psi^\mu(w)\cdot e^{-\tphi/2}\gamma_{\mu\alpha\dot\alpha}\epsilon^{\dot\alpha\dot\beta}
\Sigma_{\dot\beta,-}( z) \to W^\kappa_\alpha(z), ~~~w\to z,\end{equation}
where 
\begin{equation}\label{urmy}W^\kappa_\alpha=e^{\tphi/2}\Sigma_{\alpha,-}.\end{equation}
All operators appearing in eqn. (\ref{curmy}) are holomorphic, and in particular $W^\kappa_\alpha$ is holomorphic and has dimension 0.

So our integral reduces to 
\begin{equation}\label{medolo}\J_{\alpha\beta}=-\frac{i}{2\,\mathrm{Im}\,\tau}\int \d\bar z\,\d z\bigl\langle W^\kappa_\alpha(z)\, 
V^\zeta_\beta(0;\neg 0)\bigr\rangle.\end{equation}
This integral is easily evaluated because the operator $W_\alpha^\kappa$ is holomorphic.
The correlator $F(z)= \bigl\langle W^\kappa_\alpha(z)\, 
V^\zeta_\beta(0;\neg 0)\bigr\rangle     $ that appears in (\ref{medolo}) is therefore a holomorphic function of $z$, apart from some poles
that can be understood using the operator product expansion.

The function $F(z)$ is not invariant under $z\to z+1$ or $z\to z+\tau$, because in general 
moving a Ramond vertex operator around a noncontractible  loop permutes the generalized spin structures\footnote{
\label{generalized} In the presence of Ramond
punctures, spin structures are replaced by generalized spin structures, defined for instance in section 4.2.4 of \cite{Surfaces}.  The distinction
is not important for what follows.}  on a super Riemann surface $\Sigma$.  We do have  $F(z)=F(z+2)=F(z+2\tau)$, since moving
around the same loop twice returns us to the original generalized spin structure.  We can sum over the generalized spin structures by simply
replacing $F(z)$ with $G(z)=\frac{1}{2}(F(z)+F(z+1)+F(z+\tau)+F(z+1+\tau))$.  (The reason for the factor of $1/2$ is that in genus $g$, the sum over
spin structures is accompanied by a factor of $2^{-g}$.) The fact that there is a meaningful way to sum over generalized spin structures
before integrating over $z$ and $\tau$ is another reflection of the fact that in this problem, because there is only 1 odd modulus, $z$ and $\tau$ are well-defined.

The singularities of $F(z)$ at $z=0$  are determined by the operator product expansion:
\begin{equation}\label{poklo} W_\alpha^\kappa(z)\cdot  V^\zeta_\beta(0;\neg 0)\sim \epsilon_{\alpha\beta}\frac{J_\ell(0)}{z} +\epsilon_{\alpha\beta}V_D(0;\neg 0)+\O(z), \end{equation}
where $V_D=J_\ell J_r$, introduced in eqn. (\ref{winzo}), is the vertex operator of the auxiliary field $D$.  
So $F(z)$ has a pole at $z=0$, and it also has poles at $z=1,\tau,$ and $1+\tau$ that are governed by the same formula with a different
generalized spin structure.  
However, the residue of the poles is given by the one-point function of the operator
$J_\ell$, which does not couple to right-moving RNS fermions.   The insertion of this operator does not disturb the GSO cancellation, and after summing
over spin structures, the contribution to $F(z)$ that is proportional to the one-point function of $J_\ell$ disappears.  After summing
over spin structures, $F(z)$ becomes a constant   $\langle V_D(0;\neg 0)\rangle$.   So 
 the only integral we really have to do is the one that computes the volume of the torus:
$-i\int_\Sigma \d\t z\,\d z=2\,\mathrm{Im}\,\tau$.   This cancels the factor of $\mathrm{Im}\,\tau$ in eqn. (\ref{medolo}), and
the final integral over $\tau$ is the same as it was in sections \ref{took} and \ref{moresplit}.

Thus the supersymmetry-violating  1-loop $\kappa\zeta$ mass term is proportional to $\langle V_D\rangle$ with a universal coefficient, just like the
other supersymmetry-violating 1-loop mass terms that we reviewed in sections \ref{took} and \ref{moresplit}.  The analysis, however, was notably
more straightforward: there were no $k^2/k^2$ terms, and no need to regularize a 
conditionally convergent integral.  The reason for this is that the computation
involved a supermoduli space with only one odd modulus.  

At least one aspect of this calculation  perhaps requires better explanation.  We have followed \cite{FMS} and represented
the $\beta\gamma$ system of the RNS model in terms of ``bosonized'' 
fields $\tphi,\xi,$ and $\eta$.   This is an extremely powerful method to describe the operators of the 
$\beta\gamma$ system, including the spin fields,
 determine their dimensions and operator product expansions, and compute correlation functions in genus 0.
In positive genus, the description of the $\beta\gamma$ system via $\tphi,\xi,$ and $\eta$ is tricky, since  
these fields have zero-modes on a surface
of positive genus whose interpretation and proper treatment
is not very transparent.  For an analysis of the $\tphi\xi\eta$ system in positive genus, see
\cite{VV}.  In the foregoing, we used the fields  $\tphi,\xi,$ and $\eta$ only to construct certain holomorphic operators and 
compute some terms
in their
operator product expansion.  For these purposes there is no problem. The analysis led to a final answer 
$\langle V_D\rangle$ which is most transparently computed using the original
$\beta\gamma$ variables.  (The role that $\beta$ and $\gamma$
 play in the 1-loop evaluation of $\langle V_D\rangle$ is very simple: their determinant cancels the determinant
of two of the RNS fermions $\psi^\mu$.)  Conceptually, one might prefer to perform
the entire computation in terms of the variables $\beta,\gamma$ whose geometrical meaning is clear; for some direct
approaches to the $\beta\gamma$ system on a surface of positive genus, see \cite{Lechtenfeld} or section 10 of \cite{Revisited}.

\subsection{More On The GSO Cancellation}\label{alphadetail}

\def\Spin{{\mathrm{Spin}}}
\def\8{\mathbf{8}}
\def\1{\mathbf{1}}
\def\3{{\mathbf 3}}
At several points in this analysis, starting in the discussion of eqn. (\ref{mizzo}), we invoked the GSO cancellation to claim that certain terms vanish upon
summing over spin structures.  Such claims are straightforward if $\XX$ is a Calabi-Yau orbifold and the vertex operators come from the untwisted sector.
Here we wish to explain why these claims hold in the case that $\XX$ is a general Calabi-Yau manifold.   In this section, $\Sigma$ is always an ordinary
Riemann surface of genus 1.  

First let us recall how one sees spacetime supersymmetry in superstring theory at 1-loop order
in light-cone gauge.  In the RNS description, the fields that are sensitive to the spin structure of $\Sigma$ are
 ten right-moving worldsheet fermions $\psi^I$, $I=0,\dots,9$ and the commuting ghosts $\beta$ and $\gamma$.  In computing
the partition function, the determinant of the $\beta\gamma$ system cancels the determinant of two of the $\psi^I$, say $\psi^0$ and $\psi^1$.
In light-cone gauge, all external string states are represented by vertex operators that do not disturb this cancellation.  Given this, the theory is described by eight RNS fermions
$\psi^I$, $I=2,\dots,9$, along with other fields not sensitive to the spin structure.  The $\psi^I$ transform in the vector representation of $SO(8)$ or $\Spin(8)$.  
We recall that the group $\Spin(8)$ has three representations of dimension eight, namely 
the vector representation, which we denote as $\8$,
and the two spinor representations of definite chirality, which we call $\8'$ and $\8''$.  We now use the following fact about two-dimensional
conformal field theory: in genus 1, eight fermions $\psi^I$ transforming in the representation $\8$ and with a 
sum\footnote{This ``sum'' requires a choice of sign: the path integral measure of the $\psi^I$ has a natural sign
if the spin structure of $\Sigma$ is even, but if it is is odd, the measure has no natural sign and one has to pick one.   
This choice, which determines the sign of parity-violating amplitudes in the string theory,
determines whether the $\Theta^\alpha$ transform in the $\8'$ or $\8''$ representation of $\Spin(8)$.}   over spin structures are equivalent
to eight fermions $\Theta^\alpha$ transforming in the $\8'$ and with an odd spin 
structure.

The $\Theta^\alpha$ are known as light-cone Green-Schwarz fermions.  Since they are coupled to an odd spin
structure on $\Sigma$,  the $\Theta^\alpha$  are completely
periodic, $\Theta^\alpha(z)=\Theta^\alpha(z+1)=\Theta^\alpha(z+\tau)$.  In particular, they have constant zero-modes.  Integration over those zero-modes causes
the partition function to vanish.  In terms of the $\psi^I$, that is the basic GSO cancellation in the sum over spin structures.  

Now let us pick an $SU(3)$ subgroup of $\Spin(8)$ (in our application, this will be the holonomy group of $\XX$) under which the representation $\8$
decomposes as $\1\oplus \1\oplus \3\oplus\bar\3$.  It is likewise true that the representation $\8'$ (or $\8''$) decomposes under the same $SU(3)$ as
$\1\oplus \1\oplus \3\oplus \bar\3$.  So two of the $\Theta^\alpha$, say $\Theta^1$ and $\Theta^2$, are $SU(3)$ singlets.  Consequently, arbitrary insertions of
$SU(3)$ currents do not disturb the fact that $\Theta^1$ and $\Theta^2$ have zero-modes.  These zero-modes ensure the vanishing of the partition function, and thus,
arbitrary insertions of $SU(3)$ currents do not disturb the GSO cancellation.

Next, let us replace $\R^{10}$ by $\R^4\times \XX$ with a general Calabi-Yau manifold $\XX$.   We want to explain why the GSO projection holds for the
one-loop partition function of such a sigma-model.  This will make it clear to what extent it holds for correlation functions.  
We consider the eight RNS fermions $\psi^I$ of the above discussion (after canceling $\psi^0$ and $\psi^1$ against $\beta$ and $\gamma$) to be
modes in a sigma-model with target $\R^2\times \XX$.  Let $\T$ denote the tangent bundle to $\R^2\times \XX$.  Its holonomy group is $SU(3)$, since $\R^2$
has trivial holonomy and $\XX$ has holonomy $SU(3)$. The bosonic fields of the sigma-model comprise a map
 $\Phi:\SIgma\to \R^2\times \XX$, and the $\psi^I$ take values in the pullback $\Phi^*(\T)$.  The $\R^2\times \XX$ partition function in genus 1
 actually vanishes
 for fixed $\Phi$ and fixed values of the left-moving fermions $\lambda$
 after integrating over the $\psi^I$ and summing over spin structures.  As we will see momentarily, that is true because, for fixed 
 $\Phi$ and $\lambda$, the $\psi^I$ are coupled
 only to $SU(3)$ currents, so the reasoning in the last paragraph applies.  The part of the action that involves the $\psi^I$ is
 \begin{equation}\label{melofx} \int_\SIgma\d^2 z\left(\frac{1}{2}\psi^I \frac{D}{D\bar z}\psi^I 
 +\frac{1}{4}\lambda^i\lambda^{\bar i}R_{i\bar i j\bar j}\psi^j\psi^{\bar j}\right).
 \end{equation} Here $R_{i\bar i j \bar j}$ is the Riemann tensor of $\R^2\times \XX$. The connection that is hidden 
 in the kinetic operator $D/D\bar z$ is the pullback by $\Phi$ of the connection on $\T$, so it has structure group $SU(3)$,
 meaning that in the kinetic energy, the $\psi^I$ couple to $\Phi$ only via their $SU(3)$ currents.  To analyze the $\lambda^2 R\psi^2$ term,
 note that the only non-trivial 
 part of the Riemann tensor of $\R^2\times \XX$ is the
 Riemann tensor of $\XX$.   The fact that $\XX$ is Kahler implies that in its last two indices the Riemann tensor of $\XX$ is of  type $(1,1)$ and so transforms
  in general  as 
 $\3\otimes \bar \3=\1\oplus \8$ of $SU(3)$.
 But Ricci-flatness implies that the Riemann tensor  of $\XX$ is traceless in its last two indices (in other words $g^{j\bar j}R_{i\bar i j \bar j}=0$), which precisely 
means that the $\1$ contribution is absent.  So the $\lambda\lambda R\psi\psi$ term in the 
 action couples to a bilinear in $\psi$
 transforming as the $\8$ or adjoint representation of $SU(3)$. In other words,  the $\psi^I$ only couple via 
 $SU(3)$ currents, ensuring the GSO cancellation.
 
Now it is clear what we can do without disturbing the GSO cancellation: we can make arbitrary insertions 
of operators that only couple to the $\psi^I$ through
$SU(3)$ currents.  This statement is enough to justify all uses we have made of the GSO cancellation. 
As an example, consider the discussion following
eqn. (\ref{poklo}) of the poles in the function $F(z)$.  In this case, the operator that is inserted is $J_\ell(0)$, 
which does not couple directly to the $\psi^I$ at all,
so it certainly does not affect the GSO cancellation.   Similarly, insertions of
the $D_j w_{i\bar i}$ term in eqn. (\ref{turkey}) do not disturb the GSO cancellation (in this case, the 
relation $g^{j\bar i}D_j w_{i\bar i}=0$ for a harmonic
form on a manifold of $SU(3)$ holonomy ensures that this term couples to $\psi$ only via $SU(3)$ currents). 
 
The above discussion is oversimplified in one important respect.  In string theory, the statement 
that $\XX$ has $SU(3)$ holonomy is only valid in the large volume limit; in sigma-model perturbation theory, there are $\alpha'$ corrections
to the metric\footnote{For heterotic string models with the spin connection embedded in the gauge group
in the usual way, the leading correction to the metric is of order $\alpha'^3$ (this correction reflects the four-loop beta function
computed in \cite{gvz}) and the leading correction
to the Riemannian connection is of order $\alpha'^4$.  This is actually too high an order to be relevant
to our analysis of the one-loop $D$-term, since that effect is of order $\alpha'^3$ (it involves a topological
invariant $\int_\XX\tr_{SU(3)} F^3$, which is of order $\alpha'^3$ relative to the volume of $\XX$ in
string units).  The argument given in the text applies more broadly to any heterotic string compactification with
$\N=1$ supersymmetry in four dimensions.}
 of $\XX$.  The exact statement is not that $\XX$ has $SU(3)$ holonomy, but that the sigma-model of
$\R^2\times \XX$ has a pair of holomorphic spin fields $\hat\Sigma_\pm$ of dimension $1/2$.   
The triality transformation
to light-cone Green-Schwarz fermions  maps the spin fields
 $\hat\Sigma_\pm$ to two of the Green-Schwarz fermions, which in the above notation are 
 $\Theta^1\pm i\Theta^2$. The
fact that $\hat\Sigma_\pm$ are holomorphic means that $\Theta^1 $ and $\Theta^2$ are free fields; their zero-modes
give the GSO cancellation.   The reasoning of the last paragraph can be restated more accurately 
in this language.
The operator $J_\ell$, being antiholomorphic, certainly does not disturb holomorphy of $\hat\Sigma_\pm$.   
And the
vertex operator $V_{T,k}$ of eqn. (\ref{turkey}) is the vertex operator of a massless chiral superfield.  At $k=0$,
to first order, turning on this field does not disturb spacetime supersymmetry.  Hence this operator, including the
 $D_j w_{i\bar i}$ term as well as $\alpha'$ corrections, commutes with the $\hat\Sigma_\pm$ and its insertion does
not disturb holomorphy of $\hat\Sigma_\pm$ or the GSO cancellation.

\section{The Two-Loop Vacuum Energy}\label{dilaton}

\subsection{Overview}\label{overview}
If supersymmetry is spontaneously broken at 1-loop order, then at 2-loop order, we expect to generate a non-trivial vacuum energy.   
To understand how
this comes about, we first review the basic framework  for computing the genus 2 vacuum amplitude in the RNS description.   

A super Riemann surface $\Sigma$ of genus 2 without punctures has a moduli space of dimension $3g-3|2g-2=3|2$, with three even moduli
and two odd ones.  The genus 2 vacuum amplitude will come from an integral over this moduli space.
As we explained for instance in discussing eqn. (\ref{tolfo}), two odd moduli is enough to produce the
characteristic subtleties of superstring perturbation theory.  
Experience has shown that it is indeed tricky 
to correctly compute the vacuum amplitude in genus 2.  However, the subtleties have been neatly resolved by D'Hoker and Phong
 in work surveyed in \cite{DPh}.
This work involved very intricate calculations, but the underlying idea was actually very simple.\footnote{As reviewed in \cite{DPh}, D'Hoker and Phong also computed certain genus 2 scattering amplitudes -- parity-conserving amplitudes with all external states
being bosons from the NS sector.  This was a much more difficult computation than the computation of the genus 2 vacuum energy.}   

First of all, let $\Sigma_0$ be an ordinary Riemann surface of genus $g$.  $\SIgma_0$ has a period matrix, whose definition we will
recall in section \ref{details}.  It is a $g\times g$ symmetric complex matrix $\Omega_{ij}$, $i,j=1,\dots,g$, with a positive-definite
imaginary part; it is uniquely determined up to the action of the symplectic group $Sp(2g;\Z)$.  
A $g\times g$ symmetric matrix has $g(g+1)/2$ independent matrix elements.  A Riemann surface of genus $g$
has $3g-3$ complex moduli.  These numbers coincide if $g=2$ or 3, and this suggests that in those two special cases it may
be possible to use the matrix elements of $\Omega$ (modulo the action of $Sp(2g;\Z)$) as moduli for $\Sigma_0$.  This is actually
true in genus 2, and something very similar is true for $g=3$.  (For $g>3$, the matrix elements of $\Omega$ are not independent
but obey the Schottky relations, so they cannot be used to parametrize the moduli space of $\Sigma_0$ in such a simple way.)

A super Riemann surface $\Sigma$ of genus $g$ with an even spin structure has a super period matrix $\hat\Omega_{ij}$ which
is a purely bosonic $g\times g$ complex matrix, again with positive-definite imaginary part.\footnote{\label{zelb}
Actually, $\hat\Omega$ is only
generically defined; it has poles with nilpotent residue along the locus where the reduced space $\Sigma_0$ of $\Sigma$
is such that $H^0(\SIgma_0,K^{1/2})\not=0$, where $K^{1/2}$ is the spin bundle of $\Sigma_0$.  (See footnote \ref{helplater} in section
\ref{splitsuper}.)
This phenomenon never occurs  in genus 2 (or less), and this is one of the simplifications
behind the calculations surveyed in \cite{DPh}.  The procedure used there can possibly be adapted to $g=3$, but one would
have to deal with the fact that for $g\geq 3$, the super period matrix does have poles, and the projection from supermoduli space to the bosonic moduli space introduces further poles. 
For details, see \cite{holomorphy}.}  The basic idea in \cite{DPh} is to use the matrix elements of $\hat\Omega$
as bosonic moduli of $\Sigma$ that are kept fixed while integrating over the fermionic moduli.

\begin{figure}
 \begin{center}
   \includegraphics[width=3.5in]{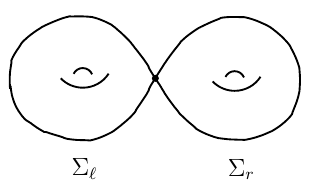}
 \end{center}
\caption{\small Splitting of a genus 2 super Riemann surface to a union of two surfaces of genus 1, joined at a point.}
 \label{Question}
\end{figure}
This gives a clear framework for computing the genus 2 vacuum amplitude, though implementing this framework requires some hard work.  However, given what
we have learned in section \ref{mass}, there is a basic question to ask.   A super Riemann surface of genus 2 can split into a union of
two surfaces of genus 1, joined at a point (fig. \ref{Question}).  When this occurs, we are not free to specify arbitrarily what bosonic variable
is kept fixed while integrating over odd moduli.  There is a distinguished parameter $\varepsilon$ that has a simple zero on the compactification
divisor $\frak D$, and this is what should be held fixed when we integrate over the odd moduli.  Is $\varepsilon$ a function of the $\hat\Omega_{ij}$
or does it differ from such a function by a bilinear in the odd moduli?

It turns out that the answer to this question is that $\varepsilon$ cannot be expressed just in terms of the $\hat\Omega_{ij}$, and 
therefore in general the procedure described in \cite{DPh} does require a correction that
is supported on the divisor $\frak D$.  However, the correction vanishes in supersymmetric models above four dimensions or with $\N\geq 2$ supersymmetry in four dimensions.
It is nonvanishing
if there is an operator $V_D$ that is in the bottom component of
a supermultiplet, has dimension $(1,1)$, and has a nonzero expectation value $\langle V_D\rangle$ 
on a super Riemann surface of genus 1 with even spin structure.
In four-dimensional $\N=1$ models that have an anomalous $U(1)$ gauge symmetry at tree level, such an operator $V_D$ exists, and in this case, the correction term
shifts the 2-loop vacuum energy by a universal multiple of $\langle V_D\rangle^2$ (where $\langle V_D\rangle$ 
is computed at 1-loop order).   It turns out that the bulk contribution to the 2-loop
vacuum energy, computed using the procedure of \cite{DPh}, vanishes in all compactifications to four dimensions that have 
spacetime supersymmetry at tree level.  This has been shown in some explicit orbifold computations in \cite{DPhlatest} and proved more generally in \cite{Wittennew}.
Because of this bulk vanishing,  the correction that we will find at infinity gives the full answer.

In section \ref{obso}, we explain how this correction arises when there is an operator with the properties of $V_D$.  (Potential corrections
associated to operators of dimension less than $(1,1)$ are discussed in section \ref{otherbound}.)
In that analysis, we make use of the detailed relationship between $\varepsilon$ and $\hat\Omega$, which is explained in 
section \ref{details}.

\subsection{The Correction At Infinity}\label{obso}

\subsubsection{Separating Degeneration For Bosonic Strings}\label{sepbo}

First let us recall the behavior of the worldsheet path integral near a separating degeneration where a Riemann surface $\Sigma$ of genus $g$
splits into a union of surfaces $\Sigma_\ell$ and $\Sigma_r$ of genera $g_\ell$ and $g_r$, 
joined at a point $p$.  (See for example section 6.4.4 of \cite{Revisited} for more
detail.) We practice first with bosonic strings, and for simplicity
we begin by considering holomorphic degrees of freedom only.  It is more or less equivalent to begin with bosonic open strings.

We use local coordinates $x$ and $y$ on $\Sigma_\ell$ and $\Sigma_r$, and we 
write the gluing formula as
\begin{equation}\label{yuggle} (x-a)(y-b) =q.\end{equation}
Thus for $q=0$, the point $x=a$ on $\SIgma_\ell$ is glued to the point $y=b$ on $\SIgma_r$.  In general, both $\SIgma_\ell$ and $\Sigma_r$ have other
moduli, but they do not play an important role in what follows.  We want to analyze the measure for integrating over $a,b$, and $q$ near $q=0$.
This measure is a sum of contributions from various string states that propagate through the narrow neck between $\Sigma_\ell$ and $\Sigma_r$.
Each such contribution can be evaluated by inserting some vertex operator $V$ on $\Sigma_\ell$ and a conjugate\footnote{The only sense in which $\h V$ is ``conjugate'' to $V$ is that the two-point function $\langle V\h V\rangle$ is nonzero in genus 0; no complex conjugation is implied.} vertex operator $\h V$ on $\Sigma_r$.
(Here and at some later points  in this paper, we use the integrated forms of the vertex operators, though it is more precise to develop analogous formulas
using the unintegrated version.)  The contribution of the string
state in question to the path integral measure is given by the expression 
\begin{equation}\label{umbro} \d a \, V(a) \cdot \d q \,F(q) \cdot\d b \, \h V(b) ,\end{equation}
where $F(q)$ is some function that we have to determine.  Here $\d a\, V(a)$ and $\d b\, \h V(b)$ are the amplitudes for the indicated states to couple
to $\Sigma_\ell$ and $\Sigma_r$, respectively, and $\d q\,F(q)$ is the amplitude for the relevant state to propagate through the narrow neck.
To compute the contribution of the chosen state propagating between $\Sigma_\ell$ and $\SIgma_r$ to the world sheet path integral, one
 has to insert the expression (\ref{umbro}) in the worldsheet path integral on $\Sigma_\ell$ and $\Sigma_r$,
calculate the path integral including any other vertex operators that may be present in addition to the ones associated to the degeneration, and then integrate
over $a$, $b$, $q$, and the other moduli of $\Sigma_\ell$ and $\Sigma_r$. 

The function $F(q)$ can be determined by requiring that the expression (\ref{umbro}) is invariant under a scaling of the coordinates.  
We can work in a basis of operators such that $V$ and its conjugate $\h V$ are eigenstates of $L_0$.
Under the scaling $x\to \lambda x$, $a\to \lambda a$, along with $y\to \h\lambda y$, $b\to \h\lambda b$ (with arbitrary nonzero parameters $\lambda,\h\lambda$), 
we see from eqn. (\ref{yuggle}) that $q$ scales
as $q\to \lambda\h\lambda q$.    On the other hand, the vertex operators $V$ and $\h V$ scale as $\lambda^{-L_0}$ and $\h\lambda^{-L_0}$.   Invariance of
(\ref{umbro}) implies that $F(q)$ is a constant times $q^{L_0-2}$.  If  the operators $V$ and $\h V$ are normalized 
to have a canonical two-point function on a two-sphere, then the coefficient of $q^{L_0-2}$ is precisely $g_\st^2$ ($g_\st$ is the string coupling constant).
Omitting this universal factor, the contribution of a string state of given $L_0$ to the measure comes from insertion of
\begin{equation}\label{zuggle}\d a\, V(a)\cdot {\d q}\,{q^{L_0-2}}\cdot\d b\, \h V(b). \end{equation}

\subsubsection{The Pole And The Tadpole}\label{tadpole}

As a function of $L_0$, the integral $\int_0^\Lambda \d q\, q^{L_0-2}\sim 1/(L_0-1)$ has a pole at $L_0=1$.  (Here $\Lambda$ is an irrelevant upper
cutoff; the pole comes from the contribution near $q=0$.)   In a typical situation (fig. \ref{Split}) with vertex operators on both $\Sigma_\ell$ and $\Sigma_r$,
one has (for open strings) $L_0=(\alpha'/2)P^2+N$, where $P$ is the momentum flowing between $\Sigma_\ell$ and $\Sigma_r$,
and $N$ is constructed from oscillator modes of the string.  The pole occurs when the string state flowing between $\Sigma_\ell$ and $\Sigma_r$ is
on-shell, and plays the same role as the pole $1/(P^2+m^2)$ of the Feynman propagator in field theory.  Thus this pole is essential to the physical
interpretation of string theory.  Similar poles arise in closed string theory (where $\d q\, q^{L_0-2}$ is replaced by $\d^2q\,\bar q^{\t L_0-2} q^{L_0-2}$)
and in superstring theory, where  an analogous pole is exhibited at the end of section \ref{srsan}.
 
\begin{figure}
 \begin{center}
   \includegraphics[width=3.5in]{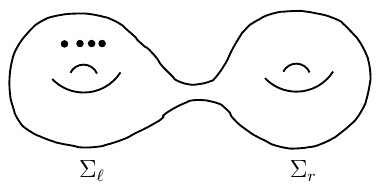}
 \end{center}
\caption{\small A separating degeneration that can lead to trouble.  A Riemann surface $\Sigma $ splits into two components $\SIgma_\ell$ and $\SIgma_r$
with all vertex operators on one side, here $\Sigma_r$.  The string state propagating between the two branches carries zero momentum in spacetime.  Unless the ``tadpole''
-- the amplitude for a zero-momentum massless scalar to disappear into the vacuum -- vanishes, the contribution of such a process is divergent. }
 \label{Tadpolefig}
\end{figure}

One of the greatest delicacies in string perturbation theory
 involves the ``tadpoles'' of massless spin-zero particles.  The tadpole problem arises
if all external vertex operators are inserted on $\Sigma_\ell$ (or $\SIgma_r$), as in fig. \ref{Tadpolefig}, in which case the string state
flowing between the two sides always has $P=0$.  As a result, if this state is a massless spin-zero particle, it is automatically on-shell;
the $q$ integral behaves as $\d q/q$ and is logarithmically divergent at $q=0$.   One has the same logarithmic divergence in closed-string theory
or in superstring theory, for similar reasons.  
From  a field theory point of view, we are sitting on the pole of the propagator $1/(P^2+m^2)$ at $P=m=0$.   

After we integrate over the moduli of $\SIgma_r$, the coefficient of the logarithmic divergence is proportional to the ``tadpole'' -- the amplitude
for the massless scalar in question to be absorbed in the vacuum (or more precisely the genus $g_r$ contribution to this tadpole).
In many supersymmetric compactifications, one can use spacetime supersymmetry to show that the integrated tadpoles vanish (in other words,
for all values of $g_r\geq 1$, the one-point function for a massless scalar vertex operator inserted on a surface $\Sigma_r$ of genus $g_r$ vanishes
after integration over the moduli of $\SIgma_r$).  In section \ref{supersym}, we explain how this is proved and also how the proof can fail
when a potential Goldstone fermion is present.

But even when the integrated tadpoles vanish for all massless particles, the integrals defining $g$-loop scattering amplitudes with $g>0$
are at best only conditionally convergent, because in the region in which $\Sigma$ degenerates in the fashion indicated in fig. \ref{Tadpolefig},
one gets one answer (the contribution of any given massless scalar is divergent) by integrating first over $q$, and a different answer (the contribution
of the massless scalar is 0) by integrating first over the moduli of $\Sigma_r$.   In some low order cases, one can find a suitable regularization of
the conditionally convergent integrals by hand, but in general one requires the procedure described in section \ref{regu}.  For analysis of
the tadpole problem using that procedure, see section 7.6 of \cite{Revisited}.

In field theory, to make sense of perturbation theory in a similar situation, one requires either vanishing of the tadpole or else a shift
in the vacuum to cancel it.  The techniques to analyze superstring perturbation theory when a shift in the vacuum is required have been developed
in \cite{ASen}.

\subsubsection{The Analog For Super Riemann Surfaces}\label{srsan}

Now we will explain the analog of equation (\ref{zuggle}) for super Riemann surfaces. We will only consider the case that the string state
propagating between $\Sigma_\ell$ and $\Sigma_r$ is in the NS sector.  (See section \ref{where} for Ramond sector gluing.)
Again we start with a holomorphic sector or with open strings.

We use local superconformal coordinates $x|\theta$ on $\Sigma_\ell$ and $y|\psi$ on $\Sigma_r$.  
We need a slight generalization of eqn. (\ref{melk}) so that at $\varepsilon=0$, the point $x|\theta=a|\alpha$ in $\Sigma_\ell$
is glued to $y|\psi=b|\beta$ in $\Sigma_r$. The resulting formula \begin{align}\label{iblondo} (x-a-\alpha\theta)(y-b-\beta\psi)& =-\varepsilon^2 \cr
                                         (y-b-\beta\psi)(\theta-\alpha) & = \varepsilon(\psi-\beta) \cr
                                          (x-a-\alpha\theta)(\psi-\beta)&= -\varepsilon(\theta-\alpha)\cr
                                                              (\theta-\alpha)(\psi-\beta)& = 0\end{align}
                                                    can be obtained from eqn. (\ref{melk}) by global supersymmetry transformations of $x|\theta$ and $y|\psi$.
                                                    But all we really need to know is the scaling behavior:
\begin{align}\label{ipooz} (x,a,\alpha)\to &(\lambda x,\lambda a,\lambda^{1/2}\alpha) \cr
                                     (y,b,\beta)\to & (\h\lambda y,\h\lambda b,\h\lambda^{1/2}\beta)\cr
                                      \varepsilon\to& (\lambda\h\lambda)^{1/2}\varepsilon. \end{align}
In particular, given the scaling of $\varepsilon$, we can see what must be the analog of eqn. (\ref{zuggle}):
\begin{equation}\label{superzuggle}\d a\,\d\alpha \,\V(a|\alpha)\cdot {\d \varepsilon}\,{\varepsilon^{2L_0-2}}\cdot \d b\,\d\beta\,\h\V(b|\beta).\end{equation} 
Here $\V$ and $\h\V$ are conjugate superfields with given $L_0$.

A physical state  from the NS sector is represented by a superconformal primary field $\V$ of $L_0=1/2$.   The
integral $\int_0^\Lambda \d\varepsilon \,\varepsilon^{2L_0-2}\sim 1/(2L_0-1)$ has a pole at $L_0=1/2$, quite analogous to the pole discussed in
section \ref{tadpole} above.  The residue of the pole comes from the insertion of the integrated vertex operator $\int \d a\,\d\alpha\, \V(a|\alpha)$ on
$\Sigma_\ell$ and of its conjugate on $\Sigma_r$.
In particular, integration over $\alpha$ and $\beta$ projects onto the top components of the vertex operators and in that sense the pole is
associated to propagation of the top component.                                   

\subsubsection{The Boundary Correction}\label{bcor}

We are, however, interested not in the pole associated to a physical state, but in a subtlety associated to the existence of the  operator
$V_D$ that is associated to spontaneous supersymmetry-breaking.  This field
is a conformal primary of dimension $(1,1)$, but not a superconformal primary.  It is the bottom component of a supermultiplet, not the top
component.  Its contribution to the measure near $\varepsilon=0$ is obtained by setting $L_0=1$ in (\ref{superzuggle}):
\begin{equation}\label{specialzuggle}\d a\,\d\alpha \,V_D(a) \cdot \d\varepsilon \cdot \d b\,\d\beta\,V_D(b). \end{equation}

We stress that as $V_D$ is the bottom component of a supermultiplet, it depends only on the bosonic coordinates $a$ and $b$ and not
on the fermionic coordinates $\alpha$ and $\beta$.  For two reasons,   the contribution of $V_D$ looks completely harmless:
the differential form written in (\ref{superzuggle}) has no singularity at all at $\varepsilon=0$, and anyway, this expression looks
like it will vanish after integration over $\alpha$ and $\beta$, since the integrand has no dependence on those odd variables.

Both of these arguments have fallacies that echo what was explained in section \ref{moresplit}.
To explain this, we need a formula more complete than (\ref{specialzuggle}) that includes the antiholomorphic degrees of freedom.
In the case of the heterotic string, the antiholomorphic variables are governed by the bosonic string formula (\ref{zuggle}).  Hence a more complete
analog of (\ref{specialzuggle}) is
\begin{equation}\label{bigzuggle}\d^2a\,\d\alpha\,V_D(\t a;a)\cdot \frac{\d \t q}{\t q}\,\d\varepsilon \cdot\d^2b \,\d\beta\, V_D(\t b; b).\end{equation}
We are in the same situation as in section \ref{moresplit}.  The integral
\begin{equation}\label{factorzuggle} \d\alpha \cdot \frac{\d \t q}{\t q} \d\varepsilon \cdot \d\beta \end{equation}
is scale-invariant, that is, it is invariant both under holomorphic scaling $\alpha\to \lambda^{1/2}\alpha$, $\beta\to \h\lambda^{1/2},$
$\varepsilon\to \lambda^{1/2}\h\lambda^{1/2}\varepsilon$, and antiholomorphic scaling $\t q\to \t\lambda \t q$. (The remaining factors $\d^2a\,V_D(\t a;a)$
and $\d^2b\,V_D(\t b;b)$ in (\ref{bigzuggle}) are also scale-invariant.)  Thus, we are in a situation very close
to that of section \ref{moresplit}.  The integral in (\ref{factorzuggle}) is only conditionally convergent.
It vanishes if we integrate first over $\alpha$ and $\beta$ keeping fixed the bosonic variables, but not if we integrate over $\alpha$ and $\beta$
keeping fixed some other combination such as $\varepsilon+\alpha\beta$.  

The procedure of \cite{DPh} for computing the 2-loop vacuum amplitude amplitude amounts to integrating over the
odd variables while holding fixed not $\t q$ and $\varepsilon$ but $\t q$ and $\varepsilon^*=\varepsilon+\alpha\beta$.   
(The reason for this is that, as we will see in eqn. (\ref{metzo}), it is not $q_\NS=-\varepsilon^2$
but $q_\NS^*=-(\varepsilon^*)^2$ that is a matrix element of the super period matrix $\hat\Omega$.)  
Likewise,  the general procedure explained in section \ref{gluehol} tells us to set $\bar{\t q}=q_\NS(1+\dots)$ near
$\t q=q_\NS=0$ (the ellipses represent arbitrary nilpotent terms), but the procedure of \cite{DPh} is slightly different.
In that formulation, antiholomorphic moduli are taken to be complex conjugates of the matrix elements of $\hat\Omega$, so in particular the relation
between $\t q$ and $q^*_\NS$ is $\bar{\t q}=q_\NS^*$.  

To compare the two approaches, we define
\begin{equation}\label{mytro}\varepsilon^\star = \varepsilon+h(\t q;\neg q_\NS)\alpha\beta,\end{equation}
where $h(\t q;\neg q_\NS)$, which plays essentially the same role\footnote{In the comparison between the two
problems, $\varepsilon$ corresponds to $\h z$ and $\varepsilon^\diamond$ to $z$.  The  function $h$ in (\ref{mytro}) really
corresponds to $1-h$ in (\ref{urok}), because we started the present analysis with the gluing parameter $\varepsilon$,
while in section \ref{moresplit}, we began the analysis with the bosonic variable $z$ rather than the gluing parameter $\h z$. }   
as the function  $h(\t z;\neg z)$ of eqn. (\ref{urok}), is any function that equals 0 for
$\t q q_\NS<\eta^2$ (for some small positive $\eta$) and 1 for, say, $\t q q_\NS>2\eta^2$.   The first condition ensures that
$\varepsilon^\star$ agrees with $\varepsilon$ near $\t q=q_\NS=0$, and the second condition ensures that except very near $\t q=q_\NS=0$,
$\varepsilon^\star$ coincides with $\varepsilon^*$, the variable effectively used in \cite{DPh}.  

Our integration procedure then is to set $\bar{\t q}$ equal to $q_\NS^\star=-(\varepsilon^\star)^2$ and to integrate over $\alpha$ and $\beta$
holding $\t q$ and $\bar{\t q}$ fixed.  This procedure is correct near $\t q=q_\NS=0$.  Away from $\t q=q_\NS=0$, it does not matter exactly what we do.  The procedure of setting $\bar{\t q}$ equal to $q_\NS^\star$
has been chosen to agree away from $\t q =q_\NS=0$  with the procedure used in \cite{DPh}.  

To actually calculate the integral, we proceed as in section \ref{moresplit}.    Eqn. (\ref{mytro})
is equivalent to $\varepsilon=\varepsilon^\star-\alpha\beta h(\t q;\neg -({\varepsilon^\star})^2)$.  (This is because the differences between $\varepsilon$,
$\varepsilon^\diamond$, and $\varepsilon^*$ are of order $\alpha\beta$, and vanish when multiplied by $\alpha\beta$.)  So
\begin{equation}\label{lunky} \d\varepsilon=\d\varepsilon^\star\left(1+2\alpha\beta\varepsilon^\star \frac{\partial}{\partial q^\star_\NS}h(\t q;\neg q_\NS^\star)
\right)+\dots , \end{equation}
where we only indicate the terms proportional to $\d\varepsilon^\star$ on the right hand side.  
Equivalently,
\begin{equation}\label{unky}\d\varepsilon=\d\varepsilon^\star-\alpha\beta \d q_\NS^\star\frac{\partial}{\partial q^\star_\NS}\left(h(\t q;\neg q_\NS^\star)-1\right)+\dots. \end{equation}
We  set $2\d\varepsilon^\star\cdot \varepsilon^\star=-\d q_\NS^\star$, and we used the fact that $h$ and $h-1$ have the same derivative.
We use this expression to substitute for $\d\varepsilon$
in favor of $\d\varepsilon^\star$ or equivalently $\d q_\NS^\star$ in the integral (\ref{factorzuggle}).  Setting also $q_\NS^\star =\bar{\t q}$, the term  in (\ref{lunky})
that will survive when we integrate
over $\alpha$ and $\beta$ at fixed $\t q$ and $\bar {\t q}$ is the term proportional to $\alpha\beta\d q^\star_\NS$.  The contribution of this term in
(\ref{factorzuggle}) is 
\begin{equation}\label{lonely} \d \t q\d \bar{\t q}\d\alpha\d\beta \, \frac{\alpha\beta}{\t q}\frac{\partial }{\partial \bar{\t q}}(h(\t q;\bar{\t q})-1).\end{equation} 
To evaluate this, we simply integrate by parts.  There is no surface term at large $\t q$ (where the approximations used in arriving at (\ref{lonely})
would not be valid) since $h-1=0$ at large $\t q$.  There is a contribution at $\t q=0$ that comes from 
\begin{equation}\label{indox}\frac{\partial}{\partial \bar{\t q}}\, \frac{1}{\t q}=2\pi \delta^2(\t q). \end{equation}  
With the help of this formula, and the fact that $h(0;\neg 0)=0$, the integral over $\t q$, $\bar {\t q}$, $\alpha$, and $\beta$ just gives $2\pi$.

We still have to integrate the remaining factors $\d^2 a V_D(\t a;a) \,\d^2 b V_D(\t b;b)$ in (\ref{bigzuggle}), and integrate over the bosonic
moduli (the $\tau$ parameters) of $\Sigma_\ell$ and $\Sigma_r$.  Since $\Sigma_\ell$ and $\Sigma_r$ each have genus 1, these integrals\footnote{As
$\Sigma_\ell$ and $\SIgma_r$ have genus 1, the $\d^2 a$ and $\d^2 b$ integrals can be factored out using the translation symmetries of $\Sigma_\ell$
and $\Sigma_r$. To this
end, one might prefer to make the whole derivation starting with a variant of  eqn. (\ref{umbro}) expressed in terms of unintegrated vertex operators. See
for instance section 6.4.4 of \cite{Revisited}.}
give two factors of the 1-loop expectation value $\langle V_D\rangle$.  Restoring also the factor of $g_\st^2$ that was suppressed in eqn.
(\ref{zuggle}), the contribution of this calculation to the two-loop vacuum energy is $2\pi g_\st^2\langle V_D\rangle^2$. 

 This correction at infinity to the D'Hoker-Phong procedure \cite{DPh} is the full answer
since, as shown in \cite{DPhlatest,Wittennew},  in a general heterotic string 
compactification to four dimensions that has spacetime supersymmetry at tree level, the D'Hoker-Phong procedure  gives 0 for the bulk contribution to the two-loop
vacuum energy.

\subsubsection{Other Boundary Corrections?}\label{otherbound}

In this derivation, starting in eqn. (\ref{bigzuggle}), we have made a simplified approximation to the measure on supermoduli
space that comes from  the worldsheet
path integral, considering only the contribution of the vertex operator $V_D$.   The full measure is of course far more complicated.
We have chosen
an integration procedure that agrees with that of \cite{DPh} except very near $\t q=q_\NS=0$, but gives a correction there
for the contribution of $V_D$.  Does this procedure lead to any other corrections?

Consider the propagation between $\Sigma_\ell$ and $\Sigma_r$ of an arbitrary NS sector state represented by a superfield
$\O(\t z;\neg z|\theta)$.  Let $\h \O(\t z;\neg z|\theta)$ be the conjugate superfield.  Suppose that the bottom component of $\O$
has holomorphic and antiholomorphic conformal dimensions $(L_0,\t L_0)$.  Then the analog of (\ref{bigzuggle}) for the contribution
of this superfield is
\begin{equation}\label{tolomo}\d^2 a \,\d \alpha\,\O(\t a;\neg a|\alpha)\d\t q\,\t q^{\t L_0-2}\,\d\varepsilon\,\varepsilon^{2L_0-2}
\h \O(\t b;\neg b|\beta)\, \d^2 b \,\d\beta,\end{equation}
with an insertion of $\O$ on one side and of $\h \O$ on the other.

 A preliminary comment is
that only the contribution of the bottom components of the superfields can lead to the sort of subtlety discussed in this paper.   (For example,
$V_D$ is such a bottom component.)  Indeed, if $\O(\t a;\neg a|\alpha)=\O_0(\t a;\neg a)+\alpha \O_1(\t a;\neg a)$,
$\h \O(\t b;\neg b|\beta)=\h \O_0(\t b;\neg b)+\beta\h \O_1(\t b;\neg b)$, then contributions involving $\O_1$ or $\h \O_1$
are unaffected by a change of variables in which a multiple of $\alpha\beta$ is added to $\varepsilon$, simply because
$\alpha^2=\beta^2=0$.
Also, the only potentially dangerous case is $L_0=\t L_0$, since otherwise, after setting $\bar{\t q}=q_\NS$ near $\t q=0$, a possible surface term vanishes upon integration
over $\mathrm{Arg}\,\t q$.  For $L_0=\t L_0\not=1$, a possible boundary contribution is multiplied by $(q_\NS \t q)^{L_0-1}$ relative
to what we had in studying $V_D$, where $q_\NS,\t q\to 0$ at the boundary. For $L_0=1$, the boundary
correction is nonzero and interesting, as we have seen.  For $L_0<1$, it would be divergent (and 
in general this will lead to spurious 
infrared divergences if one uses the wrong integration procedure at infinity).  For $L_0>1$, the boundary correction 
vanishes.
Finally, obviously the splitting of a genus 2 surface to a pair of genus 1 surfaces with insertions of $\O_0$ and $\h\O_0$
on the two branches can only lead to a boundary term if $\O_0$ and $\h\O_0$ have nonvanishing 1-point functions\footnote{With an integration
procedure that preserves all the conformal symmetry, the only operators whose 1-point functions can be defined and can play a role are conformal vertex operators
(see section \ref{cvo}).  With a more general integration procedure, more general 1-point functions can enter.}   in genus 1.

The existence of potentially troublesome operators is model-dependent.  Let us discuss the one such operator
that always exists; 
this is the identity operator, with $L_0=\t L_0=0$. (In many simple models, such as toroidal compactifications, it is the only
operator that satisfies the criteria.)
The corresponding contribution to the measure is
\begin{equation}\label{wolomo}\d^2 a \,\d\alpha \, \frac{\d \t q}{\t q^2}\frac{\d\varepsilon}{\varepsilon^2}\d^2 b\,\d\beta.\end{equation}

The integral over $\t q $ and $\varepsilon$ looks divergent. This apparent divergence
is the contribution of the NS sector tachyon, whose vertex operator at zero momentum is the identity operator.  The NS sector tachyon is not a physical state of superstring theory, so we do not expect a tachyon divergence in the integration over moduli.   It is tempting to argue that the tachyon divergence is eliminated because, as the expression (\ref{wolomo})
does not depend on $\alpha$ and $\beta$, the Berezin integral over those parameters, keeping fixed $\t q $ and with  $\bar {\t q}$ set to $-\varepsilon^2$,
 vanishes.  (The expression explicitly written in eqn. (\ref{wolomo}) should be multiplied by other factors that depend only on the moduli of $\Sigma_\ell$ and $\Sigma_r$ and not on $\alpha$ and $\beta$; this does not affect the suggestion that was just made.)  The trouble with this argument
is that in general, because of the dependence of the gluing operation (\ref{iblondo}) on choices of local parameters $x|\theta$ and $y|\psi$, $\varepsilon$ is 
really  only defined modulo \begin{equation}\label{zinbo}\varepsilon\to 
\varepsilon'=\varepsilon e^w,\end{equation}  where $w$
is a function of the other moduli (in general including $\alpha$ and $\beta$) that is holomorphic at $\varepsilon=0$.  Any function $\varepsilon'=\varepsilon
e^w$ has a simple zero along the divisor at infinity and (if $w$ is constrained by GSO symmetry as in footnote \ref{zone} below) is as natural as any other such function.   Integration over $\alpha$ and $\beta$ keeping fixed $\t q$ and with $\bar{\t q}$
set to $-(\varepsilon')^2$ does not necessarily eliminate the tachyon contribution.

What really does eliminate the tachyon contribution is the GSO projection\footnote{\label{zone} The function $w$ in eqn. (\ref{zinbo}) can be somewhat constrained by GSO symmetry; at $\varepsilon=0$, one can require $w$ to be invariant
under a sign change of all odd moduli of $\SIgma_\ell$, or all odd moduli of $\SIgma_r$.  In general (if $\Sigma_\ell$ and $\Sigma_r$ have additional
odd moduli), this condition allows $w$ to depend nontrivially
on $\alpha$ and $\beta$ and hence does not affect the discussion surrounding eqn. (\ref{zinbo}).  However, for the genus 2 vacuum amplitude, with $\alpha$ and $\beta$
being the only odd moduli, GSO symmetry implies that $w$ is independent of $\alpha$ and $\beta$ at $\varepsilon=0$ and hence does not play any
important role.  So in that particular case, there is a natural sense in which integration over $\alpha$ and $\beta$ kills the tachyon contribution
at a separating degeneration without reference to the GSO projection.  (For a corresponding study of a nonseparating degeneration,
see section 5.3 of \cite{holomorphy}.)} 
 \cite{GOS}, which roughly is the sum over the two possible signs of $\varepsilon$
for given $q_\NS=-\varepsilon^2$.  

 However, let us see what happens if we follow a similar procedure, but after
replacing $\varepsilon$ with $\varepsilon^*=\varepsilon+\alpha\beta$ (which notably  is not of the form $\varepsilon e^w$, so it is not a good parameter
defining the divisor at infinity). In terms of $\varepsilon^*$, (\ref{wolomo}) becomes
\begin{equation}\label{zolomo}\d^2 a \,\d\alpha \frac{\d\t q}{\t q^2}\frac{\d\varepsilon^*}
{(\varepsilon^*-\alpha\beta)^2}\d^2 b\,\d\beta.\end{equation}
After integrating over $\alpha$ and $\beta$ and setting $q=-(\varepsilon^*)^2 =\bar{\t q}$, we get
\begin{equation}\label{polomo} \d^2a \,\frac{\d\t q\cdot \d q}{\t q^2q^2}\,\d^2b. \end{equation}
Comparing to the bosonic string formula (\ref{zuggle}), we see that this is the contribution one would expect from the identity
operator (of $L_0=\t L_0=0$) in bosonic string theory.  It is not a natural behavior in superstring theory.  This singular behavior was found\footnote{See eqn. (10.4) in that paper.
In that equation, $\d^3\tau$ corresponds to, in our notation of section \ref{splitor}, $\d\tau_{\ell\ell}\d\tau_{rr}\d\tau_{\ell r}$;
and $\tau=\tau_{\ell r}=q$.
So $\d^3\tau/\tau^2\sim \d q/q^2$ (times $\d\tau_{\ell\ell}\d\tau_{rr}$), which is the behavior claimed in (\ref{polomo}).} in the work
reviewed in \cite{DPh}.  This caused no difficulty because the unwanted term (\ref{polomo}), with
a natural regularization, canceled upon summing over spin structures on $\SIgma_\ell$ and $\Sigma_r$.

At least near the large volume limit of the sigma-model, similar reasoning applies for all bosonic
operators of $L_0<1$.  Such operators are constructed from
the bosonic fields of the sigma-model only, since a fermion bilinear would contribute 1 to $L_0$.
Insertion on a genus 1 surface of an operator constructed from bosonic fields only does not disturb the GSO cancellation.

\subsection{Details Concerning The Super Period Matrix}\label{details}

Let us first recall the definition of the period matrix of an ordinary Riemann surface $\Sigma_0$ of genus $g$.  $\Sigma_0$ has a $g$-dimensional
space of holomorphic 1-forms $\omega_1,\dots,\omega_g$.  We pick a symplectic basis of 1-cycles $A^i$, $B_j$, $i,j=1,\dots ,g$, and normalize
the $\omega_i$ so that
\begin{equation}\label{tendox}\oint_{A^i}\omega_j=\delta^i_j. \end{equation}
Then the period matrix is defined by
\begin{equation}\label{mendox}\Omega_{ij}=\oint_{B_i}\omega_j. \end{equation}
Since the $\omega_i$, being holomorphic, are closed, the condition (\ref{tendox}) and the definition (\ref{mendox}) of $\Omega_{ij}$ depend
only on the homology classes of the cycles $A^i$ and $B_j$.  

Almost the same definition makes sense on a super Riemann surface $\Sigma$, with some qualifications (see \cite{RSV,DPhtwo}, and section 8
of \cite{Surfaces}).  First of all, the closest analog of the
theory of the ordinary period matrix arises if the spin structure of $\Sigma$ is even.  In this case, generically there is a $g$-dimensional space
 of
closed\footnote{On a super Riemann surface, as opposed to an ordinary one, a holomorphic 1-form is not always closed. For
brevity, we omit an alternative description of the super period matrix, explained in the references, in terms of holomorphic sections of the Berezinian of $\Sigma$.} holomorphic one-forms $\h\omega_i$, $i=1,\dots,g$ on $\Sigma$.  (This fails when the reduced space $\SIgma_0$ of $\Sigma$ has $H^0(\SIgma_0,K^{1/2})\not=0$; then the super period
matrix acquires a pole with nilpotent residue, as explained in footnote \ref{helplater}.)  We pick a symplectic basis of cycles $A^i$ and $B_j$ in $\Sigma$ of dimension
$1|0$ (one can take any cycles of dimension $1|0$ that can be deformed to ordinary $A$- and $B$-cycles in the reduced space $\Sigma_0$ of $\Sigma$) and
after normalizing the $\h\omega_i$ so that 
\begin{equation}\label{tendlox}\oint_{A^i}\h\omega_j=\delta^i_j, \end{equation}
we define the super period matrix  by
\begin{equation}\label{endox}\hat\Omega_{ij}=\oint_{B_i}\h\omega_j. \end{equation}
It can be shown to be symmetric, just like the classical period matrix defined in (\ref{mendox}).

Now we consider the case of a Riemann surface or super Riemann surface of genus 2 that is splitting into a union of two components $\Sigma_\ell$
and $\Sigma_r$ of genus 1,
meeting at a point (fig. \ref{Question}).  We want to show that for ordinary Riemann surfaces, the gluing parameter $q$ can be expressed as
a matrix element of the period matrix, while for a super Riemann surface, the gluing parameter $\varepsilon$ cannot be expressed in terms of
the super period matrix.  This is the key point that led to the boundary correction in section \ref{bcor}.

\subsubsection{Period Matrix And Gluing Parameter Of An  Ordinary Riemann Surface}\label{splitor}

For an ordinary Riemann surface $\Sigma_0$ of genus 2 
that has split into two genus 1 components $\Sigma_{0,\ell}$ and $\Sigma_{0,r}$ joined at a point, we can be very explicit
about the period matrix. Let $\tau_\ell$ and $\tau_r$ be the modular parameters of $\SIgma_{0,\ell}$ and $\Sigma_{0,r}$.  We describe
$\Sigma_{0,\ell}$ as  the quotient of the complex $z_\ell$-plane by
\begin{equation}\label{dumbry} z_\ell\cong z_\ell+1\cong z_\ell+\tau_\ell \end{equation}
and similarly $\Sigma_{0.r}$ as the quotient of the complex $z_r$-plane by
\begin{equation}\label{umbry} z_r\cong z_r+1\cong z_r+\tau_r. \end{equation}
We let $A^\ell$ and $B_\ell$ be standard $A$- and $B$-cycles in $\Sigma_{0,\ell}$: $A^\ell$ is the image in $\Sigma_{0,\ell}$ of a straight line
from $z_\ell=0$ to $z_\ell=1$, and $B_\ell$ is the image of a straight line from $z_\ell=0$ to $z_\ell=\tau_\ell$.  We define $A^r$ and $B_r$ in
a completely analogous way as standard $A$- and $B$-cycles in $\Sigma_{0,r}$.    

Suppose that $\Sigma_0$ is built by gluing the point $z_\ell=a$ in $\Sigma_{0,\ell}$ to the point $z_r=b$ in $\Sigma_{0,r}$. 
(We pick $a$ and $b$ to not lie on any of the chosen  $A$- or $B$-cycles or alternatively we deform the cycles to avoid $a$ and $b$. By translation symmetry, the choices of $a$ and $b$ do not matter.)
On $\Sigma_0$, we can take a basis of holomorphic differentials $\omega_\ell=\d z_\ell$ and $\omega_r=\d z_r$.  Thus $\omega_\ell=0$
on $\Sigma_{0,r}$ (since $z_\ell$ is constant there) and likewise $\omega_r=0$ on $\Sigma_{0,\ell}$.
The periods of $\omega_\ell$ are
\begin{align} \oint_{A^\ell}\omega_\ell & = 1, ~~~~                        \oint_{B_\ell}\omega_\ell = \tau_\ell \cr
                         \oint _{A^r}\omega_\ell& =\int_{B_r}\omega_\ell=0. \end{align}
Similarly,
\begin{align} \oint_{A^r}\omega_r & = 1 ,~~~~~
                        \oint_{B_r}\omega_r = \tau_r \cr
                         \oint _{A^\ell}\omega_r& =\int_{B_\ell}\omega_r=0. \end{align}
These formulas show that the period matrix of $\Sigma_0$ in the basis $\begin{pmatrix}A_\ell\cr A_r\end{pmatrix}$ is
\begin{equation}\label{periodmatrix} \Omega=\begin{pmatrix}\tau_\ell & 0 \cr 0 & \tau_r \end{pmatrix}.\end{equation}  
In other words, $\Omega_{\ell\ell}=\tau_\ell$, $\Omega_{rr}=\tau_r$, and $\Omega_{\ell r}=\Omega_{r\ell}=0$. 

Now let us perturb $\Sigma_0$ slightly so that $\Sigma_{0,\ell}$ and $\Sigma_{0,r}$ are joined through a very narrow neck.
Near $z_\ell=a$, $z_r=b$, we glue the two branches  by
\begin{equation}\label{refo} (z_\ell-a)(z_r-b)=q. \end{equation}
When modifying $\Sigma_0$ in this way, we want to modify $\omega_\ell$ and $\omega_r$ so that they continue to have canonical $A$-periods:
\begin{align}\label{nope}\oint_{A^\ell}\omega_\ell & = \oint_{A^r}\omega_r=1\cr
                                           \oint_{A^r}\omega_\ell& = \oint_{A^\ell}\omega_r=0 .\end{align}
Their $B$-periods will then give the deformed period matrix.   We are primarily interested in the off-diagonal component of the deformed
period matrix, which will be non-zero because for $q\not=0$, $\omega_\ell$ is non-zero on $\Sigma_{0,r}$.                                           
From (\ref{refo}), we have 
\begin{equation}\label{momf}\d z_\ell = \d(z_\ell-a)=q\,\d\frac{1}{z_r-b}=-\frac{q \,\d z_r }{(z_r-b)^2}.\end{equation}  
This implies that a form that at $q=0$ is simply $\omega_\ell=\d z_\ell$ and  vanishes on $\Sigma_{0,r}$ will, in linear order in $q$,
become non-zero on $\Sigma_{0,r}$ with the double pole indicated in (\ref{momf}) near $z_r=b$.
The most general 1-form on $\Sigma_{0,r}$ that is holomorphic except for such a double pole is
\begin{equation}\label{delfry}\omega_\ell^{(1)}=  q\cdot     \d z_r\bigl(-\,P(z_r-b;1,\tau_r) + w\bigr) \end{equation}
where $P$ is the Weierstrass $P$-function and $w$ is a constant.   $P(z_r-b;1,\tau)$ is a doubly-periodic
 function that is holomorphic away from $z_r=b$ and behaves for $z_r\to b$ as
\begin{equation}\label{gomely} P(z_r-b;1,\tau)\sim \frac{1}{(z_r-b)^2}+   \O((z_r-b)^2).\end{equation}
These conditions characterize it uniquely.  To compute the correction to the period matrix, we are supposed to adjust the constant
$w$ so that
\begin{equation}\label{omely}\int_{A^r}\omega^{(1)}_\ell=0,\end{equation}
and then  (up to further corrections of higher order in $q$) the off-diagonal matrix element of the period matrix is
\begin{equation}\label{zomely}\Omega_{lr}=\int_{B_r}  \omega^{(1)}_\ell=q\int_{B_r}  \d z_\ell(-P(z_r-b;1,\tau) +w). \end{equation}

This formula shows that to first order in $q$, the off-diagonal matrix element of the period matrix is a multiple of $q$.  We will now determine the precise coefficient in this relation.  We do the following
computations on the unperturbed surface $\Sigma_{0,r}$ defined in eqn. (\ref{umbry}); we have already an explicit factor of $q$
in  $\omega_\ell^{(1)}$, and we can set $q$ to zero elsewhere.

It is not quite true that the form $\omega_\ell^{(1)}$  is closed.  The familiar relation 
\begin{equation}\label{mexico}\frac{\partial}{\partial\bar z_r}\frac{1}{z_r-b}=2\pi \delta^2(z_r-b) \end{equation}
implies by differentiating with respect to $z_r$ that
\begin{equation}\label{exico}\frac{\partial}{\partial\bar z_r}\frac{1}{(z_r-b)^2}=-2\pi\partial_{z_r} \delta^2(z_r-b) \end{equation}
and hence, because of the double pole in $\omega_\ell^{(1)}$,  that 
\begin{equation}\label{flexico}\d \omega_\ell^{(1)}=\d\bar z_r \wedge \d z_r 2\pi q\partial_{z_r}\delta^2(z_r-b)=-\d(\d\bar z_r \cdot 2\pi q\delta^2(z_r-b)). \end{equation}
An ``improved'' version of $\omega_\ell^{(1)}$ which can equally well be used in (\ref{zomely}) and which is closed is
\begin{equation}\label{rexoc}\omega_\ell^{(1*)}= \omega_\ell^{(1)}+\d\bar z_r\cdot 2\pi q\delta^2(z_r-b).\end{equation}
Since $B_r$ does not pass through $b$, the extra term that we have added does not contribute in (\ref{zomely}).

Now if $s$ and $t$ are any two closed one-forms on the two-torus $\Sigma_{0,r}$, we have the topological formula
\begin{equation}\label{motoro}\int_{\Sigma_{0,r}}s\wedge t = \oint_{A^r}s\oint_{B_r}t-\oint_{B_r}s\oint_{A^r}t. \end{equation}
We apply this formula with $s=\d z_r$, $t=\omega_\ell^{(1*)}$.  The left hand side receives a contribution only from the delta function term
in $\omega_\ell^{(1*)}$, and so equals $2\pi q$.   Since the $A^r$ period of $s$ is 1 and the $A^r$ period of $t$ is 0, the right hand side of (\ref{motoro})
is equal to the $B_r$ period of $t$, which by definition is
 $\Omega_{\ell r}$.  So we get the precise relation between the gluing parameter $\Omega_{\ell r}$ and $q$:
\begin{equation}\label{turnkey}\Omega_{\ell r}=2\pi q. \end{equation}

\subsubsection{Period Matrix And Gluing Parameter Of A Super Riemann Surface}\label{splitsuper}

Starting with an ordinary Riemann surface $\Sigma_0$ with a choice of spin structure, one can build a super Riemann surface $\Sigma$
in a natural way.  If $\Sigma_0$ is covered by open sets $U_\alpha$ with local coordinates $z_\alpha$, such that $z_\alpha=u_{\alpha\beta}(z_\beta)$
in intersections $U_\alpha\cap U_\beta$, then $\SIgma$ is covered by the same open sets $U_\alpha$ with local superconformal coordinates
$z_\alpha|\theta_\alpha$, and gluing rules
\begin{align} z_\alpha & = u_{\alpha\beta}(z_\beta)\cr
                    \theta_\alpha & = \left(\frac{\partial u_{\alpha\beta}}{\partial z_\beta}\right)^{1/2}\theta_\beta. \end{align}
(The role of the spin structure on $\Sigma_0$ is to determine the signs of the square roots.)  From this, it is clear that there is a holomorphic
embedding $i:\Sigma_0\to\Sigma$ that maps $z_\alpha$ to $z_\alpha|0$, and a holomorphic projection $\pi:\Sigma\to\Sigma_0$
that maps $z_\alpha|\theta_\alpha$ to $z_\alpha$.  Moreover, $\pi\circ i=1$.   The super Riemann surface $\Sigma$ is said to be ``split'' and
$\Sigma_0$ is called its reduced space.  From a supergravity point of view, the gravitino field vanishes in a split super Riemann surface; its odd moduli are zero.

To develop the theory of super period matrices, we assume that the spin structure of $\Sigma_0$ is
even and that $\Sigma_0$ is sufficiently generic that $H^0(\Sigma_0,K^{1/2})=0$.  These conditions ensure that the space of closed
holomorphic 1-forms on $\SIgma$ has dimension $g|0$.

Given this,   the super period matrix $\hat\Omega$ of a split super Riemann surface $\Sigma$ equals the ordinary period matrix $\Omega$ of $\SIgma_0$.
Indeed, if $\omega_1,\dots,\omega_g$ are a basis of holomorphic 1-forms on $\Sigma_0$, then their pullbacks to $\Sigma$ (via the projection $\pi$) give  a basis
of closed holomorphic 1-forms $\h\omega_1,\dots,\h\omega_g$ on $\Sigma$.  $A$- and $B$-cycles on $\SIgma_0$ can be embedded in $\SIgma$ via
the embedding $i$.  Then the periods of the $\h\omega_i$ coincide with the periods of the $\omega_i$, so  
the definition (\ref{endox}) of the super period matrix reduces to the definition (\ref{mendox}) of the ordinary period
matrix.                 

Once one turns on the odd moduli of $\SIgma$, the $\h\omega_i$ need to be modified and
it is no longer true that $\hat \Omega$ coincides with $\Omega$.  The difference was computed
in \cite{DPhtwo}; see also section 8.3 of \cite{Surfaces}. For us, it will suffice to consider the case of two odd moduli (though the general case
is not much more complicated).  From a supergravity point of view, having two odd moduli means that we take the gravitino field to be
\begin{equation}\label{mexo}\chi_{\t z}^\theta=\sum_{s=1}^2\eta_s f^\theta_{s\,\t z}, \end{equation}
where $\eta_s$, $s=1,2$ are the odd moduli, and the gravitino wavefunctions $f_{s\,\t z}^\theta$, $s=1,2$, are $(0,1)$-forms on $\Sigma_0$ valued in $T^{1/2}$ (the inverse of the line
bundle $K^{1/2}\to\Sigma_0$ that defines the spin structure).   The difference
between $\hat\Omega$ and $\Omega$ is then given by an integral over a product $\SIgma_0\times \Sigma_0'$ of two copies of $\SIgma_0$,
which we parametrize respectively by $\t z,z$ and by $\t z',z'$:
\begin{equation}\label{turnok}\hat\Omega_{ij}-\Omega_{ij}=-\frac{1}{2\pi}\sum_{s,t=1}^2\eta_s\eta_t\int_{\SIgma_0\times \Sigma_0'}
\omega_j(z) f_{s\,\t z}^\theta(\t z;\neg z)\d\t z\, S(z,z') f_{t\,\t z'}^\theta\d\t z'\, \omega_i(z'). \end{equation}
(In this section, the difference between $\t z$ and $\bar z$ is generally not important, because all integrals will be taken
on ordinary Riemann surfaces.)  Here $S(z,z')$ is the Dirac propagator\footnote{\label{helplater}
This  propagator only exists if the Dirac equation has no zero-modes or in other words if $H^0(\Sigma_0,K^{1/2})=0$. As one varies $\Sigma_0$ in its moduli space $\M_g$, there is a divisor  in $\M_g$ along which $H^0(\SIgma_0,K^{1/2})\not=0$.
Along this divisor, $S(z,z')$ has a pole and hence $\hat\Omega_{ij}$ has a pole, with nilpotent residue.}   on $\SIgma_0$ (for the chosen spin structure); it obeys $S(z,z')=-S(z',z)$, and for fixed $z'$, it satisfies the Dirac
equation on $\Sigma_0$ with a simple pole of residue 1 at $z=z'$:
\begin{equation}\label{elbow} S(z,z')\sim \frac{1}{z-z'}\sqrt{\d z}\sqrt{\d z'},~~z\to z'. \end{equation}
The integral in (\ref{turnok}) is well-defined because $\omega_i$ and $\omega_j$ are $(1,0)$-forms, the $f$'s are $(0,1)$-forms valued in $T^{1/2}$, and as a function of either $z$ or $z'$,
$S$ is valued in $K^{1/2}$, which is the dual of $T^{1/2}$.

We want to apply this to the case that $\Sigma_0$ is a surface of genus 2 built by gluing together two genus 1 surfaces $\SIgma_{0,\ell}$
and $\SIgma_{0,r}$, each with an even spin structure.  In this case, the genus 2 super Riemann surface $\SIgma$ is likewise built
by gluing together genus 1 super Riemann surfaces $\Sigma_\ell$ and $\SIgma_r$.  $\Sigma_\ell$ is parametrized by superconformal
coordinates $z_\ell|\theta_\ell$ with equivalences 
\begin{equation}\label{hro}z_\ell|\theta_\ell\cong z_\ell+1|\pm \theta_\ell \cong z_\ell+\tau_\ell|\pm \theta_\ell\end{equation}
(where the signs depend on the spin structure and are not both positive, since the spin structure is even), 
and $\SIgma_r$ is parametrized by $z_r|\theta_r$ with analogous equivalences.
$\SIgma$ is built by gluing a marked
point $z_\ell|\theta_\ell=a|\alpha$ in $\SIgma_\ell$ to a marked point $z_r|\theta_r=b|\beta$ in $\SIgma_r$.  $\Sigma_\ell$ with its marked point has one odd modulus,
namely $\eta_\ell=\alpha$, and likewise $\Sigma_r$ with its marked point has one odd modulus, namely $\eta_r=\beta$.    The odd modulus of $\SIgma_\ell$ is
associated to a gravitino wavefunction $f_{\ell \,\t z_\ell}^{\theta_\ell}$ that is supported on $\SIgma_\ell$,
and the odd modulus of $\SIgma_r$ is associated to a gravitino
wavefunction $f_{r,\,\t z_r}^{\theta_r}$ that is supported on $\SIgma_r$.  These wavefunctions have special properties: the odd moduli
in question are trivial if one forgets about the marked points in $\SIgma_\ell$ and $\SIgma_r$, and this means that $f_{\ell\,\t z_\ell}^{\theta_\ell}$
and $f_{r\,\t z_r}^{\theta_r}$ can be gauged away, but not by gauge parameters that vanish at $z_\ell=a$ or $z_r=b$.   Rather
\begin{align}\label{believe} f_{\ell,\t z_\ell}^{\theta_\ell} & =\frac{\partial}{\partial \t z_\ell}w_\ell,~~~w_\ell(\t a;\neg a)=1 \cr
                    f_{r,\t z_r}^{\theta_r} & =\frac{\partial}{\partial \t z_r}w_r,~~~w_r(\t b;\neg b)=1 .\end{align}
However, we will postpone using this fact.  

We want to use  eqn. (\ref{turnok}) to compute the off-diagonal matrix element $\hat\Omega_{\ell r}$.  If $\SIgma$ is obtained by
gluing together $\SIgma_\ell$ and $\SIgma_r$, then we can take $\omega_\ell=\d z_\ell$, $\omega_r=\d z_r$, as in section \ref{splitor}.
In this case, since $\omega_\ell$ is supported on $\SIgma_\ell$ and $\omega_r$ on $\SIgma_r$, the integral in (\ref{turnok})
over $\Sigma_0\times\Sigma_0'$ becomes an integral over $\SIgma_{0,\ell}\times \SIgma_{0,r}$:
\begin{equation}\label{mizno} \hat\Omega_{\ell r}-\Omega_{\ell r}=-\frac{\alpha\beta}{\pi}  \int_{\Sigma_{0,\ell}\times \Sigma_{0,r}}
\omega_\ell f_{\ell\,\t z_\ell}^{\theta_\ell} (\t z_\ell; z_\ell)\,\d\t z_\ell\, S(z_\ell, z_r)f_{r\,\t z_r}^{\theta_r}(\t z_r;\neg z_r)\,\d \t z_r\, \omega_r. \end{equation}  
This vanishes as long as $\Sigma$ is made by simply gluing together $\Sigma_\ell$ and $\SIgma_r$ at a point, because in this case
the Dirac propagator $S(z,z')$ vanishes for $z\in \Sigma_\ell$, $z'\in \SIgma_r$.  

To get a nonzero result, we have to deform away from the case that $\Sigma_\ell$ and $\Sigma_r$ are simply glued together at a point,
and let the gluing parameter $\varepsilon$ become nonzero. At $\varepsilon=0$, $\SIgma_{0,\ell}$ has a Dirac propagator $S_\ell(z_\ell, z_\ell')$ and $\SIgma_{0,r}$ has a Dirac propagator
$S_r(z_r,z_r')$.    We claim that the small $\varepsilon$ behavior of the full Dirac propagator $S(z,z')$ for $z\in \SIgma_\ell$, $z'\in \SIgma_r$
is 
\begin{equation}\label{enzdob} S(z,z')=\varepsilon S_\ell(z,a)S_r(b,z')+\O(\varepsilon^3). \end{equation}
This formula arises as follows.\footnote{As will be clear from the derivation, the formula is the first term in an expansion in powers of $\varepsilon^2/(z_\ell-a)(z_r-b)$. The
correction to the leading term is of order $\varepsilon^3$, since the operator associated to the next-to-leading contribution is the descendant $\partial\psi$,
of dimension $3/2$.}  We interpret $S(z,z')=\langle\psi(z)\psi(z')\rangle$ as
 a two-point function in the conformal field theory of a free
fermion $\psi$.  In conformal field theory, propagation from $\Sigma_\ell$ to $\SIgma_r$ through the
narrow neck joining them can be expressed as a sum of contributions obtained by inserting an operator $\O$ at the point $a\in \SIgma_\ell$
and a conjugate operator $\h\O$ at the point $b\in \Sigma_r$, and multiplying by an $\varepsilon$-dependent factor that accounts for propagation
through the neck.  The insertion is thus $\O(a)\cdot F(\varepsilon)\cdot \h\O(b)$ for some $F(\varepsilon)$; this
is analogous to eqn. (\ref{umbro}), with the difference that we now considering a correlation function rather than a measure on moduli
space.   Scale-invariance determines that if $\O$ and $\h\O$ have a given value of $L_0$, then $F(\varepsilon)$ is
a multiple of $\varepsilon^{2L_0}$. The multiple is 1 if the genus zero  two-point function  $\langle\O(z) \h\O(z')\rangle$ is canonical.  For
small $\varepsilon$, the dominant contribution comes from the operators of lowest dimension that  contribute.  
In the case of free
fermion conformal field theory, this means that the dominant contribution to $\langle \psi(z)\psi(z')\rangle$, with $z\in \Sigma_\ell$, $z'\in \SIgma_r$, comes from $\O=\h\O=\psi$. This operator 
   has $L_0=1/2$,
so $\varepsilon^{2L_0}=\varepsilon$, accounting for the factor of $\varepsilon$ in (\ref{enzdob}); and $\psi$  has a canonical two-point function,
so the coefficient of $\varepsilon$ is 1. 
The upshot of all this is that the correlation function $\langle \psi(z)\psi(z')\rangle$ is,
to first order in $\varepsilon$, equal to $\varepsilon$ times 
 the correlation function
$\langle\psi(z)\psi(z')\psi(a)\psi(b)\rangle$ computed at $\varepsilon=0$.  That last correlation function
is the product of the relevant matrix elements of $S_\ell$ and $S_r$, and this leads to eqn. (\ref{enzdob}).
  
Now we insert (\ref{believe}) and (\ref{enzdob}) in the formula (\ref{mizno}) for $\h\Omega_{\ell r}-\Omega_{\ell r}$.  We also
set $\omega_\ell=\d z_\ell$, $\omega_r=\d z_r$.  We get
\begin{equation}\label{tenzo}\hat\Omega_{\ell r}-\Omega_{\ell r}=-\frac{\alpha\beta\varepsilon}{\pi} \int_{\SIgma_\ell} \d z_\ell \,\d\t z_\ell
\frac{\partial}{\partial \t z_\ell}w_\ell \cdot S(z_\ell,a)\cdot \int_{\SIgma_r} \d z_r \,\d\t z_r
\frac{\partial}{\partial \t z_r}w_\ell \cdot S(z_r,b) . \end{equation}
Upon integrating by parts in both integrals, using
\begin{equation}\label{wenzo}\frac{\partial}{\partial \t z_\ell} S(z_\ell,a)=2\pi \delta^2(z_\ell-a),~~ 
\frac{\partial}{\partial \t z_r} S(z_r,a)=2\pi \delta^2(z_r-a),\end{equation}
along with $w_\ell(\t a;\neg a)=w_r(\t b;\neg b)=1$, we get finally
\begin{equation}\label{genzo}\h\Omega_{\ell r}-\Omega_{\ell r}=-4\pi\varepsilon\alpha\beta. \end{equation}

To turn this into a formula for $\h\Omega_{\ell r}$, we also need to compute $\Omega_{\ell r}$.  This was evaluated, modulo higher
order corrections, 
 in eqn. (\ref{turnkey}), where we should now interpret $q$ as $q_\NS=-\varepsilon^2$.
 Combining these results, we get a formula for $\h\Omega_{\ell r}$:
\begin{equation}\label{yenzo}\h\Omega_{\ell r} =-2\pi (\varepsilon^2+2 \varepsilon\alpha\beta)=-2\pi(\varepsilon+\alpha\beta)^2.\end{equation}
So it is not $\varepsilon$ that can be expressed in terms of the super period matrix, but $\varepsilon+\alpha\beta$:
\begin{equation}\label{metzo}\varepsilon+\alpha\beta = \left(-\frac{\hat\Omega_{\ell r}}{2\pi}\right)^{1/2}. \end{equation}
This is the reason for the correction at infinity that was analyzed in section \ref{bcor}.  The procedure of \cite{DPh} involves holding
$\Omega_{\ell r}$ fixed when integrating over the odd variables, and therefore it involves holding $\varepsilon+\alpha\beta$ fixed;
but near $\varepsilon=0$, it is important to hold $\varepsilon$ fixed instead.

\section{The Supersymmetric Ward Identity}\label{supersym}

\subsection{Review Of Bosonic Symmetries}\label{bosreview}

Our final goal is to clarify at a fundamental level how it is possible for loop corrections in a given vacuum to spontaneously
break supersymmetry even if supersymmetry is unbroken at tree level in that vacuum.  (As usual, we do not consider the effects
of shifting the vacuum to try to restore supersymmetry.)
  It is essential to understand first why this is actually not possible for bosonic 
symmetries of oriented closed-string theories.\footnote{A rough analog of supersymmetry breaking by loops
does exist in open and/or unoriented
string theories: bosonic symmetries that hold at the closed-string tree level can be broken by open-string
boundary conditions and/or orientifold projections.}
Let us consider two rather different examples: momentum conservation and the anomalous $U(1)$ of the 
$SO(32)$ heterotic string on a
Calabi-Yau manifold.  Like all continuous bosonic symmetries of closed-string theories, these are associated 
to conserved currents on the
string worldsheet.  In bosonic string theory, momentum conservation is associated to the worldsheet current 
$J_\mu^I=\partial X^I/\partial\sigma^\mu$
(where the $\sigma^\mu,\,\mu=1,2$ are worldsheet coordinates and $X^I$ is a free field on the worldsheet 
representing motion of the string in the $x^I$
direction in spacetime).  In superstring theory, there is an analogous formula; the relevant conserved 
current is now part of a superfield on
the worldsheet, but this does not modify what follows in an essential way.  For the anomalous $U(1)$, 
the conserved current is the antiholomorphic
current $J_\ell =g^{i\bar i}\lambda_i\lambda_{\bar i}$ that is familiar from section \ref{mass}.  

\begin{figure}
 \begin{center}
   \includegraphics[width=3.5in]{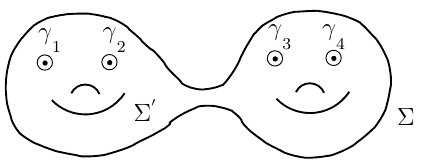}
 \end{center}
\caption{\small To derive a Ward identity from a conserved current in closed string theory, we omit a small open ball around each vertex operator
in the string worldsheet $\SIgma$ to make a two-manifold $\Sigma'$ with boundary, over which we then integrate
the divergence of the  current. The
 open balls are bounded by circles $\gamma_1,\dots,\gamma_n$ -- sketched here for $n=4$.}
 \label{omitting}
\end{figure}
Once we have a conserved current $J^\mu$, or equivalently a closed operator-valued
one-form ${\eusm J}=\epsilon_{\mu\nu}J^\mu \d\sigma^\nu$,
we can derive a Ward identity.  Suppose that we are given vertex operators $\V_1,\dots,\V_n$ that
have definite charges $q_1,\dots,q_n$ in the sense that
\begin{equation}\label{monk}\oint_{\gamma_i} {\eusm J}\cdot \V_i=q_i\V_i, \end{equation}
where $\gamma_i$ is a small closed circle  that wraps counterclockwise once around  $\V_i$.   Now consider the correlation function 
$\langle \V_1\V_2\dots\V_n\rangle$ on a string worldsheet $\SIgma$.  Here we make ghost insertions\footnote{When we say that a worldsheet current is ``conserved,'' we mean in 
particular that it is  anomaly-free, and hence truly is conserved even on a curved worldsheet.   We also assume that it remains conserved in the presence of  the ghost insertions that are needed in defining
 superstring scattering amplitudes.  
In particular, the ghost number current does not qualify: it has an anomaly on a curved worldsheet, and does not commute with the usual ghost insertions.}  as necessary so that this correlation function is not trivially
zero, but we do not integrate over any moduli. To prove a Ward identity, we proceed in the standard fashion.  We let
$\Sigma'$ be the complement in  $\SIgma$ of the interiors of the $\gamma_i$ (fig. \ref{omitting}). By integrating the conservation
law $0=\d\eusm J$ over $\Sigma'$ and then integrating by parts to pick up surface terms, which we evaluate using
(\ref{monk}), we deduce the Ward identity:
\begin{equation}\label{zonder}0=\int_{\Sigma'} \left\langle\d{\eusm J}\cdot
\V_1\dots \V_n\right\rangle=\sum_{i=1}^n q_i\cdot \bigl\langle \V_1\dots \V_n\bigl\rangle. \end{equation}
Thus, the correlation function $\langle \V_1\dots\V_n\rangle $ vanishes unless $\sum_iq_i=0$.

This is our Ward identity, and since  it  holds without any integration over moduli, there is no room for any subtlety.  The
contribution to a scattering amplitude with $\sum_iq_i\not=0$ vanishes before any integration over moduli, so it certainly vanishes
after any  such integration. 

We expressed the computation in eqn. (\ref{zonder})
 in terms of a string worldsheet with bosonic coordinates only, but including fermionic
coordinates on the worldsheet changes nothing essential: a conserved current leads to a conservation law
on a fixed worldsheet and this conservation law remains valid after integration over moduli.  In practice, this means that
in closed oriented
superstring theory, all bosonic symmetries that hold at tree level are also valid in perturbation theory. 

This is so  even for the anomalous $U(1)$
symmetries that can arise in heterotic string compactification to four dimensions.  
An anomalous $U(1)$ gauge boson gets mass at 1-loop order via
a Higgs mechanism, but the associated {\it global} conservation law -- which is what we proved using the Ward identity --
remains valid in perturbation theory. This was explained in section \ref{effreview}: the scalar field $a$ that is important in the
Higgs mechanism decouples in perturbation theory.

\subsection{How Supersymmetry Is Different}\label{susydifferent}

Spacetime supersymmetry is {\it not} associated to a conserved current in this sense.  
The supersymmetry generator is the
fermion vertex operator  of \cite{FMS,K}, taken at zero spacetime momentum.  
Because it is a Ramond vertex operator,
it is not really a conserved worldsheet current in the traditional sense assumed in section \ref{bosreview}.

 For heterotic strings in $\R^{10}$, the fermion vertex operator is customarily written
\begin{equation}\label{yedly}S_A=e^{-\tphi/2}\Sigma_A,\end{equation}
where $\Sigma_A$ is the spin operator of the matter system, which transforms as a positive chirality spinor
of $SO(1,9)$.  We will take the basic object to be not $S_A$ but its unintegrated counterpart 
\begin{equation}\label{medly}\S_A=c S_A,\end{equation}
which we call the spacetime supersymmetry generator.
Importantly, $\S_A$ has a factor of $c$ and no corresponding $\t c$, so its ghost number is less by 1 than
that of the vertex operator for a physical state of the heterotic string.  The reason for emphasizing $\S_A$ rather
than $S_A$ will hopefully become clear.  We sometimes omit the $A$ index and write just $\S$ for a generic
linear combination of the $\S_A$.

The  operator $\S_A$ is holomorphic, in the sense that it varies
holomorphically with the moduli of the superstring worldsheet $\Sigma$,
and is on-shell, in the sense that it obeys the holomorphic part of the physical state conditions
(its antiholomorphic part is the identity operator, which is not an on-shell vertex operator).  
A Neveu-Schwarz vertex operator
with those properties would be associated to a conserved worldsheet current that could be used to generate a Ward
identity along the standard lines that were reviewed in section \ref{bosreview}.  But the fermion
vertex operator  is a Ramond sector vertex operator, and the usual framework for deriving a Ward
identity {\it on a fixed worldsheet}, without integrating over moduli, does not make sense for 
Ramond sector vertex operators.

This is because a Ramond vertex operator is inserted at a singularity in the superconformal structure of
$\Sigma$.  (The usual explanation of this is the assertion that fermi fields have square root branch points
near a Ramond vertex operator insertion.  One can eliminate the branch points in favor of a more subtle
singularity in the superconformal structure; see for example section 4 of \cite{Surfaces} or section 
\ref{where} below.)  It does not make
sense to move this singularity while keeping the other moduli of $\Sigma$
fixed; there is no notion of two super Riemann surfaces being the same
except for the location of a Ramond singularity.  So the usual procedure of deriving
a Ward identity by integration over $\Sigma$ does not apply for spacetime
supersymmetry.  This is true even for superstring theory in $\R^{10}$.

At string tree level, it is possible to put the discussion of spacetime
supersymmetry in the framework of a ``conserved worldsheet  current.''
To do this, one absorbs the odd moduli in the definitions
of the vertex operators by using vertex operators of appropriate picture numbers.
Once the odd moduli are hidden in this way, one can treat $\S_A$ rather like conventional
conserved currents.  In loops, this procedure leads to what technically have been called  ``spurious singularities.''
Trying to express loop amplitudes in a framework that really does not quite apply made the literature of the
1980's cumbersome in places.

\subsubsection{Closed Form On Supermoduli Space}\label{closedsuper}

If we cannot interpret the supersymmetry generator $\S_A$ as a conserved current on the worldsheet, how can we use it to
derive a Ward identity and why is spacetime supersymmetry ever valid?  What follows is only an overview;
much more can be found in \cite{Revisited}. 
Some of the necessary ideas were  developed in the 1990's
in work that has unfortunately remained little-known \cite{Belopolsky}.  

To derive the Ward identity for an $n$-particle scattering amplitude in the case of a bosonic symmetry, we started
with an $n+1$-point correlation function 
\begin{equation}\label{uggly}\langle J^\mu\,\V_1\dots\V_n\rangle.\end{equation}
The analog for supersymmetry is the correlation function
\begin{equation}\label{juggly}F_{\S_A\V_1\dots\V_n}=\langle \S_A\V_1\dots\V_n\rangle.\end{equation}
But what kind of object is $F_{\S_A\V_1\dots\V_n}$?  In the case of a bosonic conserved current $J^\mu$,
to derive the Ward identity, we varied the insertion point of $J^\mu$ but kept fixed $\Sigma$ and the insertion points
of all other vertex operators.  We cannot do this for the Ramond  vertex operator $\S_A$,
since there is no natural operation of varying the position of a Ramond vertex operator in a super Riemann surface $\Sigma$ without
varying all of the moduli of $\Sigma$.   The only natural operation is to vary all of the moduli of $\Sigma$, and to interpret
$F_{\S_A\V_1\dots\V_n}$ as an object of some kind on $\MM_{g,n+1}$, the moduli space of super Riemann surfaces
of genus $g$ with $n+1$ punctures (here our notation is oversimplified: we should specify the separate numbers
of NS and Ramond punctures, but to keep the notation simple we only indicate the total number of punctures).

What sort of object
on $\MM_{g,n+1}$ is $F_{\S_A\V_1\dots\V_n}$?
 It is not a {\it measure} that can be integrated over $\MM_{g,n+1}$ to get a scattering amplitude
-- since $\S_A$ is not the vertex operator of a physical state.  Indeed, the ghost number of
$\S_A$ is less by 1 than the ghost number of a physical state vertex operator; as
a result, $F_{\S_A\V_1\dots\V_n}$ is a  form of codimension 1 on $\MM_{g,n+1}$. It is a closed form, obeying
$\d F_{\S_A\V_1\dots\V_n}=0$, if the vertex operators $\V_1\dots\V_n$ are all annihilated by the BRST operator $\sQ$.
(In a dual language, one would call this correlation function a conserved current on $\MM_{g,n+1}$ rather than a closed form.)
For background on forms and exterior
derivatives on a supermanifold and the supermanifold version of Stokes's theorem, see for example \cite{Supermanifold}.
For other statements made in this paragraph, see \cite{Revisited}, starting with section 3.

The upshot is that
 we can derive a conservation law, not by integrating the equation $\partial_\mu J^\mu=0$ over $\Sigma$ but by integrating 
the equation   $\d F_{\S_A\V_1\dots\V_n}=0$ over $\MM_{g,n+1}$.  
Upon using the supermanifold version of Stokes's theorem, we get
\begin{equation}\label{reffo} 0=\int_{\MM_{g,n+1}}\d F_{\S_A\V_1\dots\V_n}=\int_{\partial\MM_{g,n+1}}
F_{\S_A\,\V_1\dots\V_n}.  \end{equation}
Here $\partial\MM_{g,n+1}$, the ``boundary'' of $\MM_{g,n+1}$, is a union of components associated to the different ways that
a narrow neck in  $\Sigma$ may collapse.  As we will see, the relation (\ref{reffo}) is the Ward identity of spacetime supersymmetry, including
a possible Goldstone fermion contribution. 
  
\subsubsection{A Bosonic String Analog}\label{bosan}

Before analyzing the Ward identity in detail, we pause to explain that some of the key points we have made actually
have bosonic string analogs.  For brevity, we consider only open strings or a chiral sector of closed strings.

A conformal vertex operator representing a physical state of the bosonic string has the form $\V=c V$, where $c$ is the usual ghost field and 
$V$ is a dimension 1 primary constructed
from matter fields only.  If $\V$ is BRST-trivial, meaning that $\V=\{\sQ,\W\}$ for some $\W$, then the string state corresponding to $\V$ is called
a null state and should decouple from scattering amplitudes.

The decoupling of 
{\it massless} null states can be proved in a particularly elementary way.  If $\V$ is massless and null, then $V=L_{-1} W$ 
for some $W$.  Equivalently, $V=\partial W$, so
 the integrated insertion of $V$ vanishes: $\int V=\int \partial W=0$.  (Here one has to verify that there are no anomalies
coming from boundary terms in this integral, but this is not a serious problem.) 

For massive null states, there is no equally elementary argument.  If $\V=cV$ is massive and null, then $\V=\{\sQ,\W\}$ for some $\W$
and furthermore  \cite{Thorn} one has $V=L_{-1}W_1+(L_{-2}+(3/2) L_{-1}^2)W_2$
for some $W_1$ and $W_2$.  But (because of the $W_2$ term) this does not make $V$ a total derivative on the string worldsheet
and there is no reason for the integral $\int V$ to vanish on a fixed string worldsheet.
Hence the only simplicity comes in the full integral over moduli space: the relation $\V=\{\sQ,\W\}$ implies that the form
on $\M_{g,n}$ that must be integrated to compute a scattering amplitude   with an insertion of $\V$ is exact.  

In other words, the decoupling of
a massive null state in bosonic string theory must be proved by integration by parts on $\M_{g,n+1}$, not on the string worldsheet.
The identity that proves decoupling of a null state has the same form as (\ref{reffo}) except that the left hand side is a scattering amplitude
with a null state included, rather than 0.  If $\V=\{\sQ,\W\}$, then
\begin{equation}\label{mozlo}F_{\V\V_1\dots\V_n}+\d F_{\W\V_1\dots\V_n}=0\end{equation}
so 
\begin{equation}\label{ozlo}-\int_{\M_{g,n+1}}F_{\V\V_1\dots\V_n}=\int_{\M_{g,n+1}}\d F_{\W\V_1\dots\V_n}=\int_{\partial\M_{g,n+1}}F_{\W\V_1\dots
\V_n}. \end{equation}
For a null state of generic momentum, it is not hard to show that the surface terms on the right hand side of eqn. (\ref{ozlo}) vanish, establishing
the decoupling of the null state $\V$.

What we learn here is that the essential subtlety of spacetime supersymmetry in superstring theory has a close analog in the procedure to prove decoupling
of massive null states of the bosonic string (or of superstring theory).
 The analogy becomes even closer if we analyze the decoupling of a longitudinal gravitino state at nonzero (but on-shell) momentum.
 Eqn. (\ref{ozlo}) applies without change for the case that $\V=\{\sQ,e^{ik\cdot\X}\mu^A\S_A\}$ is the vertex operator of a longitudinal gravitino of lightlike
 momentum $k$ (here
 $\mu^A$ is a $c$-number spinor that obeys the Dirac equation $k\cdot \Gamma\,\mu=0$).  What is special about spacetime supersymmetry is merely that in the case of the gravitino, $\V=0$ at $k=0$ (since $\{\sQ,\S_A\}=0$).  As a result, in the case of the gravitino, upon setting $k=0$,
  we get an identity (\ref{reffo})
 with the same
form as (\ref{ozlo}) except that the left hand side vanishes, leading to a conservation law (rather than the decoupling of a null state).
There is no conservation law associated to decoupling of massive null states in bosonic string theory or superstring theory, 
since there is no value of the momentum
at which the vertex operator of a massive null state vanishes.  This is usually described by saying that gauge invariances of massless null states
(including the gravitino if it is massless) lead to conservation laws in spacetime, while the gauge invariances of massive string states are spontaneously broken.

\subsubsection{Contributions To The Ward Identity}\label{wardit}  
  
Now we return to the analysis of the Ward identity (\ref{reffo}).  
In general,  $\partial\MM_{g,n+1}$ has many components -- associated to  various separating and nonseparating degenerations.  But many of these boundary components do not contribute to the Ward identity.  
The only ones that do contribute are   those in which, for kinematic reasons, the spacetime  momentum
flowing through a narrow neck in the string worldsheet is forced to be on-shell.  For example, a nonseparating degeneration does not contribute
to the Ward identity since in this case the momentum flowing through the neck is one of the integration variables in the
path integral and is generically not on-shell.
Likewise a separating degeneration with multiple vertex operators on both sides does not contribute.  

\begin{figure}
 \begin{center}
   \includegraphics[width=3.5in]{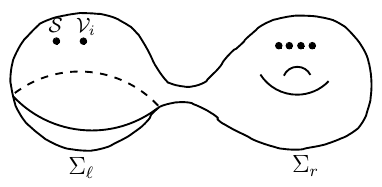}
 \end{center}
\caption{\small The Ward identity always receives contributions from separating degenerations of this kind in which  one
component $\Sigma_\ell$ contains a supersymmetry generator $\S$ and precisely
one more vertex operator. If these are the only contributions, then the Ward identity expresses the invariance of the $S$-matrix under spacetime supersymmetry.  In the example sketched,
$\Sigma_\ell$ has genus 0.  This leads to the familiar tree-level expressions for the supercharges.  }
 \label{Supersymmetry}
\end{figure}

One type of boundary component that always contributes is sketched in fig. \ref{Supersymmetry}.  The string worldsheet 
$\Sigma$, of genus $g$, degenerates
to a union of two components $\Sigma_\ell$ and $\Sigma_r$, of genera $g_\ell$ and $g_r$,
where $\Sigma_\ell$ contains precisely two vertex operators:
the supersymmetry generator $\S_A$ and  one of the others, say $\V_i$.  $\Sigma_r$ contains
the other vertex operators $\V_1\dots \hat \V_i \dots \V_n$.   Since $\S_A$
carries zero momentum in spacetime, the momentum flowing through the neck is equal to the momentum carried by $\V_i$
and in particular is on-shell for some string states.   The contribution of this type of degeneration is as follows.
$\Sigma_\ell$ with the two vertex operators $\S_A$ and $\V_i$ that it contains, and the instruction to extract
a boundary contribution, can be replaced with a physical state vertex operator $\O_{i,A}$ that should be inserted on  $\Sigma_r$.
The contribution of this boundary component to the Ward identity is then given by a path integral on $\Sigma_r$; this
path integral is the genus $g_r$ contribution to a scattering amplitude $\langle \V_1\dots\V_{i-1}\O_{i,A}\V_{i+1}\dots\V_n\rangle$.

The operator $\O_{i,A}$, since it is produced by a path integral on $\Sigma_\ell$ with insertions of $\S_A$ and $\V_i$,
is linear in $\S_A$ and also linear in $\V_i$.  So we can define a linear transformation $Q^{(g_\ell)}_A$ of the space of physical
vertex operators such that $\O_{i,A}=\{Q^{(g_\ell)}_A,\V_i\}$.  This $Q^{(g_\ell)}_A$ is the spacetime supersymmetry charge, or more
precisely it is the genus $g_\ell$ contribution to it.  The full spacetime supersymmetry
charge is $Q_A=\sum_{\ell=0}^\infty g_\st^{2g_\ell}Q_A^{(g_\ell)}$, with $g_\st$ the string coupling.  The $g_\ell=0$ contribution to $Q_A$ coincides with the spacetime supersymmetry charge
as traditionally defined \cite{FMS,K}.  This is so because a degeneration with $g_\ell=0$ simply results from a collision between
two operators, and its effects, in the present context, are captured by the leading behavior in the  $\S_A\cdot \V_i$ operator product.
The operators $Q_A^{(g_\ell)}$, $g_\ell>0$, represent loop corrections to the supersymmetry charges; these have not been investigated,
but 
probably are entirely determined by the loop corrections to particle masses and possible loop corrections to
central charges in the supersymmetry algebra.

If  the boundary contributions just analyzed are the only ones, then the Ward identity, after summing
over the choice of the  vertex operator $\V_i$ that is contained in $\Sigma_\ell$ along with the supersymmetry generator, takes a familiar form:
\begin{equation}\label{pokkol} 0=\sum_i\langle \V_1\dots \V_{i-1}\{Q_A,\V_i\}\V_{i+1}\dots\V_n\rangle. \end{equation}
This is the standard form of the identity expressing invariance of the $S$-matrix under a conserved charge.  It says
that the $Q_A$ generate symmetries of the $S$-matrix.  These are the spacetime supersymmetries.

\begin{figure}
 \begin{center}
   \includegraphics[width=3.5in]{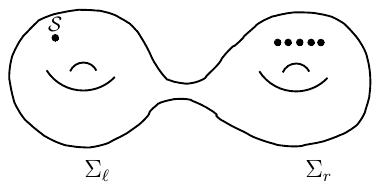}
 \end{center}
\caption{\small One other degeneration may contribute to the Ward identity.  This is
the Goldstone fermion contribution.  It represents spontaneous breaking of spacetime supersymmetry.
This contribution can exist only when the genus of $\Sigma_\ell$ is positive.}
 \label{Goldstone}
\end{figure}

However, as one might surmise based on experience in field theory, there is also a possible  boundary
contribution to the Ward identity associated to spontaneous symmetry breaking. This arises (fig. \ref{Goldstone}) when one component, say $\Sigma_\ell$, contains
the supersymmetry generator $\S_A$ and no other vertex operators.  Since $\S_A$ carries zero momentum,
the momentum flowing between $\Sigma_\ell $ and $\Sigma_r$ is zero, which can be the momentum of an on-shell
string state -- a massless fermion.   The contribution of this kind of boundary component to the $S$-matrix can be evaluated
by inserting on $\Sigma_r$ a physical state vertex operator $\O(\S_A)$ that reproduces the effect of the path integral
on $\Sigma_\ell$.  Thus the boundary contribution to the Ward identity is the genus $g_r$ contribution to an $n+1$-particle
scattering amplitude $\langle  \O(\S_A)\V_1\dots\V_n\rangle$.   Comparing to field theory, the interpretation is clear:
when not zero, $\O(\S_A)$  is the vertex operator of a Goldstone fermion.  

When $\O(\S_A)$ is nonzero,
we no longer get a Ward identity (\ref{pokkol}) for unbroken supersymmetry; there is an additional Goldstone fermion
contribution, just as occurs in a field theory model with spontaneously broken supersymmetry.  The existence in perturbative string theory of a massless dilaton means that once a Goldstone fermion is generated,
one expects to find an instability in higher order, and one does not expect the $S$-matrix to exist to all orders.
Conversely, when supersymmetry is valid to all orders, part of the proof of this involves an inductive argument to show the
vanishing of massless tadpoles.

\subsubsection{The Dilaton Tadpole And The Mass Splitting}\label{tadmass}

To provide a context for the discussion of Goldstone fermion contributions, let us return to the heterotic string on a Calabi-Yau, and 
consider in this light the two supersymmetry-breaking effects that we studied in sections \ref{mass} and \ref{dilaton}
-- the 1-loop mass splittings and the 2-loop vacuum energy.  In this case, only the four-dimensional Lorentz
group $SO(1,3)$ (and not its ten-dimensional cousin $SO(1,9)$) acts on the unbroken supersymmetries.
We write $\S_\alpha$ and $\S_{\dot\alpha}$, $\alpha,\dot\alpha=1,2$ for supersymmetry generators of positive
and negative chirality, and $Q_\alpha$, $Q_{\dot\alpha}$ for the corresponding supercharges.

\begin{figure}
 \begin{center}
   \includegraphics[width=6.5in]{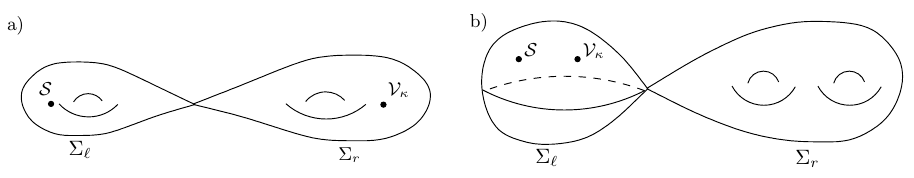}
 \end{center}
\caption{\small For the case of only two vertex operators, namely a supersymmetry generator $\S$ and the vertex operator
$\V_\kappa$ for the dilatino, there are only two possible contributions
to the Ward identity: a Goldstone fermion contribution (a), and a second contribution (b) that is proportional to the dilaton
tadpole.  The dilaton tadpole vanishes if and only if there is no Goldstone fermion contribution.}
 \label{DT}
\end{figure}

The usual way to use supersymmetry to analyze the vacuum energy is to observe that the $g$-loop vacuum energy is essentially
equivalent to the $g$-loop tadpole of the dilaton vertex operator $\V_\phi$.  To   analyze
this tadpole using supersymmtry, one considers a Ward identity involving the dilatino
(the massless fermion $\kappa$ that is in the same supersymmetry multiplet with the dilaton). Let $\V_{\kappa,\beta}$ be the
dilatino vertex operator 
or more precisely a particular spinor component of this operator, and let  $\S_\alpha$ be a supersymmetry generator such
that $\epsilon_{\alpha\beta}\V_\phi=\{Q_\alpha,\V_{\kappa,\beta}\}$.  Consider the supersymmetric Ward identity (\ref{reffo})
derived from a two-point function $F_{\S_\alpha \V_{\kappa,\beta}}=\langle \S_\alpha\V_{\kappa,\beta}\rangle$.  
There are only two 
possible boundary
contributions
(fig. \ref{DT}).   One contribution, shown on the left of the figure, is the Goldstone fermion contribution; the other contribution,
shown on the right, is the tadpole of $\V_\phi$.  
The dilaton tadpole vanishes if and only if there is no Goldstone fermion contribution.

It will probably come as no surprise to the reader who has gotten this far that  heterotic string compactifications with
an anomalous $U(1)$ do develop a Goldstone fermion at 1-loop level. Demonstrating this explicitly will be the goal of
section \ref{goldstonefermion}.   The Goldstone fermion is the gaugino $\zeta_\alpha$,
the fermion that is in the vector multiplet that contains the anomalous $U(1)$ gauge field. 
So in fig. \ref{DT}(a), we can replace $\Sigma_\ell$ by an insertion on $\Sigma_r$ of the gaugino vertex operator $\O(\S_\alpha)=\V_{\zeta_\alpha}$.
The resulting path integral on $\Sigma_r$ is then  the two-point function $\langle\V_{\zeta,\alpha}\V_{\kappa,\beta}\rangle$; in
other words, it is the 1-loop contribution to the $\zeta\kappa$ mass term.  In section \ref{simplesplit}, we showed that this quantity
is nonzero, and therefore the vanishing of the sum of the boundary
contributions in fig. \ref{DT} 
implies that there must be a 2-loop dilaton tadpole.  

\begin{figure}
 \begin{center}
   \includegraphics[width=4.5in]{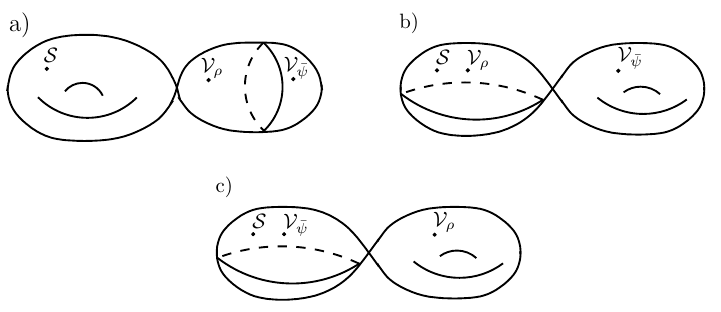}
 \end{center}
\caption{\small The Ward identity governing the 1-loop mass splittings studied in section \ref{mass}
has these three contributions.  The 1-loop fermion and boson mass splittings appear in (b) and (c),
respectively, while the Goldstone fermion contribution to the Ward identity appears in (a).}
 \label{underwood}
\end{figure}

We can proceed in the same way  to analyze the 1-loop mass shift in a charged chiral multiplet.   In this case, suppressing
some indices,
we consider a Ward identity derived from a correlation function 
\begin{equation}\label{boxxo}F_{\S\,\V_\rho \V_{\bar \psi}}=\langle \S\,\V_\rho\,\V_{\bar\psi}\rangle \end{equation}
of three vertex operators (fig. \ref{underwood}), 
 namely a supersymmetry generator $\S$,
the vertex operator $\V_\rho$ for a boson $\rho$ in a chiral multiplet, and the vertex operator 
$\V_{\bar \psi}$ for a fermion $\bar\psi$ in the conjugate
antichiral multiplet.  The Ward identity (\ref{reffo}) now has three terms (fig. \ref{underwood}).  One term is the Goldstone fermion contribution
and the other two involve the 1-loop bose and fermi mass shifts.  The bose and fermi mass shifts fail to be equal
if and only if the Goldstone fermion contribution is nonvanishing.  

\subsection{The Goldstone Fermion}\label{goldstonefermion}

It is rather tricky to show explicitly that in heterotic string compactifications with an anomalous $U(1)$ gauge symmetry,
the gaugino becomes a Goldstone fermion at the 1-loop level.  
This amounts, roughly speaking, to evaluating
a two-point function in genus 1, but not quite a standard one. 

To understand exactly what we have to calculate, let us start with a concrete example such
as that of fig. \ref{underwood} and take a close look at the supersymmetry-violating contribution of 
fig. \ref{underwood}(a).  This contribution is associated to the splitting of a super Riemann surface $\Sigma$
into two components $\Sigma_\ell$ and $\Sigma_r$, joined at a Ramond degeneration.  In other words, the string
state propagating between $\Sigma_\ell$ and $\SIgma_r$ is a state in the Ramond sector (namely the Goldstone fermion).
By contrast, previous
degenerations considered  in this paper have always been NS (or bosonic string) degenerations.    

In the analysis of the Goldstone fermion contribution to the Ward identity,
we are supposed to keep $\Sigma_r$ fixed.   The idea of spontaneous supersymmetry breaking is that no matter
what $\Sigma_r$ may be, the contribution of the degeneration of fig. \ref{underwood}(a) to the Ward identity is
an ordinary scattering amplitude evaluated on $\Sigma_r$ with $\Sigma_\ell$ replaced by the vertex operator of
the Goldstone fermion.  Since this statement is supposed to hold for any $\Sigma_r$, we can consider $\Sigma_r$ to be
fixed and arbitrary
in the analysis.

A minor simplification in the example of fig. \ref{underwood} is that, as $\Sigma_r$ is a super Riemann surface
of genus 0 with 1 NS puncture and 2 Ramond punctures, it has no even or odd moduli and its moduli space $\M_r$ is
a point.  That makes it particularly straightforward to keep $\Sigma_r$ fixed in this example.

\subsubsection{Where To Integrate}\label{where}

Near a separating Ramond degeneration, the (holomorphic) moduli of $\Sigma$ can be factored as follows:
\begin{itemize}\item One factor is the moduli space $\MM_\ell$ associated to $\Sigma_\ell$.  Here $\Sigma_\ell$ is
a super Riemann surface of genus $g_\ell$ (for us, $g_\ell=1$) with 2 Ramond punctures.
\item A second factor is the moduli space  $\MM_r$ associated to $\Sigma_r$.  This depends on the specific process considered,
but will play no important role since $\Sigma_r$ is held
fixed in the whole analysis.
\item Finally, there are moduli associated to the gluing.  There is a bosonic gluing parameter 
$q_\Ra$ associated to a Ramond degeneration; it is quite analogous to the bosonic and NS sector
gluing parameters $q$ and $q_\NS$ that are familiar from section \ref{genless}. But there is also
a fermionic gluing parameter $\alpha$ that is special to the Ramond sector.  Its existence is 
related to the fact that, in the Ramond sector,
the worldsheet supercurrent has a zero-mode $G_0$. \end{itemize}

To understand technically why there is a fermionic gluing parameter (see \cite{Surfaces}, especially sections 4
and 6.2, for much more), we must recall that a Ramond vertex
operator is associated to a singularity of the superconformal structure of $\Sigma$.  In general, a super Riemann
surface is a $1|1$ complex supermanifold that is endowed with a superconformal structure.  Such a structure is
a rank $0|1$ subbundle  $\D$ of the tangent bundle $T\Sigma$ of $\Sigma$ that is generated, in local 
superconformal coordinates $x|\theta$, by $D_\theta
=\partial_\theta+\theta\partial_x$.  The key property of this operator is that $D_\theta$ and $D_\theta^2=\partial_x$
are everywhere linearly independent and generate the full tangent bundle $T\Sigma$.  A Ramond
vertex operator is inserted on a divisor (that is, a submanifold of $\Sigma$ of dimension $0|1$) on which this linear
independence fails.  The local structure is that $\D$ is generated in suitable coordinates by $D^*_\theta=\partial_\theta
+\theta x \partial_x$, obeying $(D_\theta^*)^2=x\partial_x$; thus, the linear independence of $D_\theta^*$ and its
square fails on the divisor $\F$ given by $x=0$, which is where a Ramond vertex operator is inserted.  We call $\F$
a Ramond divisor.

Now let us describe the gluing of two super Riemann surfaces $\Sigma_\ell$ and $\Sigma_r$ along Ramond divisors.
We suppose that in local coordinates $x|\theta$, $\Sigma_\ell$ has superconformal structure defined by
$D^*_\theta=\partial_\theta+\theta x\partial_x$, with a Ramond divisor $\F_\ell$ at $x=0$; and similarly that in local
coordinates $y|\psi$, $\Sigma_r$ has superconformal structure defined by $D^*_\psi=\partial_\psi+\psi y\partial_y$, with
a Ramond divisor $\F_R$ at $y=0$.
Then -- ignoring for the moment the fermionic gluing parameter -- $\Sigma_\ell$ and $\SIgma_r$ can be glued by
\begin{align}\label{medzo} xy& = q_\Ra \cr
                                         \theta&=\pm \sqrt{-1}\psi .\end{align}
Here $q_\Ra$ is the analog of the familiar bosonic and NS gluing parameters $q$ and $q_\NS$.  The constant of
proportionality between 
$\theta$ and $\psi$ must be $\pm\sqrt{-1}$  so that the gluing map is superconformal, or in other words
so that  $D^*_\psi$ is proportional to $D^*_\theta$ and generates the same line bundle $\D$.  The sum over the sign
in the gluing map leads to the GSO projection.

The fermionic gluing parameter can be included by generalizing (\ref{medzo}) by first making a superconformal transformation of the local superconformal
coordinates $x|\theta$
of $\Sigma_\ell$ (or a similar transformation of the local superconformal coordinates of $\Sigma_r$) whose restriction to $\F_\ell$ (or $\F_r$) is non-trivial.  Such a transformation is
$\theta\to\theta-\alpha$, $x\to x+\alpha\theta x$, with $\alpha$ an odd parameter.  
All that matters here is how this transformation acts on
  $\F_\ell$.  In particular, for $\alpha\not=0$, the
 gluing of the two Ramond divisors    $\F_\ell$ and $\F_r$ after the change of coordinates takes the form
\begin{equation}\label{edzo}\theta-\alpha=\pm \sqrt{-1}\psi. \end{equation}

At $q_\Ra=0$, keeping fixed $\Sigma_\ell$ and  $\Sigma_r$ and thus the definitions of $\theta$ and $\psi$,
the fermionic gluing parameter $\alpha$ is the choice of a point $\theta=\alpha$ on $\F_\ell$ that must be glued to the
point $\psi=0$ on $\Sigma_r$. (This interpretation of $\alpha$ is only precise at $q_\Ra=0$.)  
For purposes of analyzing the Goldstone fermion contribution to the Ward identity,
it is useful to include $\alpha$ as an extra modulus of $\Sigma_\ell$.  To do this, we define a supermanifold
$\MM_\ell'$ that is fibered over $\MM_\ell$ with fiber $\F_\ell$.  The calculation that we will eventually perform
is best understood in terms of integration  over $\MM_\ell'$ rather than over $\MM_\ell$.   This will gradually become clear. 

To evaluate the boundary contribution in fig. \ref{underwood}(a), in addition to integrating over $\MM'_\ell$, we also
have to treat properly the bosonic gluing parameter $q_\Ra$.  Roughly speaking, this will mean setting $|q_\Ra|=\eta$,
with $\eta$ a small positive constant, and integrating over $\mathrm{Arg}\,q_\Ra$.  The integral over $\mathrm{Arg}\,q_\Ra$
just gives a factor of $2\pi$; however, the operation that naively consists of setting $|q_\Ra|=\eta$ is subtle and will
be analyzed in section \ref{cohoform}.

Returning to the fermionic gluing parameter, it may be most familiar in the following guise.  The bosonic string propagator
is\footnote{See for example section 6 of \cite{Revisited} for the following formulas.
For brevity, we write these propagators for open strings or a chiral sector of closed strings. The symbols
$b_0, \beta_0$, $L_0$, and $G_0$ are the usual zero-modes of antighosts and Virasoro or super Virasoro generators.  }
  $b_0/L_0$, and the superstring propagator in the NS sector is given by the same formula.  The factor of $1/L_0$,
which comes by integration over the bosonic gluing parameter, is the analog of the usual bosonic propagator
$1/(P^2+M^2)$ of field theory, where $P$ is the momentum and $M$ is the mass operator.  In the Ramond sector,
the propagator is instead
\begin{equation}\label{dobzo}b_0\delta(\beta_0) \frac{G_0}{L_0}, \end{equation}
where here the field theory limit of $G_0/L_0=1/G_0$ is the usual Dirac propagator $1/(\Gamma\cdot P+iM)$.                                   
The factor $G_0$ in the numerator in (\ref{dobzo}) comes from integration over the fermionic gluing parameter $\alpha$.
In our problem, the integration over  $\alpha$ cannot be treated as simply as that.   The reason
is that, rather than a pole associated to an on-shell state propagating between $\Sigma_\ell$ and $\Sigma_r$, we are
trying to evaluate a boundary contribution to a Ward identity, which is a more subtle matter.

\subsubsection{Conformal Vertex Operators}\label{cvo}

To proceed, we require a fact about string perturbation theory that is not new though perhaps also not
well-appreciated.\footnote{For original references, see \cite{VafaMany}, eqn. (5.18), and \cite{Nelson}.  For a recent
treatment, see \cite{Revisited}, starting with section 2.4.1.}  Most of the following does not depend on worldsheet or spacetime supersymmetry,
and for brevity, we mostly use the language of bosonic string theory.

To compute scattering amplitudes in a conformally-invariant fashion,
it does not suffice to represent external states by vertex operators that are conformal primaries of the appropriate
dimension, annihilated by the BRST operator $\sQ$.  The vertex operators  must obey an additional condition, which for unintegrated vertex operators of bosonic
string theory\footnote{Using unintegrated vertex operators in the following discussion enables us to treat
bosonic strings and the NS and Ramond sectors of superstrings in the same way.} is that they must be annihilated
by $b_n,$ $n\geq 0$.  In superstring theory, unintegrated vertex operators must also be annihilated by $\beta_n$, $n\geq 0$.  It is convenient
to refer to vertex operators that are conformal or superconformal 
primaries of dimension 0 and obey these conditions as conformal or superconformal vertex operators.

To explain briefly how this condition comes about, recall that in using 
the worldsheet path integral to construct a measure on
the moduli space of Riemann surfaces (which is then integrated to compute a scattering 
amplitude), one makes antighost insertions,
that is insertions of
\begin{equation}\label{microb}w_h=\int_\Sigma \d^2z\,h_{\t z}^z b_{zz}, \end{equation}
where $h_{\t z}^z$ is a $c$-number wavefunction -- often called a Beltrami differential -- 
that represents a deformation of the complex
structure of $\Sigma$.  For diffeomorphism invariance and conformal invariance of the 
formalism, one needs to know that $w_h$ is
unchanged if $h$ is changed in a manner induced by a diffeomorphism.  The change in $h$ 
under an infinitesimal diffeomorphism generated
by a vector field $v^z\partial_z$ is
\begin{equation}\label{icrob} h_{\t z}^z\to h_{\t z}^z+\partial_{\t z}v^z,\end{equation}
and the corresponding change in $w_h$ is
\begin{equation}\label{crob}\delta_v w_h=\int_\Sigma\d^2z \partial_{\t z}v^zb_{zz}=-\int_\Sigma\d^2z v^z\partial_{\t z}b_{zz},\end{equation}
where in the last step we integrate by parts.  We would like to claim that this last expression vanishes using the antighost equation
of motion $\partial_{\t z}b_{zz}=0$, but there is a potential for delta function contributions at positions of vertex operators.

In fact, if an unintegrated  vertex operator $\V$ is inserted at a point $p\in\Sigma$, we want to constrain the gauge parameter $v$ by
\begin{equation}\label{metto}v^z(p)=0,\end{equation}
since in working with unintegrated vertex operators, we do not regard diffeomorphisms that change $p$ as symmetries.
But we do not want to impose any further conditions on $v^z$.  Given this,
the condition
that $\delta_v w_h$ does not receive a delta function contribution at $p$ is that the $b\cdot \V$ operator product has at most a simple pole at $p$, or in other words that $b_n\V=0$, $n\geq 0$.  

As an example of a consequence of this fact, we consider physical states of the bosonic string.  (For brevity we consider open
strings or a chiral sector of closed strings.)  Every physical state of the bosonic string can be represented by a vertex operator
$\V=cV$, where $V$ is a matter primary of dimension 1.  These are conformal vertex operators, and if we use them, we can compute
scattering amplitudes in a completely conformally-invariant fashion.  The operator $\V'=c\partial c V$ is also a $\sQ$-invariant
primary of dimension 0, just like $\V$, but it is not a conformal vertex operator as it is not annihilated by $b_0$.  For $\delta_v w_h$
to vanish in the presence of an insertion of $\V'$, we would require
\begin{equation}\label{etto}v^z(p)=\partial_z v^z(p)=0. \end{equation}
In other words, the diffeomorphism generated by $v^z$ would have to act trivially not only at the point $p$ but also on the tangent
space to that point in $\Sigma$.  Equivalently, to compute  an amplitude with an insertion of $\V'$ at $p$, we would
have to be given (up to a constant independent of all moduli\footnote{\label{besure}  This constant comes in because
eqn. (\ref{etto}) ensures that we know how to keep a local parameter fixed when we change the moduli, but does not
determine an overall normalization of the local parameter.})
a local parameter $z$ vanishing at $p$, modulo terms of order $z^2$.  A local parameter $z$ that is defined modulo $z^2$ is what
we will call a first-order local parameter.

We do not usually carefully consider the consequences of inserting an operator such as $\V'$ in a scattering amplitude, because there is a more trivial
reason that this will not work.  We will express the following reasoning for bosonic closed strings.  A physical state of bosonic
closed strings at non-zero spacetime momentum can be represented by the conformal vertex operator $\V=\t c c V$, where $V$ is a matter
primary of dimension $(1,1)$.  $\V$ has ghost number 2 (holomorphic and antiholomorphic ghost numbers $(1,1)$).   A correlation
function $\langle \V_1\dots\V_n\rangle$, with all $\V_i$ having ghost number 2, and with the appropriate
antighost insertions so that the correlation function is not trivially zero, leads to 
 a form of top degree on moduli space,
which can be integrated to get a number.  If one of the vertex operators has ghost number greater than 2, for example 
$\V'=\t c c(\partial c+\t\partial\t c) V$, then to get a top-form on moduli space, 
we must  compensate by taking one vertex operator to have ghost number less than 2.  In a scattering
amplitude, we cannot do this, since at nonzero momentum, the
bosonic string has no physical states of ghost number less than 2.

However, when we calculate not a scattering amplitude but a boundary contribution to a Ward identity, one of the
vertex operators that we insert is a symmetry generator at zero momentum -- for example, the supercurrent $\S$.
The ghost number of such a symmetry generator is less by 1 than that of the usual physical state vertex
operators.  So if -- as the reasoning in section \ref{where} suggests -- we are going to get a number by integrating
over $\MM_{\ell}'$ a correlation function with an insertion of $\S$, one operator in this correlation function
will necessarily have a ghost number greater than the standard value.  This operator will be a superstring analog of
$\V'=\t c c (\partial c +\t\partial \t c)V$ -- a $\sQ$-invariant superconformal primary of dimension 0, but not a 
superconformal vertex
operator.  So we will have to understand what role such an operator can play in a superconformally-invariant formalism.

\begin{figure}
 \begin{center}
   \includegraphics[width=2.5in]{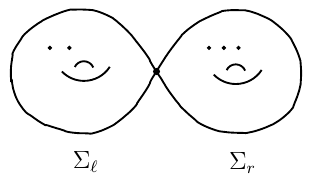}
 \end{center}
\caption{\small  This separating degeneration contributes a pole to the indicated scattering amplitude when the string state flowing between
$\Sigma_\ell$ and $\Sigma_r$ is on-shell.}
 \label{simpler}
\end{figure}
\subsubsection{The Boundary Formula}\label{expole}
In this section, we continue to use the language of the bosonic string.
Before explaining the general formula for the boundary contribution in a Ward identity, let
us first recall a simpler problem of  understanding the pole that arises in a scattering amplitude
when an on-shell string state flows between 
the two branches $\Sigma_\ell$ and $\Sigma_r$ of a separating degeneration (fig. \ref{simpler}).
General considerations of conformal field theory tell us that the effect of a 
string state propagating between $\Sigma_\ell$ and $\Sigma_r$
can be expressed via the insertion of a vertex operator $\O$ on $\Sigma_\ell$ and 
a conjugate vertex operator $\h \O$ on
$\SIgma_r$.  In conformal field theory on a fixed string worldsheet, $\O$ and $\h\O$ would be 
conjugate in the sense of having
a nonzero two-point function on a two-sphere $S^2$; a possible example would be 
$\O=\t c c\partial c V$, $\h\O=\t c\t\partial \t c c \h V$, where 
$V$ and $\h V$ are conjugate operators in the matter system.  However,  to extract 
the pole in a string scattering amplitude,
we have to integrate over the gluing parameters\footnote{The considerations in this section apply
for a variety of string theories, so we use generic names $\t q, \,q$ for gluing parameters.  When we specialize
to our problem involving the heterotic string on a Calabi-Yau manifold, the gluing parameters will be $\t q$ and $q_\Ra$.}  $\t q, \, q$ (the form 
which is integrated over these parameters is written below in eqn. (\ref{minco})).  This integration is 
associated with antighost insertions $b_0$ and $\t b_0$ 
in the narrow neck between $\Sigma_\ell$ and $\Sigma_r$.  
These insertions, which are constructed
as in eqn. (\ref{microb}),
reduce the total ghost number of $\O$ and $\h\O$ by 2 and ensure that they 
are annihilated by $b_0$ and $\t b_0$.  As a result,
it is possible for both of these operators to be conformal vertex operators: $\O=\t c c V$ and $\h\O=\t c c \h V$.  

Not only is this possible, but the residue of a pole in a scattering amplitude can be 
computed entirely from contributions of this kind.  Indeed, if we use conformal vertex operators to compute a 
scattering amplitude, then the amplitude is
determined by integrating a completely natural measure on moduli space.  The residue of any pole must
be equally natural, and this means that it must be given by insertions of conformal vertex 
operators $\O$, $\h\O$ on the two
branches.  (Moreover,  the residue of the pole can be computed entirely from a subset of conformal vertex operators
that give a basis for the BRST cohomology;
the contributions of other
operators to the residue  cancel in pairs.)

\begin{figure}
 \begin{center}
   \includegraphics[width=2.5in]{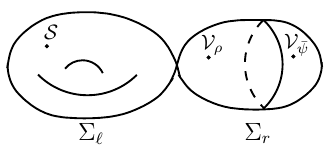}
 \end{center}
\caption{\small A typical situation in which one wishes to extract a supersymmetry-violating boundary contribution.}
 \label{bunderwood}
\end{figure}

Now let us see how this changes if instead of a pole in a scattering amplitude, we are 
trying to compute a boundary contribution to a Ward
identity (fig. \ref{bunderwood}).  
In this case, the path integral on $\SIgma_r$ is an ordinary scattering amplitude 
with insertion of $\h\O$, so in a conformally
invariant formalism, $\h\O$ will again be a conformal vertex operator.  But the path 
integral on $\Sigma_\ell$ is something less
familiar, since one of the operator insertions is a supersymmetry generator $\S$ rather 
than the vertex operator of a physical
state.  It turns out that $\O$ will not be a conformal vertex operator.

In fact, to compute the boundary contribution, we want to integrate over the argument of the gluing parameter $q$,
but not over its modulus.  The effect of this is that we still have in the narrow neck an insertion of $b_0-\t b_0$,
so we can assume that the vertex operators $\O$ and $\h\O$ are annihilated by $b_0-\t b_0$.  But we no longer
have an insertion of $b_0+\t b_0$, so we cannot assume that $\O$ and $\h\O$ are annihilated by $b_0+\t b_0$.
On the contrary, if one of them is annihilated by $b_0+\t b_0$, then the other is proportional to 
$\partial c+\t\partial \t c$ and is definitely not annihilated by $b_0+\t b_0$.  

As already explained, the boundary contribution that we want comes entirely from the case that the operator inserted on $\Sigma_r$
is a conformal vertex operator $\h\O=\t c c \h V$ (or the superstring analog of this). The operator
inserted on $\Sigma_\ell$ is then not the conjugate  conformal vertex operator $\O=\t c c V$, but
is $\O'=-(\partial c+\t\partial \t c)\O$.  It therefore seems that the
boundary contribution we want will come from the insertion 
\begin{equation}\label{mezzox}  -(\partial c+\t\partial \t c)\O\cdot \h\O \end{equation}
on the two sides.  Thus in fig. \ref{bunderwood}, the path integral on $\SIgma_r$ will be an ordinary scattering amplitude with insertion of $\h\O$,
and the one on $\Sigma_\ell$  will involve a two-point function 
$\langle \S \cdot \O'\rangle_{\Sigma_\ell}$.  The ghost numbers are such that with the usual antighost
insertions, this correlation function could give a top-form on moduli space (as explained at the end of section \ref{cvo},
$\S$ has ghost number 1 less than the usual value, and $\O'$ has ghost number 1 greater).  But there is a fundamental
problem: the operator $\O'$ is not a conformal vertex operator, and therefore the two-point function
$\langle \S \cdot \O'\rangle_{\Sigma_\ell}$ cannot be the full answer to any question in a conformally-invariant formalism.

The remedy for this was explained in section 7.7 of \cite{Revisited}.  We must study the behavior near $q=\t q=0$ of
form $F_{\S\V_1\dots\V_n}$ that
appears in the supersymmetric Ward identity (\ref{reffo}). 
The most singular contribution to this form  near $q=\t q=0$ is associated to an insertion of
\begin{equation}\label{minco} \O\cdot \frac{-i \d\t q\, \d q}{\t q q}\cdot \h O.\end{equation}
(What we call $-i \d\t q\,\d q$ is usually written $\d^2q$.)  If we were computing a scattering amplitude, we would have to integrate
over $q$ and $\t q$, and since we have taken $\O$ and $\h O$ to be on-shell conformal vertex operators, the integration would diverge
logarithmically near $q=\t q=0$.  This would be the usual singularity associated to an on-shell physical state.  Actually, we are computing 
a boundary contribution in a Ward identity.  This means that we want to integrate over $\mathrm{Arg}\,q$, but not over $|q|$.  We can
factor
\begin{equation}\label{xobbo}\frac{-i \d\t q\, \d q}{\t q q}=\frac{1}{2i}\left(\frac{\d q}{q}-\frac{\d\t q}{\t q}\right) \frac{\d(\t q q)}{\t q q }. \end{equation}
We integrate over $\mathrm{Arg}\,q$ with the aid of the 1-form $(1/2i)(\d q/q-\d{\t q}/\t q)$; this gives a factor of $2\pi$. We suppress
this factor in what follows (the same factor coming from integration over $\mathrm{Arg}\,q$ has been omitted in  (\ref{mezzox})).
After integrating out $\mathrm{Arg}\,q$, we are left with a singular contribution 
\begin{equation}\label{linco}\O \cdot \frac{\d (\t q q)}{\t q q}\cdot \h\O. \end{equation}
To evaluate the contribution of a given degeneration in the fundamental supersymmetric Ward identity (\ref{reffo}), 
we are not interested in integrating over the modulus $\t q q$ of $q$, but rather, roughly speaking, in setting it to a constant 
to define the relevant component of $\partial\MM_{g,n+1}$.  
Naively speaking, since $\t q q$ is supposed to be ``constant'' along $\partial\MM_{g,n+1}$, $\d(\t q q)$ vanishes when restricted to
$\partial\MM_{g,n+1}$, and therefore the term (\ref{linco}) does not contribute.  The trouble with this reasoning is that there is not
a conformally-invariant notion of setting $\t q q$ to a constant.    What it really means to set $\t q q$ to a ``constant'' is that we define
it to be a function of the other moduli, after which $\d(\t q q)/\t q q$ becomes a 1-form on the moduli space\footnote{\label{yrf} We can keep
$\SIgma_r$ fixed in the discussion, so we do not have to think of $\d(\t q q)/\t q q$ as a 1-form on $\MM_r$.  More fundamentally,
the insertion on $\Sigma_r$ of the operator $\h\O$ that appears
in eqn. (\ref{linco}) already leads to a top-form on $\MM_r$, so the part of $\d(\t q q)/\t q q$
that is a 1-form on $\MM_r$ does not contribute.}  $\MM_\ell'$ that parametrizes
$\SIgma_\ell$.  
There is no conformally-invariant way to get rid of this term; rather,  the conformally-invariant extension of the naive formula (\ref{mezzox}) is
obtained by including it:
\begin{equation}\label{ggg}\left(-\O'+\frac{\d(\t q q)}{\t q q} \O\right)\cdot\h\O=\left(-\partial c-\t\partial\t c +\frac{\d(\t q q)}{\t q q}\right) \O\cdot\h\O .\end{equation}

Here is a partial explanation of the conformal invariance of this combined formula.  Let us recall from eqn. (\ref{pelmo})
that to define what we mean by $q$, we need a first-order  holomorphic local parameter at the point  $p\in \Sigma_\ell$ at which
the gluing occurs (we also need such a parameter on $\Sigma_r$, but the dependence on this is irrelevant for the same reason as in
footnote \ref{yrf}).  To define the product $\t q q$, we need a product of holomorphic and antiholomorphic first-order local parameters,
and to define the 1-form $\d(\t q q)/\t q q$, we need such a product up to a multiplicative constant independent of all moduli.  
This is precisely the same data needed to define the insertion of $(\partial c+\t\partial \t c)\O$.
Indeed,   as the operator $\O'=(\partial c+\t\partial\t c)\O$ is annihilated
by $b_0-\t b_0$, and by $b_n,\t b_n$, $n>0$, the analog of eqn. (\ref{etto}) for insertion of $\O'$ at $p$ is 
$v^z(p)=v^{\t z}(p)=\partial_z v^z(p)+\partial_{\t z}v^{\t z}(p)=0$.  These conditions mean that $v$ leaves fixed the product of
holomorphic and antiholomorphic first-order local parameters at $p$; and accordingly, up to a constant independent of all moduli,
there is a well-defined product of local parameters at $p$.  
    So the $\O'$ and $\d(\t q q)/\t q q \cdot \O$ terms in (\ref{ggg}) violate conformal symmetry in the same way.
Hopefully, this makes it plausible that their sum is conformally-invariant.

We can understand a little more as follows. Split $\O'=(\partial c+\t\partial\t c)\O$ as the sum of two contributions
$\partial c \O$ and $\t\partial\t c\O$. The former insertion requires a holomorphic first-order local   parameter and the latter one requires
an antiholomorphic one.  Similarly, split  $\d(\t q q )/\t q q$ as the sum of $\d q/q$ and $\d\t q/\t q$,
where again the first term depends on a holomorphic first-order parameter and the second on an antiholomorphic one.
Let us just look at the terms in (\ref{ggg}) that depend on the holomorphic local parameter, namely
\begin{equation}\label{cormaly} \left(-\partial c+\frac{\d q}{q}\right)\O'. \end{equation}
As explained at the end of section \ref{condcon},  $q$ is not a complex-valued function
but a linear function on a holomorphic line bundle over $\MM_\ell'$ -- the normal bundle $\frak N$.  Thus $q$ is 
a section of the dual $\frak N^\vee$ of the normal bundle (also called the conormal bundle).  In differential geometry in
general, if one is given a section $q$ of a line bundle $\frak N^\vee$, then to define a 1-form $\d q/q$, one needs
a connection on $\frak N^\vee$.  In the present context, we do not have such a connection (until we pick local
parameters), so $\d q/q$ cannot
be defined in a conformally-invariant way.

However, using the complex structure of\footnote{{\it A priori}, this
 computation should really be done not on $\MM_{\ell}'$,
but on the corresponding heterotic string integration cycle $\varGamma_\ell'$, constructed along
lines described in section \ref{gluehol}.  In the present example, as explained in section \ref{cohoform}, 
$\MM_{\ell}'$ is naturally
split and there is no need to make this distinction.}
 $\MM_\ell'$, we can decompose the exterior derivative on $\MM_\ell'$ as a sum of
pieces of type $(1,0)$ and $(0,1)$:  $\d=\partial+\t\partial$.  Though it does not have a connection, $\frak N^\vee$ does
have a holomorphic structure that is perfectly natural and conformally-invariant.  This means that $\t\partial q/q$,
which is the 
$(0,1)$ part of $\d q/q$, is well-defined, independent of any local parameters.  It will turn out that  the 
Goldstone fermion contribution to the Ward identity comes entirely from this.
The $(1,0)$ part of $\d q/q$ is not conformally-invariant by itself, though the combination 
$(-\partial c+ \partial q/q)\O'$ that appears in (\ref{cormaly}) is conformally-invariant.

\subsubsection{The Vertex Operators}\label{operators}

Let us now specialize to the heterotic string with an anomalous $U(1)$ gauge symmetry.  What vertex operators
$\O$ and $\h\O$ shall we use?  We want to test the hypothesis that the gaugino $\zeta$
becomes a Goldstone fermion
at 1-loop order.  So $\h\O$ will be a gaugino vertex operator at zero spacetime momentum.  
For a gaugino with positive chirality in $\R^4$,
the vertex operator at zero momentum was described in eqn. (\ref{numbo}):
\begin{equation}\label{golp}V_\beta^\zeta= J_\ell \cdot e^{-\tphi/2} \Sigma_{\beta,+}. \end{equation}
This is the integrated vertex operator without the usual $\t c c$ factor.
The conjugate operator in the conformal field theory of the matter fields and the $\beta\gamma$ ghosts
(but without the $bc$ ghosts)   is
\begin{equation}\label{golvo} W^{\zeta}_\beta=J_\ell \cdot e^{-3\tphi/2} \Sigma_{\beta,-}.  \end{equation}
These operators are conjugate in the sense that in genus 0, 
they obey $\langle W^{\zeta} _\beta V^\zeta_\gamma \rangle
\sim \epsilon_{\beta\gamma}$, where $\epsilon_{\beta\gamma}$ is the Lorentz-invariant antisymmetric tensor.
In conventional language, the vertex operator $W^{\zeta}_\beta$ 
is a zero momentum gaugino vertex operator at picture
number $-3/2$ (with negative chirality; see the explanation of picture-changing in section \ref{ferglu}).

Let us now include the Virasoro ghosts and put these operators in the framework of section \ref{expole}. 
To compute the matrix element of
 a supercurrent $\S_\alpha$ to create a gaugino of positive chirality with spinor index $\beta$,
 we must take $\h\O=\t c c V_\beta^\zeta$, and then in eqn. (\ref{ggg}), we must take
 $\O=\t c c W^{\zeta;\,\beta}=\epsilon^{\beta\gamma}\t c c W^{\zeta}_\gamma$.  
 Accordingly, we finally get a formula for the matrix element for $\S_\alpha$ to create the gaugino of polarization
 $\beta$ from the vacuum:
 \begin{equation}\label{mellowbook} \I_\alpha^\beta= \int_{\MM_\ell'}\biggl\langle \S_\alpha \cdot \left(-(\partial c+\t\partial
 \t c)+\frac{\d (\t q q_\Ra)}{\t q q_\Ra }\right) \t c c W^{\zeta;\,\beta}\biggr\rangle. \end{equation}
 (By Lorentz invariance, $\I_\alpha^\beta$ is  a multiple of $\delta_\alpha^\beta$.)
 In writing this formula, since we are now considering specifically the case of a Ramond sector degeneration of
 the heterotic string, we denote the holomorphic gluing parameter as $q_\Ra$ rather than $q$.
We aim to convert this formula to something more concrete.
 
\subsubsection{The Role Of The Fermionic Gluing Parameter}\label{ferglu} 

We come next to a crucial part of this problem.  $\MM_{\ell'}$ is fibered over $\MM_\ell$
with a fiber of dimension $0|1$ that is parametrized by the fermionic gluing parameter $\alpha$.   One approach
to integrating over $\MM_{\ell'}$ is to first integrate over $\alpha$, so as to reduce to an integral over $\MM_\ell$.
By thinking about what will happen if we do this, we can learn something essential.

Integration over $\alpha$ kills what one might call the obvious part of  (\ref{mellowbook})
-- the part that involves the insertion of the operator $\O=(\partial c+\t\partial \t c)\t c c W^\zeta$.  The main reason
is that this operator describes a gaugino at zero momentum in spacetime.  

If we were evaluating the gaugino contribution to a pole
in a scattering amplitude rather than a boundary contribution in a Ward identity, the gaugino would
carry a nonzero (but almost lightlike) spacetime momentum $k$.  In this case, as remarked in relation to
eqn. (\ref{dobzo}), integration over the fermionic gluing parameter  converts a boson propagator $1/L_0$ into a Dirac-like propagator $G_0/L_0$.  For massless
fermions, $G_0$ can be replaced by the Dirac operator $\Gamma\cdot k$.  This vanishes for an on-shell
gaugino, but the propagator also has a factor $1/L_0$ that comes from integration over the magnitude $|q|$ of the
bosonic gluing parameter $q$.  Since $G_0/L_0=1/G_0\sim 1/\Gamma\cdot k$, 
the net effect of this is that an
on-shell gaugino propagating between $\Sigma_\ell$ and $\Sigma_r$ produces a Dirac-like pole.

In computing a boundary contribution to a Ward identity, there is no integral over $|q|$, 
so there is no factor $1/L_0$.  Any term with a factor of 
$G_0$ in the numerator will vanish, since $G_0\sim \Gamma\cdot k$ vanishes for massless states at $k=0$.  (Essentially
this observation has been made in the study of supersymmetric Ward identities in the early literature; see eqn. (6.42) of \cite{HVerlinde}.)

Another perspective on what we have just explained is as follows.  As described in \cite{Surfaces}, section 4.3,
a Ramond vertex operator of picture number $-3/2$ is inserted at a point $f$ on a Ramond divisor $\F$,
while a Ramond vertex operator of picture number $-1/2$ is associated to the whole divisor.  The picture-changing
operation from picture number $-3/2$ to picture number $-1/2$ is integration over the point $f\in\F$.  In our
context, we are gluing $\Sigma_\ell$ to $\Sigma_r$ by gluing a Ramond divisor $\F_\ell\subset \Sigma_\ell$
to a Ramond divisor $\F_r\subset\Sigma_r$.  The fermionic gluing parameter is precisely the choice of a point
in $\F_\ell$ that is glued to a given point in $\F_r$, and integration over this gluing parameter is the picture-changing
operation.   This integration actually produces in eqn. (\ref{dobzo}) not just the factor $G_0$ that we discussed above
but the combination $\delta(\beta_0)G_0$.  This combination is the picture-changing operation that at momentum
$k$ maps
a picture number $-3/2$ gaugino vertex operator $W^{\zeta;\,\beta}$ to 
$(\Gamma\cdot k)^{\beta\dot\beta}V^\zeta_{\dot\beta}$, where 
\begin{equation}\label{delft}\V^\zeta_{\dot\beta}=J_\ell e^{-\tphi/2}\Sigma_{\dot\beta,-} \exp(ik\cdot X)\end{equation}
 is a gaugino vertex operator of picture number $-1/2$ and negative chirality.  
 In our problem, the momentum $k$ vanishes, so $\Gamma\cdot k=0$ and picture-changing simply annihilates
$W^{\zeta;\,\beta}$. Hence integration over the fermionic gluing parameter annihilates the operator
$\O=(\partial c+\t\partial \t c)\t c c W^\zeta$.

In view of what will be explained shortly, one should wonder if  $\O$
has a hidden dependence on the fermionic gluing parameter because it is not a conformal vertex operator
and depends on choices of local parameters at its insertion point.  What prevents this is as follows.  One contribution
in $\O$, namely $\partial c \cdot\t c c W^\zeta$, can be dropped for the trivial reason that its holomorphic and
antiholomorphic ghost numbers imply the vanishing of its contribution in (\ref{mellowbook}).  (With $\S=c S$,
this contribution has ghost quantum numbers 
$c^3 \t c$, while a nonzero contribution would come from $c^2\t c^2$.)  The
other contribution, namely $\t\partial\t c\cdot \t c c W^\zeta$, depends on an antiholomorphic local parameter at $p$,
not a holomorphic one.  Roughly speaking, this operator has no hidden dependence on the fermionic
gluing parameter  because the antiholomorphic local parameter can be defined to never depend
on the fermionic moduli, which are holomorphic.  More accurately, this is true if we follow the procedure
explained in section \ref{cohoform} (see especially footnote \ref{alto}), in which a natural projection of $\MM_\ell'$ is used in evaluating the boundary
contribution to the Ward identity.   

So we can drop the more obvious term in (\ref{mellowbook}), and we are left with the slightly exotic-looking 
term involving
$\d(\t q q_\Ra)/\t q q_\Ra$:
\begin{equation}\label{prto} \I_\alpha^\beta=-\int_{\MM_\ell'}\frac{\d(\t q q_\Ra)}{\t q q_\Ra}\biggl\langle \S_\alpha
\cdot \t c c W^{\zeta;\,\beta}\biggr\rangle
.\end{equation}
Just as before, integration over the fermionic gluing parameter will annihilate the operator $ \t c c W^{\zeta;\,\beta}$,
so to get a nonzero result, the fermionic gluing parameter will have to somehow be hidden in the expression
$\d(\t q q_\Ra)/\t q q_\Ra$. 
We can actually be slightly more precise.  With $\S_\alpha =c S_\alpha$, the ghost quantum numbers
in  $\bigl\langle \S_\alpha
\cdot \t c c W^{\zeta;\,\beta}\bigr\rangle$ are $c^2\t c$; this
is missing one $\t c$ factor relative to a correlation function that would
give a top-form on moduli space.  Hence it produces a form of codimension $(0,1)$.  This means
that only the part of $\d(\t q q_\Ra)/\t q q_\Ra$ that is of type $(0,1)$ is relevant.  We can thus replace (\ref{prto}) with
\begin{equation}\label{parto} \I_\alpha^\beta=-\int_{\MM_\ell'}\frac{\t\partial(\t q q_\Ra)}{\t q q_\Ra}\biggl\langle \S_\alpha
\cdot \t c c W^{\zeta;\,\beta}\biggr\rangle, \end{equation}
and we will have to find the fermionic gluing parameter in the $(0,1)$-form $\t\partial(\t q q_\Ra)/\t q q_\Ra$. 

We further have
\begin{equation}\label{minzo} \frac{\t\partial(\t q q_\Ra)}{\t q q _\Ra}
=\frac{\t\partial q_\Ra}{ q _\Ra}+\frac{\t\partial\t q }{\t q}.\end{equation}
As we have explained at the end of  section \ref{expole}, $\t\partial q_\Ra/q_\Ra$ is completely
well-defined and conformally-invariant, for any section $q_\Ra$ of the holomorphic conormal bundle $\frak N^\vee$,
simply because $\frak N^\vee$ is a holomorphic line bundle.  By contrast, $\t q$ is a section of the antiholomorphic
conormal
bundle $\t{\frak N}^\vee$, so $\t\partial\t q/\t q$ cannot be defined in a conformally-invariant way.  But if we follow
the procedure explained in section \ref{cohoform}, using a hermitian metric that does not depend on the odd
moduli to define a connection on $\t\fN^\vee$, then this term does not contibute.  So with this sort of procedure
(which is also the procedure that ensures vanishing of the $\t\partial\t c \cdot \t c c W^\zeta$ contribution to $\I$), (\ref{parto}) really only 
receives a contribution
 $\partial q_\Ra/q_\Ra$.  Still,  it is most convenient to leave the formula for $\I$ in the form given in (\ref{parto}),
without dropping the $\t\partial \t q/\t q$ term.   This will slightly shorten the explanation in section \ref{cohoform}.

We will see next what feature of supergeometry can force  $\t q q_\Ra$ to 
have an unavoidable dependence on the fermionic
gluing parameter.  The non-zero value of the integral $\I$ comes entirely from this, since as we have
seen, integration over the fermionic gluing parameter annihilates everything else.

\subsubsection{The Holomorphic Projection}\label{holproj}

We defined $\MM_\ell'$ as a fiber bundle, with fibers of dimension $0|1$ parametrized by the fermionic gluing
parameter, over $\MM_\ell$.  In particular, there is a natural projection $\MM_\ell'\to \MM_\ell$.  In turn, in our main
example, $\MM_\ell$ is the moduli space of genus 1 super Riemann
surfaces with no NS punctures and 2 Ramond punctures, which we might call $\MM_{1,0,2}$.  The dimension
of $\MM_\ell$ is $2|1$.   (We always intend the Deligne-Mumford compactifications of  these moduli
spaces, though we do not indicate this in the notation.)

Now we require a slight extension of what was explained in section \ref{gluehol}.
A complex supermanifold $M$, say of dimension $p|q$, has a reduced space $M_\red$ of dimension $p|0$, obtained
by reducing modulo the odd variables.
If $M$ has local coordinates $z_1,\dots,z_p|\theta_1\dots\theta_q$, then $M_\red$ has local
coordinates $z_1,\dots,z_p$.
There is always a natural embedding $i:M_\red\to M$, which in local coordinates maps $z_1,\dots ,z_p$
to $z_1,\dots,z_p|0,\dots,0$.  $M$ is said to be holomorphically projected if there is also a holomorphic
projection $\pi:M\to M_\red$, obeying $\pi i=1$; the last condition means that $\pi=1$ when restricted
to $M_\red$. 
For example, if the odd dimension of $M$ is 1, there is always a unique projection $M\to M_\red$.    In local coordinates,
this map takes $z_1,\dots,z_p|\theta$ to $z_1,\dots,z_p$.  
With two or more odd coordinates, a holomorphic projection is not unique locally  (for example, in dimension $1|2$, 
$z|\theta_1,\theta_2$ could be mapped to $z$ or to $z+\theta_1\theta_2$), and
may not exist globally.    An example important for superstring perturbation theory is that
in general the moduli space of super Riemann surfaces
is not holomorphically projected \cite{DW}. Holomorphic projections from supermoduli space to ordinary moduli space,
when they exist, are a powerful tool in superstring perturbation theory, exploited notably in \cite{DPh}.

Returning to $\MM_\ell$, since its odd dimension is 1,  it has
a unique projection to its reduced space $\MM_{\ell,\red}$.
Composing the natural projection $\MM_\ell'\to \MM_\ell$ with the unique projection $\MM_{\ell}\to \MM_{\ell,\red}$,
we get a natural projection $\pi:\MM_\ell'\to \MM_{\ell,\red}$.   Actually, $\MM'_\ell$ and $\MM_\ell$ have the same
reduced space (since the fibers of $\MM'_\ell\to\MM_\ell$ are purely odd), so $\pi$ is a holomorphic projection of
$\MM_\ell'$ to its reduced space $\MM'_{\ell,\red}$.

Thus, the subtleties that were discussed in section \ref{gluehol} have no analog for $\MM_{\ell}'$, even though
its odd dimension is 2, which in general is enough to produce such subtleties.  Local holomorphic coordinates
$m_\alpha$ on $\MM_{\ell,\red}$ can be pulled back in a natural way to bosonic local  coordinates $\pi^*(m_\alpha)$ on $\MM_{\ell}'$. Integration over $\MM_{\ell}'$ can be performed in a natural fashion by first integrating over the fibers of
$\pi:\MM_{\ell}'\to \MM_{\ell,\red}$.

The heterotic string on $\Sigma_\ell$ has antiholomorphic as well as holomorphic moduli.  The antiholomorphic moduli space 
that is ``seen'' by the antiholomorphic variables of the heterotic string is simply
 the complex conjugate of $\MM_{\ell,\red}$.    So once one takes the holomorphic even 
moduli $m_\alpha$ of $\SIgma_\ell$ to be pullbacks from $\MM_{\ell,\red}$, 
one can take the antiholomorphic moduli $\t m_\alpha$ of $\Sigma_{\ell}$ to be simply their complex conjugates
\begin{equation}\label{ozmo} \bar{\t m_\alpha}=m_\alpha,\end{equation}
without the nilpotent terms of eqn. (\ref{mezmo}).  Those terms are inescapable in the absence of a holomorphic
projection, but with such a projection, there is no need for them. 
Accordingly, there is no need to distinguish between $\MM_\ell'$ and a corresponding heterotic string 
integration cycle $\varGamma'_\ell$, and the integral (\ref{parto}) for the Goldstone fermion contribution to the Ward
identity really is properly understood as an integral over $\MM_\ell'$.    

The reduced space $\MM'_{\ell,\red}$ parametrizes an ordinary Riemann surface $\Sigma_0$ of genus 1 with
some additional data (two punctures and a generalized spin structure).  Forgetting the additional data,
we get a holomorphic map from $\MM'_{\ell,\red}$ to $\M_1$, the moduli space 
of ordinary Riemann surfaces of genus 1.  Composing this with $\pi:\MM'_\ell\to \MM'_{\ell,\red}$, we get
a holomorphic fibration of $\MM'_{\ell,\red}$ over $\M_1$.  Let
$\MM'_{\ell,\tau}$ be the fiber of this map lying over the point in $\M_1$ that corresponds to an elliptic curve with
modular parameter $\tau$:
\begin{equation}\label{okely}  \begin{matrix} \MM'_{\ell,\tau}& \to & \MM'_{\ell}\cr
   & & \downarrow \cr
      & & \M_1.\end{matrix}                \end{equation}
  We can integrate over $\MM_{\ell}'$ by  integrating first over $\MM_{\ell,\tau}'$ and
then  over $\tau$.  This is analogous to the procedure followed in section \ref{mass}, and just
as there, the interesting phenomena occur in the integral over $\MM_{\ell,\tau}'$, at fixed $\tau$.

\subsubsection{The Cohomological Formula}\label{cohoform}

In proceeding, it helps slightly to consider the basic case (fig. \ref{bunderwood}) of the supersymmetric
Ward identity  that governs a 1-loop supersymmetry-violating mass shift.  
In this case, $\Sigma$ is a super Riemann surface of genus 1 with 1 NS puncture and 2 Ramond punctures.
$\Sigma$ is parametrized by a  moduli space $\MM_{1,1,2}$ of dimension $3|2$.  $\MM_\ell'$ is a divisor ``at infinity''
in $\MM_{1,1,2}$ describing the splitting of $\Sigma$ to components $\Sigma_\ell$ and $\Sigma_r$ of genera 1 and 0, 
respectively.   $\MM_r$ is a point and plays no role.  In the case of a more general supersymmetric Ward identity,
the discussion would proceed in the same way except that everything would be fibered over $\MM_r$, which would play
no essential role.

To understand the $(0,1)$-form 
$\lambda=(\t q q_\Ra)^{-1}\t\partial (\t q q_\Ra)$ that appears in (\ref{parto}), we need to understand the 
holomorphic and antiholomorphic
normal bundles $\fN$ and $\t{\fN}$ to $\MM_\ell'$ in $\MM_{1,1,2}$.

First of all, let $\fN_0$ be the restriction of $\fN$ to $\MM'_{\ell,\red}\subset \MM'_\ell$.  Using the holomorphic
projection $\pi:\MM'_\ell\to \MM'_{\ell,\red}$, we can pull back $\fN_0\to \MM'_{\ell,\red}$ to $\pi^*(\fN_0)\to \MM'_\ell$.
$\fN$ is not necessarily equivalent to $\pi^*(\fN_0)$.  All that we know {\it a priori} is that they coincide when restricted
to $\MM'_{\ell,\red}$, so if we define another
line bundle $\fN_1\to\MM'_{\ell}$ by
\begin{equation}\label{zembo} \fN=\pi^*(\fN_0)\otimes \fN_1,\end{equation}
then $\fN_1$ is canonically trivial when restricted to $\MM'_{\ell,\red}$.  Since the odd directions in a supermanifold are
infinitesimal, a line bundle over any complex supermanifold $M$ that is trivial when restricted to $M_\red$
is always topologically trivial, but it may be holomorphically non-trivial.  In the present example, it turns out
that $\fN_1$ is holomorphically non-trivial.

Once we restrict to $\MM'_{\ell,\red}$, we can consistently replace a super Riemann surface by its 
reduced space, and the 
Ramond sector gluing law (\ref{medzo}) reduces to the bosonic gluing law $xy=q_\Ra$.  
$\MM'_{\ell,\red}$ is  a divisor in 
 $\MM_{1,1,2,\red}$ (that is, in the reduced space of
$\MM_{1,1,2}$) and $\fN_0$ is its normal bundle.
If we reverse the complex structures of all objects mentioned in the last sentence, 
$\MM_{1,1,2,\red}$ becomes the moduli space appropriate for antiholomorphic degrees of freedom of the heterotic
string; it contains the divisor $\MM'_{\ell,\red}$ with normal bundle $\fN_0$, both with their complex structures reversed.
So the antiholomorphic normal bundle $\t\fN$ to the divisor $\MM'_\ell\subset \MM_{1,1,2}$  is just $\pi^*(\fN_0)$ with its
complex structure reversed.  We write this as $\pi^*(\bar\fN_0)$.    We combine this  result
with (\ref{zembo}):
\begin{equation}\label{wembo}\t\fN\otimes \fN=\pi^*(\bar \fN_0\otimes \fN_0)\otimes \fN_1. \end{equation}

Taking the duals gives an equivalent formula for the tensor product of conormal bundles:
\begin{equation}\label{rembo}\t\fN^\vee\otimes \fN^\vee=\pi^*(\bar\fN_0^\vee\otimes \fN_0^\vee)
\otimes \fN_1^\vee. \end{equation}
For any line bundle $\L$, the tensor product $\bar\L\otimes \L$ is always trivial: it can be trivialized by the choice
of a hermitian metric on $\L$.  (We call a section of $\bar\L\otimes \L$  positive if it is associated to a hermitian metric on $\L$.) And we know already that $\fN_1$, and hence its dual $\fN_1^\vee$, is topologically
trivial.  So $\t\fN^\vee\otimes \fN^\vee$ is topologically trivial.  Therefore, it is possible to pick 
a smooth trivialization of this line bundle -- 
a section of
it  that is everywhere nonzero.   Such a section is what we mean by $\t q q_\Ra$ in the formula (\ref{parto}).  
Locally but not globally, one can assume that the everywhere
nonzero section of 
$\t\fN^\vee\otimes \fN^\vee$ that we call $\t q q_\Ra$ is the product of a section $\t q$ of $\t\fN^\vee$ and
a section $q_\Ra$ of $\fN^\vee$.  (As explained in section \ref{expole},
 our main formulas require only a trivialization of $\t\fN^\vee\otimes \fN^\vee$
and not separate trivializations of the two factors because the operator $\O'$ is annihilated by $b_0-\t b_0$.) 

We can take $\t q q_\Ra = TU$, where $T$ is a trivialization of $\pi^*(\bar\fN_0^\vee\otimes \fN_0^\vee)$
and $U$ is a trivialization of $\fN_1$.  Moreover, since $\pi^*(\bar\fN_0^\vee\otimes \fN_0^\vee)$ is a pullback
from $\MM'_{\ell,\red}$, we can assume that $T$ is also such a pullback.  (More specifically, we can assume that $T$ is positive, and require that $U=1$
when restricted to $\MM'_{\ell,\red}$.)    Then $T$ is independent of
the fermionic gluing parameter, and so does not contribute in the formula (\ref{parto}) for the Goldstone
fermion contribution to the Ward identity.\footnote{\label{alto} We cannot simply argue -- as we do below for $U$ -- that the choice
of $T$ does not affect the integral we are trying to evaluate; the argument does not work since $T$ trivializes a line bundle that does not have a natural
holomorphic structure.  However, defining $T$ as a pullback from $\MM'_{\ell,\red}$ ensures that
$\t q q_\Ra$ is the product of such a pullback with a section of a holomorphic line bundle (namely $\fN^\vee_1$).
This makes it possible to drop the term $(\partial c+\t\partial \t c)\O'$ in the main formula (\ref{mellowbook}), as we did in section \ref{ferglu}, by ensuring that
that operator has 
no hidden dependence on  the fermionic gluing parameter via its dependence on  local
parameters; a pullback from $\MM'_{\ell,\red}$ does not depend on the fermionic gluing parameter, and a section of a holomorphic line
bundle  affects only  $\partial c \O'$, which has the wrong holomorphic and antiholomorphic ghost  numbers to be relevant. 
In other words, taking $T$ to be a pullback from $\MM'_{\ell,\red}$ causes the contribution
of $T$ and that of $(\partial c+\t\partial\t c)\O'$ to vanish separately and justifies our claims in section \ref{ferglu}.}

So (\ref{parto}) can be expressed entirely in terms of the trivialization $U$ of the holomorphic line bundle
$\fN_1^\vee$:
\begin{equation}\label{karto} \I_\alpha^\beta=-\int_{\MM_\ell'}\frac{\t\partial_{\fN_1^\vee} U}{U}\biggl\langle \S_\alpha
\cdot \t c c W^{\zeta;\,\beta}\biggr\rangle
.\end{equation}
Here $\t\partial_{\fN_1^\vee}$ is the $\t\partial$ operator on the line bundle $\fN_1^\vee$; we usually omit
this subscript (as the line bundle is generally clear from the context) but here we include it for emphasis.

The reason that this is an advance is that a trivialization of a holomorphic line bundle has a cohomological
meaning.  Over any complex manifold or supermanifold $M$, a holomorphic line bundle $\L$ that is topologically
trivial is associated to a natural class  $\Lambda(\L)\in H^1(M,\eO)$, where $\eO$ is the sheaf of 
holomorphic functions on $M$.
Indeed, if $\L$ is topologically trivial, let $U$ be a smooth trivialization of $\L$, and consider
the $(0,1)$-form $\lambda=U^{-1}\t\partial_\L U$, where
$\t\partial_\L$ is the $\t\partial$ operator of $\L$.   The class $\Lambda(\L)$ associated to $\L$ is simply
the cohomology class of $\lambda$ in $H^1(M,\eO)$.
To show that this class does not depend on the choice of $U$, one simply observes that any other trivialization would
be $e^fU$, for some function $f$.  But $(e^fU)^{-1}\t\partial_\L(e^fU)=\lambda+\t\partial_\eO f$, where now $\t\partial_\eO$
is the $\t\partial $-operator on functions (sections of $\eO$).  By the definition of $\t\partial$-cohomology,
the class of a $(0,1)$-form $\lambda$ in $H^1(M,\eO)$ is invariant under $\lambda\to\lambda+\t\partial_\eO f$
for any function $f$.  A standard argument shows that the correspondence between $\L$ and $\Lambda(\L)$ is
a 1-1 correspondence between topologically trivial line bundles and classes in $H^1(M,\eO)$. 

So we  can rewrite our basic formula (\ref{karto}):
\begin{equation}\label{arto} \I_\alpha^\beta=-\int_{\MM_\ell'}\Lambda(\fN_1^\vee)\biggl\langle \S_\alpha
\cdot \t c c W^{\zeta;\,\beta}\biggr\rangle.\end{equation}
To understand this better, we should give a cohomological interpretation to the correlation function
$F_{\S\cdot \t c c W}=\bigl\langle \S_\alpha
\cdot \t c c W^{\zeta;\,\beta}\bigr\rangle$. 
From a holomorphic point of view, $F_{\S\cdot \t c c W}$ is a top-form.  On a complex supermanifold $M$, a top-form in the
holomorphic sense is a section of $\BBer(M)$, the Berezinian of $M$ in the holomorphic sense. 
 From an antiholomorphic point of view, 
$\MM'_\ell$ has dimension $2|0$ and $F_{\S\cdot \t c c W}$ is a form of codimension 1, and hence a $(0,1)$-form. 
Combining these facts, $F_{\S\cdot \t c c W}$ is an element of $H^1(\MM'_\ell,\BBer(\MM'_\ell))$.
  The cup product of $\Lambda(\fN_1^\vee)\in H^1(\MM'_\ell,\eO)$
and  $F_{\S\cdot \t c c W}$ gives a class in $H^2(\MM'_\ell,\BBer(\MM'_\ell))$.  Such a class is a top-form 
both holomorphically and antiholomorphically; equivalently, it is a section of $\mathrm{Ber}(\MM'_\ell)$, which is the
Berezinian of  $\MM'_\ell$ in the smooth sense, with $\MM'_\ell$ viewed
as a smooth supermanifold of dimension $4|2$.  So the product $\Lambda(\fN_1^\vee) F_{\S\cdot \t c c W}$
 can be integrated, and this integral
is what appears on the right hand side of eqn. (\ref{arto}).

\subsubsection{Overview Of The Remaining Steps}\label{teldor}

The formula  (\ref{arto}) expresses as the product of two factors the measure that must be integrated over $\MM'_\ell$ to determine
if supersymmetry is spontaneously broken by 1-loop effects.  The first factor $\Lambda(\fN_1^\vee)$ is universal, independent
of the specific string theory compactification.  It purely reflects properties of the moduli space of super Riemann surfaces.  The information
involving the choice of a specific compactification is entirely contained in the correlation function $\bigl\langle \S_\alpha
\cdot \t c c W^{\zeta;\,\beta}\bigr\rangle$
that comprises the second factor in
eqn. (\ref{arto}). 

If the universal factor  $\Lambda(\fN_1^\vee)$
were zero, 1-loop effects would never trigger the spontaneous breaking of supersymmetry, 
irrespective of the details of a specific
model.   Actually, this cohomology class does not vanish, but to show this 
one must go in somewhat greater depth with super Riemann surfaces
than has been necessary in this paper.
We defer this to elsewhere \cite{DWtwo}, and here we merely explore the consequences of a nonzero $\Lambda(\fN_1^\vee)$.

A simple observation is that $\Lambda(\fN_1^\vee)$ vanishes if restricted to $\MM'_{\ell,\red}$. This is because the line bundle $\fN_1$ is trivial
when restricted to $\MM"_{\ell,\red}$.   Concretely, we can choose $U$ so
that $U=1$ on $\MM'_{\ell,\red}$, in which case the form $\lambda=\t\partial_{\fN_1^\vee} U/U$ is identically zero when restricted to $\MM'_{\ell,\red}$.
(More generally, for any $U$, the cohomology class  of this form is zero when restricted to $\MM'_{\ell,\red}$.)
We recall now that  $\MM'_{\ell,\red}$ has precisely two odd moduli -- the fermionic
gluing parameter $\alpha$, and one more odd modulus, which we will call $\eta$.    With our choice of $U$, since $\lambda$ vanishes when $\alpha$ and $\eta$
are zero (and since it is a $(0,1)$-form  that is valued in even functions), $\lambda$ is proportional
to $\alpha\eta=\delta(\alpha)\delta(\eta)$.  

Now we return to eqn. (\ref{karto}).
The delta functions in  $\lambda$ mean that we can set $\alpha=\eta=0$ when we study the correlation function $F_{\S\cdot \t c c W}=\bigl\langle \S_\alpha
\cdot \t c c W^{\zeta;\,\beta}\bigr\rangle$.   Hence we do not have to worry about changes of variables such as $m\to m+\alpha\eta$
where $m$ is an even modulus.    Accordingly, most of the subtleties of superstring perturbation theory
become irrelevant.
 $\MM'_{\ell,\red}$ has 2 odd moduli, which  is enough to bring into play the subtleties of super Riemann surface theory;
 but they are all contained in the cohomology class $\Lambda(\fN_1^\vee)$.  

In particular, we can use the picture-changing approach of \cite{FMS}, with an important modification coming from the fact
that $\Lambda(\fN_1^\vee)$ is proportional to $\delta(\alpha)\delta(\eta)$.   As explained in
\cite{VV}, the  picture-changing operator $\YY(p)$ reflects the effect of integrating over an odd modulus
$\trho$ that represents the coefficient of a delta-function term in the gravitino field $\chi_{\t z}^\theta=\trho \delta^2(z-p)+\dots$. 
(In our application, the two odd moduli $\alpha$ and $\eta$ can both be represented in terms of such delta function gravitinos.)
The picture-changing operator is $\YY=\delta(\beta)S_{z\theta}$, 
where $S_{z\theta}$ is the worldsheet supercurrent and $\delta(\beta)$  
is usually written as $e^\tphi$ in the bosonized description of the $\beta\gamma$ system.  In 
the approach described in section 3.6 of \cite{Revisited}, the factors
of $S_{z\theta}$ and $\delta(\beta)$ come, respectively, from integration over $\trho$ and $\d\trho$:
\begin{equation}\label{tozzo}\int \D(\trho,\d\trho)  \exp\left(\d\trho \beta(p)+\trho S_{z\theta}(p)\right)=\delta(\beta(p))S_{z\theta}(p).   \end{equation}
However,  in our application, this integral multiplies the factor $\Lambda(\fN_1^\vee)$ that is explicitly proportional to $\delta(\trho)$ (where $\trho$ is
a linear combination of $\alpha$ and $\eta$), so we do not want to integrate
over $\trho$ with the help of the term $\trho S_{z\theta}(p)$ in the exponent; on the contrary, we can drop this term, because
of the delta function $\delta(\trho)$.  The integral over
$\d\trho$ still gives a factor of $\delta(\beta)$.  

The upshot is that we can use the picture-changing formalism, but we must use partial picture-changing operators $\delta(\beta)$
rather than the full picture-changing operator $\YY(p)=\delta(\beta)S_{z\theta}$.  Accordingly, the correlation function that we have
to evaluate is
\begin{equation}\label{irrox} \biggl\langle cS_\alpha(z) \delta(\beta(p))\delta(\beta(p')) \t c c W^{\zeta;\beta} (0)   \biggr\rangle,\end{equation}
with two arbitrary points $p$ and $p'$.  We have made the $c$ ghosts explicit by replacing $\S_\alpha$ with $c S_\alpha$,
but we do not indicate explicitly the corresponding antighost insertions (2 insertions of $b$ and 1 of $\t b$, since 
 the correlation
function is supposed to be a $(2,1)$-form on $\MM'_{\ell,\red}$ from a bosonic point of view).  
This correlation function is independent of $p$ and $p'$ and of the choices of $b$ and $\t b$ insertions
 if properly understood as
a $(0,1)$-form valued in   $\BBer(\MM'_\ell)$ (in particular, its dependence on $p$ and $p'$ comes entirely from the dependence on $p$ and $p'$
of the cohomology classes of the gravitino deformations $\delta^2(z-p)$ and $\delta^2(z-p')$;  the usual complications of the picture-changing formalism are absent, because we  compute the correlation function at $\alpha=\eta=0$).  Rather as in section
\ref{simplesplit}, we can take $p,p'\to 0$ and replace $W^{\zeta;\beta}$ with $W^{'\zeta;\beta}=e^{\tphi/2}J_\ell 
\epsilon^{\beta\gamma}\Sigma_{\gamma,-}$.  After also evaluating the $bc$ and $\t b \t c$ correlation functions, 
and recalling that $S_\alpha=e^{-\tphi/2}\Sigma_{\alpha,+}$, we reduce to a two-point function
\begin{equation}\label{noxo}
\bigl\langle e^{-\tphi/2}\Sigma_{\alpha,+}(z)\cdot e^{\tphi/2}J_\ell \epsilon^{\beta\gamma}\Sigma_{\gamma,-}(0)\bigr\rangle.\end{equation}

This two-point function is very similar to the one that we encountered in eqn. (\ref{medolo}).  It is completely determined
by the operator product expansion, because the operator $e^{-\tphi/2}\Sigma_{\alpha,+}(z)$ varies  holomorphically with $z$.
The salient fact is the appearance in the OPE of the vertex operator $V_D$ of the $D$ auxiliary field:
\begin{equation}\label{oxo}e^{-\tphi/2}\Sigma_{\alpha,+}(z)\cdot e^{\tphi/2}J_\ell \epsilon^{\beta\gamma}\Sigma_{\gamma,-}(0)\sim\delta^\beta_\alpha V_D(0).
\end{equation}
As a result, the correlation function under study and hence also the matrix element for the supercurrent to create the gaugino
from the vacuum is proportional to $\langle V_D\rangle$, with a universal coefficient.

\vskip1cm

 \noindent {\it {Acknowledgements}}  Research supported in part by NSF Grant PHY-0969448. I thank K. Becker, P. Deligne,  R. Donagi, G. Moore,
 N. Seiberg,  and H. Verlinde  for useful advice and comments.

\bibliographystyle{unsrt}

\end{document}